\documentclass[preprint,3p,times,a4paper,oneside,onecolumn]{elsarticle}

\usepackage[utf8]{inputenc} 
\usepackage[T1]{fontenc} 
\usepackage{array}
\usepackage{hyperref} 
\usepackage{lineno}
\modulolinenumbers[10]
\usepackage{nomencl} 
\usepackage{framed} 
\usepackage{multicol}
\usepackage{etoolbox} 
\usepackage{amsmath} 
\usepackage{graphicx} 
\usepackage{tabularx}
\newcolumntype{Y}{>{\centering\arraybackslash}X}
\usepackage{multirow}
\usepackage{enumitem} 
\usepackage{subcaption}
\usepackage{bm} 
\usepackage{float} 
\usepackage{xcolor}
\usepackage{pdflscape}

\makenomenclature
\renewcommand*\nompreamble{\begin{multicols}{2}} 
\renewcommand*\nompostamble{\end{multicols}}

\newcommand{\Nafion}{Nafion$^{\mbox{\scriptsize{\textregistered}}}$}
\journal{Journal of The Electrochemical Society}
\title{A critical review of proton exchange membrane fuel cells matter transports and voltage polarisation for modelling}

\author[FEMTO,LIS]{Raphaël Gass \corref{mycorrespondingauthor}}
\ead{raphael.gass@femto-st.fr}
\cortext[mycorrespondingauthor]{Corresponding authors. \\ E-mail addresses: raphael.gass@femto-st.fr (R. Gass) and zhongliang.li@univ-fcomte.fr (Z. Li)}

\author[FEMTO]{Zhongliang Li \corref{mycorrespondingauthor}}
\ead{zhongliang.li@univ-fcomte.fr}
\author[LIS]{Rachid Outbib}
\author[FEMTO]{Samir Jemei}
\author[FEMTO,Institut]{Daniel Hissel}

\address[FEMTO]{Université de Franche-Comté, UTBM, CNRS, institut FEMTO-ST, FCLAB, Belfort, France}
\address[LIS]{Aix Marseille Univ, CNRS, LIS, Marseille, France}
\address[Institut]{Institut Universitaire de France}

\begin{document}

\begin{frontmatter}
	
	\begin{abstract}
		Technologies based on the use of hydrogen are promising for future energy requirements in a more sustainable world. Consequently, modelling fuel cells is crucial, for instance, to optimize their control to achieve excellent performance, to test new materials and configurations on a limited budget, or to consider their degradation for improved lifespan. To develop such models, a comprehensive study is required, encompassing both well-established and the latest governing laws on matter transport and voltage polarisation for Proton Exchange Membrane Fuel Cells (PEMFCs). Recent articles often rely on outdated or inappropriate equations, lacking clear explanations regarding their background. Indeed, inconsistent understanding of theoretical and experimental choices or model requirements hinders comprehension and contributes to the misuse of these equations. Additionally, specific researches are needed to construct more accurate models. This study aims to offer a comprehensive understanding of the current state-of-the-art in PEMFC modeling. It clarifies the corresponding governing equations, their usage conditions, and assumptions, thus serving as a foundation for future developments. The presented laws and equations are applicable in most multi-dimensional, dynamic, and two-phase PEMFC models.
	\end{abstract}
	
	\begin{keyword}
		Polymer electrolyte membrane fuel cell (PEMFC)\sep Modelling\sep Water management\sep Hydrogen transport\sep Oxygen transport\sep Voltage polarisation
	\end{keyword}
	
\end{frontmatter}


\renewcommand\nomgroup[1]{
	\item[\bfseries
	\ifstrequal{#1}{A}{Physical quantities}{
		\ifstrequal{#1}{B}{Mathematical symbols}{
			\ifstrequal{#1}{C}{Subscripts and superscripts}{
				\ifstrequal{#1}{D}{Abbreviation}{}}}}
	]}

\nomenclature[C]{\(H_{2}\)}{dihydrogen}
\nomenclature[C]{\(O_{2}\)}{dioxygen}
\nomenclature[C]{\(N_{2}\)}{dinitrogen}
\nomenclature[C]{\(v\)}{vapor}
\nomenclature[C]{\(liquid\)}{liquid}
\nomenclature[C]{\(mem\)}{membrane}
\nomenclature[C]{\(sorp\)}{sorption}
\nomenclature[C]{\(eq\)}{equilibrium}
\nomenclature[C]{\(prod\)}{production}
\nomenclature[C]{\(fc\)}{fuel cell}
\nomenclature[C]{\(cap\)}{capillarity}
\nomenclature[C]{\(sat\)}{saturated}
\nomenclature[C]{\(conv\)}{convective}
\nomenclature[C]{\(dif\)}{diffusion}
\nomenclature[C]{\(eff\)}{effective}
\nomenclature[C]{\(a\)}{anode}
\nomenclature[C]{\(c\)}{cathode}
\nomenclature[C]{\(in\)}{inlet}
\nomenclature[C]{\(out\)}{outlet}

\nomenclature[D]{\(CL\)}{catalyst layer}
\nomenclature[D]{\(ACL\)}{anode catalyst layer}
\nomenclature[D]{\(CCL\)}{cathode catalyst layer}
\nomenclature[D]{\(GC\)}{gas channel}
\nomenclature[D]{\(AGC\)}{anode gas channel}
\nomenclature[D]{\(CGC\)}{cathode gas channel}
\nomenclature[D]{\(GDL\)}{gas diffusion layer}
\nomenclature[D]{\(AGDL\)}{anode gas diffusion layer}
\nomenclature[D]{\(CGDL\)}{cathode gas diffusion layer}
\nomenclature[D]{\(EOD\)}{electro-osmotic drag}


\section{Introduction}

As a carbon-free, efficient, and widely applicable disruptive innovation, decarbonized hydrogen technologies have garnered increased attention \cite{FutureHydrogenSeizing2019, wangReviewPolymerElectrolyte2011}. In the twenty-first century, there is a race against time to limit global warming to 1.5°C above pre-industrial levels, as agreed upon in the Paris Agreement by 192 Parties in December 2015 \cite{ParisAgreementUnited}. Decarbonized hydrogen stands out as a highly promising candidate that can significantly impact various polluting sectors through relevant developments. For instance, it plays a crucial role in the decarbonization of industries like steel and fertilizer manufacturing, in electricity storage for the development of renewable energies, and in transportation electrification \cite{StrategieNationalePour2020}. As a result, the development of hydrogen technology has become a national priority for many countries \cite{FutureHydrogenSeizing2019}.

Currently, proton-exchange membrane fuel cell (PEMFC) stands as the most widely adopted technology in fuel cells \cite{dicksFuelCellSystems2018}. Nevertheless, it faces challenges such as low power densities, high costs, and limited lifespan, factors that have impeded its broader adoption in the global market \cite{jiaoDesigningNextGeneration2021}. Enhancing power density can be handled with more effective water management, especially at high current density, where insights into water activity within each cell is valuable for refining their control. Introducing novel materials into the stack could potentially reduce production costs. Improving the lifespan involves considering degradation processes at a mesoscopic scale. In this regard, and in many other situations as well, fuel cell modeling emerges as an invaluable tool. Indeed, models provide insights that sensors might be unable to capture, particularly in cells where their slim thickness hinders sensor integration — a crucial aspect for control purposes. Virtual exploration of numerous materials and configurations via models significantly reduces the costs and time associated with experimental trials. Predicting stack degradation and evaluating the stack's remaining useful life relies on understanding the underlying physical phenomena, aspects effectively addressed through modeling.

One of the prerequisites for developing such models is a thorough understanding of the physics governing water, hydrogen, and oxygen transport in the stack, as well as voltage polarization. While Jiao et al. \cite{jiaoWaterTransportPolymer2011}, O'Hayre et al. \cite{ohayreFuelCellFundamentals2016}, and Dicks et al. \cite{dicksFuelCellSystems2018} provided comprehensive reviews on matter transport and voltage polarization phenomena in 2011, 2016, and 2018, respectively, there is a need for a review that encompasses recent developments. Certain noteworthy and recent governing equations have been overlooked in these reviews and must be acknowledged. Indeed, many recent articles do not incorporate the latest propositions, even though they offer greater precision in results or increased algorithmic stability. Disruptive errors are also spreading in the literature and need to be identified. Additionally, a more in-depth explanation of the background concerning governing laws and equations is essential. Often, multiple equations model the same phenomenon. Therefore, it is beneficial to synthesize them in a single article, providing a means to differentiate their usage based on the physics, experimental choices, and modeling considerations. Furthermore, a synthetic compilation of key constant values found throughout the literature is considered to gain insights into commonly accepted values, points of disagreement, and poorly considered values.

\begin{figure*}[htbp]
	\centering
	\includegraphics[width=\linewidth]{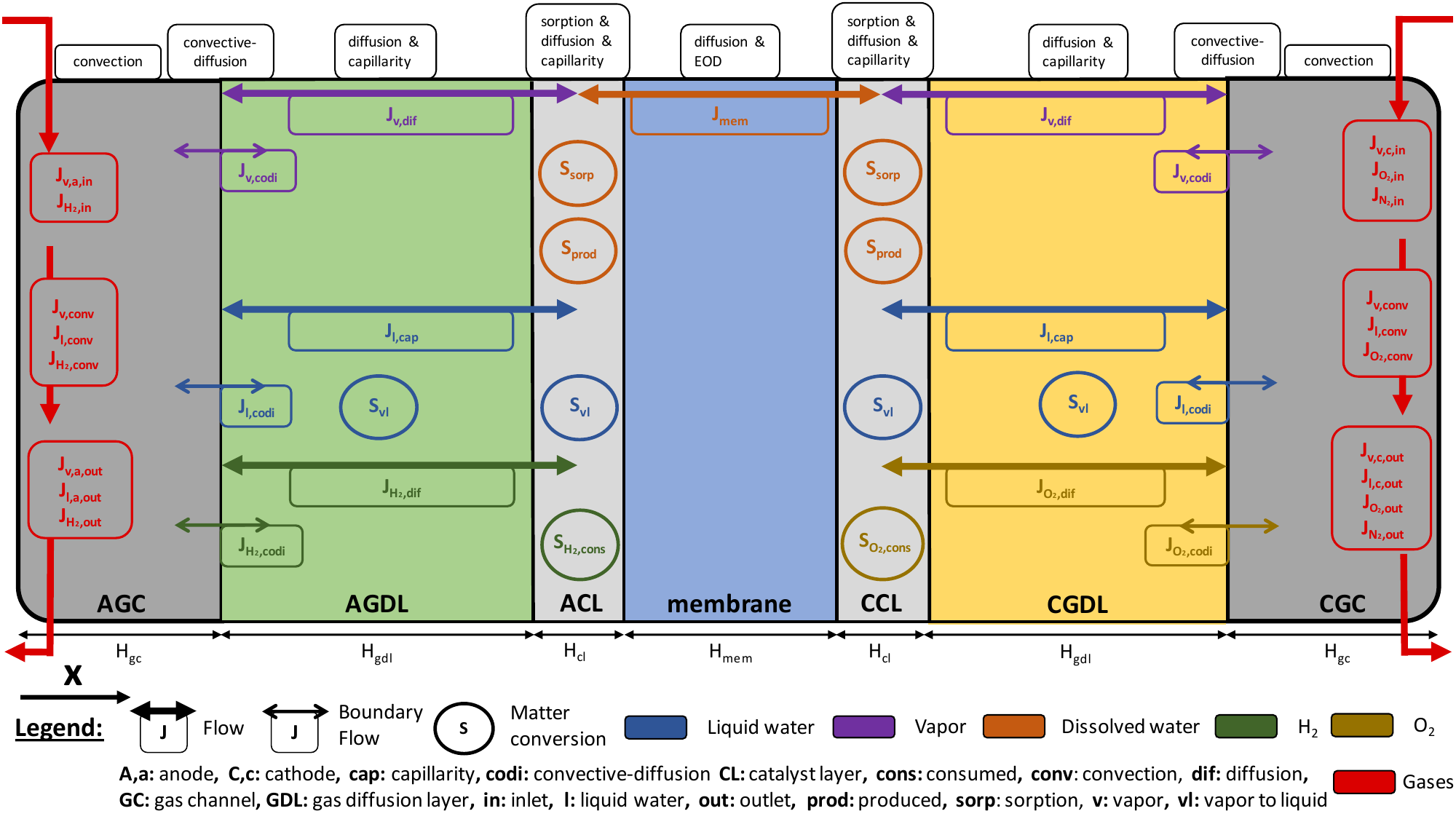}
	\caption{Schematic of a single PEMFC with the matter flows illustrated}
	\label{fig:matter_transport_in_PEMFC}
\end{figure*}

In this study, novel equations were introduced. Firstly, several expressions were combined to formulate new equations that exhibit enhanced stability, precision, or encompass a broader range of phenomena. Secondly, in an effort to simplify model generation and inspired by Pukrushpan et al.'s work \cite{pukrushpanControlOrientedModelingAnalysis2004}, we proposed simplified boundary conditions at the gas channel's inlet and outlet. These conditions aim to yield preliminary results prior to modeling auxiliary systems. Lastly, we suggested new paths for future research. All provided governing laws and equations are adaptable to any dynamic multi-dimensional two-phase model. Figure \ref{fig:matter_transport_in_PEMFC} illustrates the matter flows considered in this study within a cell. The directional flows align with the stack's thickness, facilitating graphical representation. It's worth noting that flows in other spatial dimensions are feasible, albeit less significant within a cell.

In this paper, transport of dissolved water within the \Nafion membrane is first examined. Subsequently, transport of liquid water in the catalyst layer (CL), gas diffusion layer (GDL), and gas channel (GC) is explored. Following this, vapor transport is delved, as well as the transport of hydrogen, oxygen, and nitrogen in all three regions. The study then addresses voltage polarization. In the appendices, additional equations are provided, constant values from the literature are synthesized, hypotheses considered in this study are outlined, and demonstrations are presented.

\section{Water transport in the membrane}
Protons can be efficiently transported across the membrane only when a sufficient amount of dissolved water is present within it. Indeed, state-of-the art membranes like \Nafion exhibit significant ionic conductivity exclusively when in a hydrated state. Therefore, membrane hydration must be considered in the models to ensure the proper functioning of the cell. 

Notably, in PEMFC studies, the active area has commonly served as the surface reference for various flows. Species evolve within different materials with distinct volume accessibility. Nonetheless, the active area, representing the surface of the MEA without the gasket, is a common surface shared by all materials, making it a reliable reference. Furthermore, it corresponds to the surface for defining current density. This choice of reference is valuable for accurately assessing molar transfer between elements, and consequently, the governing equations in this study incorporate constants such as porosity ($\varepsilon$) that enable to always consider the active area.

\subsection{Water content: $\lambda$}
\label{subsec:lambda}
In the \Nafion membrane, water is present in an unusual form. It is absorbed by the sulfonated side-chains ($-SO_{3}H$) in liquid phase \cite{dicksFuelCellSystems2018}. Thus, it is interesting to quantify it using the water content variable ($\lambda$) which corresponds to the number of water molecules per charged site $SO_{3}^{-}H^{+}$ in the membrane.
\nomenclature[A,2]{\(\lambda\)}{water content}

\begin{equation}
	\lambda \overset{\vartriangle}{=} \frac{n}{n_{SO_{3}^{-}H^{+}}}
\end{equation}
\nomenclature[A,1]{$n$}{number of moles $( mol )$ }
\nomenclature[B,3]{$\overset{\vartriangle}{=}$}{equality by definition}
where $n$ $( mol )$ is the number of moles of water, $n_{SO_{3}^{-}H^{+}}$ $( mol )$ is the number of moles of sulfonic acid group, and $\overset{\vartriangle}{=}$ refers to an equality by definition. 

Furthermore, $\lambda$ must be considered in the catalyst layer (CL). A thin layer of ionomer adheres to the catalyst metal particles \cite{dicksFuelCellSystems2018}, and consequently, a fraction of the CL volume is comprised of the electrolyte. This necessitates the use of $\varepsilon_{mc}$ in the governing equations, representing the ionomer volume fraction in the CL as defined in \eqref{eq:epsilon_mc}. It is then noteworthy to denote the location of the water content $\lambda_{mem}$ in the membrane or $\lambda_{cl}$ in the catalyst layer with an index. Although both are continuously linked, the governing equations differ, and this notation would simplify the writing of the differential equation for the dynamic behavior of $\lambda$. However, the omission of this index allows for a collective designation of both locations.

\begin{equation}
	\varepsilon_{mc} \overset{\vartriangle}{=} \frac{V_{\text{ionomer in CL}}}{V_{\text{CL}}}
	\label{eq:epsilon_mc}
\end{equation}
\nomenclature[A,2]{\(\varepsilon\)}{porosity}

\subsection{Schroeder's paradox}
\label{subsec:Schroeder_paradox}
In a Membrane Electrode Assembly (MEA), the membrane and catalyst layers are closely intertwined at their interface, allowing water to move between them. However, the paradox arised when considering the varying amounts of water absorption by the membrane, depending on whether the water is in a saturated vapour or in the pure liquid phase within the catalyst layer.	

In the liquid phase, the membrane absorbs a significantly higher quantity of water. This apparent contradiction was puzzling because, in theory up to this point in time, dissolved water in the membrane should reach an equilibrium with water activity, which remains constant at 1 for both saturated vapour and pure liquid water. Consequently, both equilibriums should have been identical. This perplexing phenomenon is known as Schroeder's paradox, named after the researcher who discovered it in 1903. It is prevalent in various polymers, including perfluorosulfonic acid (PFSA) polymer, exemplified by \Nafion.

\begin{figure} [H]
	\centering
	\includegraphics[width=14cm]{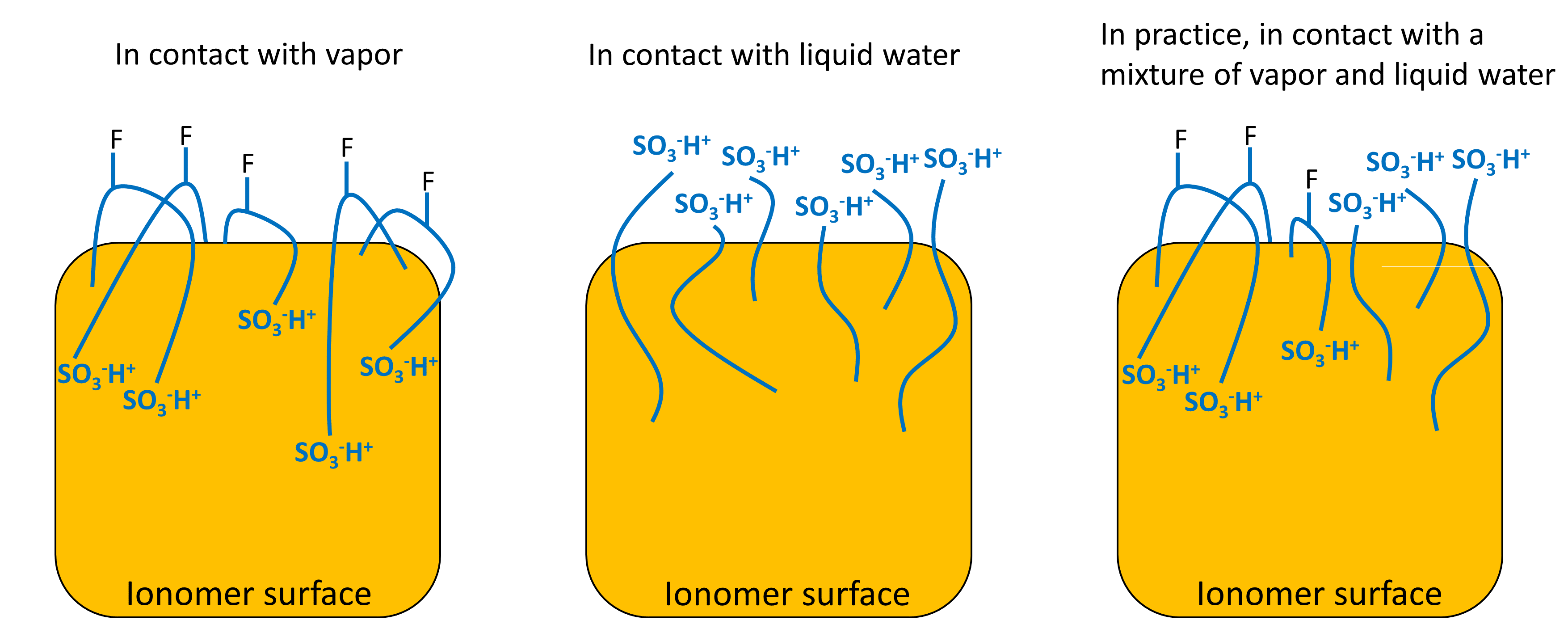}
	\caption{Illustration of PFSA membrane surface morphology when it is in contact with vapour and liquid water \cite{jiaoWaterTransportPolymer2011}.}
	\label{fig:Schroeder_membrane_surface_morphology}
\end{figure}

Although not fully understood, numerous studies have delved into this paradox. The presence of liquid water alters the morphology of the polymer, transitioning it from a strongly hydrophobic state to a hydrophilic one. In the case of liquid water, the hydrophilic sulfonated side-chains ($-SO_{3}H$) within \Nafion, initially located inside the material, can migrate to the membrane's surface, as depicted in Figure \ref{fig:Schroeder_membrane_surface_morphology}, facilitating the attraction and absorption of water.
However, the absorption of vapour entails an additional step, requiring vapour condensation at the CL ionomer interface \cite{jiaoWaterTransportPolymer2011}. This phenomenon allows the understanding of three additional phenomena. Firstly, less water is absorbed at equilibrium in the vapour phase: $\lambda_{l,eq} > \lambda_{v,eq}$, as discussed in section \ref{subsec:equilibrium_water_content}. Secondly, it is easier for water to exit the membrane (desorption) than to enter (absorption), affecting the associated time constants. Finally, the liquid water sorption flow into the membrane is much faster than that of water vapour. These latter points are addressed in section \ref{subsec:Water_flows_at_the_ionomer/CL_interface}.

While partially explained, this paradox continues to pose theoretical challenges for modeling, specifically when vapour and liquid water coexist, as stated in sections \ref{subsec:equilibrium_water_content}, \ref{subsec:water_activity}, \ref{subsec:another_way_lambda_eq} and \ref{subsec:Water_flows_at_the_ionomer/CL_interface}.

\subsection{Water flow in the membrane: $J_{mem}$}
\label{subsec:water_flow_membrane}
There are two dominant water transport mechanisms in the membrane: a diffusive flow and an electro-osmotic drag (EOD). The mathematical description of these two flows is initially expressed by the model of Weber and Newman \cite{weberTransportPolymerElectrolyteMembranes2004,futterPhysicalModelingPolymerelectrolyte2018} from the concentrated solution theory \cite{newmanElectrochemicalSystems3rd2004}. However, their formulations involve complex mathematical expressions, incorporating theoretical variables like the chemical potential of water ($\mu$), which are not practical for models at a mesoscopic scale. Consequently, their expressions have evolved into more functional forms while retaining all pertinent information. 

Diffusive flow is expressed as a Fick-like expression \cite{springerPolymerElectrolyteFuel1991}, using the gradient of $\lambda$ and an associated diffusion coefficient $D$, function of $\lambda$. 
EOD corresponds to the water molecule drag which is done by protons transport in the membrane. Protons travel in the membrane by hopping between adjacent water molecules (Grottus mechanism) or in the form of hydronium complexes $H_{3}O^{+}$ that cause them to drift (vehicle mechanism). Through this second phenomenon, protons carry water with them from the anode to the cathode \cite{jiaoWaterTransportPolymer2011,kulikovskyQuasi3DModelingWater2003}. Springer et al. assumed in 1991 \cite{springerPolymerElectrolyteFuel1991} that EOD is proportional to the current density and to the water content. Then, they found the corresponding constant, named EOD coefficient, and based on measurements in \Nafion 117. Their work, shown in the expression of $J_{mem}$ as \eqref{eq:water_flow_membrane}, has been extensively used in the literature \cite{jiaoWaterTransportPolymer2011,springerPolymerElectrolyteFuel1991,xuReduceddimensionDynamicModel2021,huAnalyticalCalculationEvaluation2016,xingNumericalAnalysisOptimum2015,yangMatchingWaterTemperature2011}. 

\begin{equation}
	\bm{J_{mem}} = \frac{2.5}{22} \frac{i_{fc}}{F} \lambda \text{ } \bm{\imath} - \frac{\rho_{mem}}{M_{eq}} D\left( \lambda\right)  \bm{\nabla \lambda}
	\label{eq:water_flow_membrane}
\end{equation}
\nomenclature[A,1]{$J$}{molar transfer flow $( mol.m^{-2}.s^{-1} ) $}
\nomenclature[A,1]{$i$}{current density per unit of cell active area $( A.m^{-2} ) $}
\nomenclature[A,1]{$F$}{Faraday constant $( C.mol^{-1} ) $}
\nomenclature[A,2]{\(\rho\)}{density $( kg.m^{-3} )$}
\nomenclature[A,1]{$M$}{molecular weight $( kg.mol^{-1} )$}
\nomenclature[A,1]{$D$}{diffusion coefficient of water in the membrane $( m^{2}.s^{-1} )$}
\nomenclature[A,1]{$x$}{space variable $( m )$}
\nomenclature[B,1]{$\bm{\imath}$}{unit vector along the x-axis}
where $J_{mem}$ $( mol.m^{-2}.s^{-1}) $ is the water flow in the membrane, $i_{fc}$ $( A.m^{-2} )$ is the current density of the fuel cell per unit of cell active area, $F$ $( C.mol^{-1} ) $ is the Faraday constant, $\rho_{mem}$ $( kg.m^{-3} )$ is the density of dry membrane, $M_{eq}$ $( kg.mol^{-1} )$ is the equivalent molar mass of ionomer calculated by its dry mass over the number of moles of $SO_{3}^{-}$, $D$ $( m^{2}.s^{-1} )$ is the diffusion coefficient of water in the membrane, and $\bm{\imath}$ is a unit vector along the x-axis, which is the space variable in the direction of the thickness of the cell, as shown in Figure \ref{fig:matter_transport_in_PEMFC}.

However, since 1991, significant enhancements in \Nafion membrane have been made \cite{karimiRecentApproachesImprove2019} and the EOD on these new membranes may differ. It would thus be interesting to reproduce the EOD calculation in order to obtain more accurate models. The literature provides an alternative formulation for the EOD \cite{jiaoWaterTransportPolymer2011}, although it is less employed and equally outdated. It is expressed as \eqref{eq:EOD_flow}.

\begin{equation}
	\bm{J_{EOD}} = 
	\begin{cases}
		\frac{i_{f c}}{F} \text{ }\bm{\imath}, & \lambda \leq 14 \\ 
		\left[ 0.1875 \lambda-1.625\right]  \frac{i_{f c}}{F} \text{ } \bm{\imath}, & \lambda>14
	\end{cases}	
	\label{eq:EOD_flow}
\end{equation}
where $J_{EOD}$ $( mol.m^{-2}.s^{-1} )$ is the EOD flow of water in the membrane.

Models other than that of Springer have been sparingly used for the flow of water through the membrane. They are mentioned in Dickinson et al. work \cite{dickinsonModellingProtonConductiveMembrane2020}.

\subsection{Diffusion coefficient: D($\lambda$)}
\begin{figure}[H]
	\centering
	\includegraphics[width=12cm]{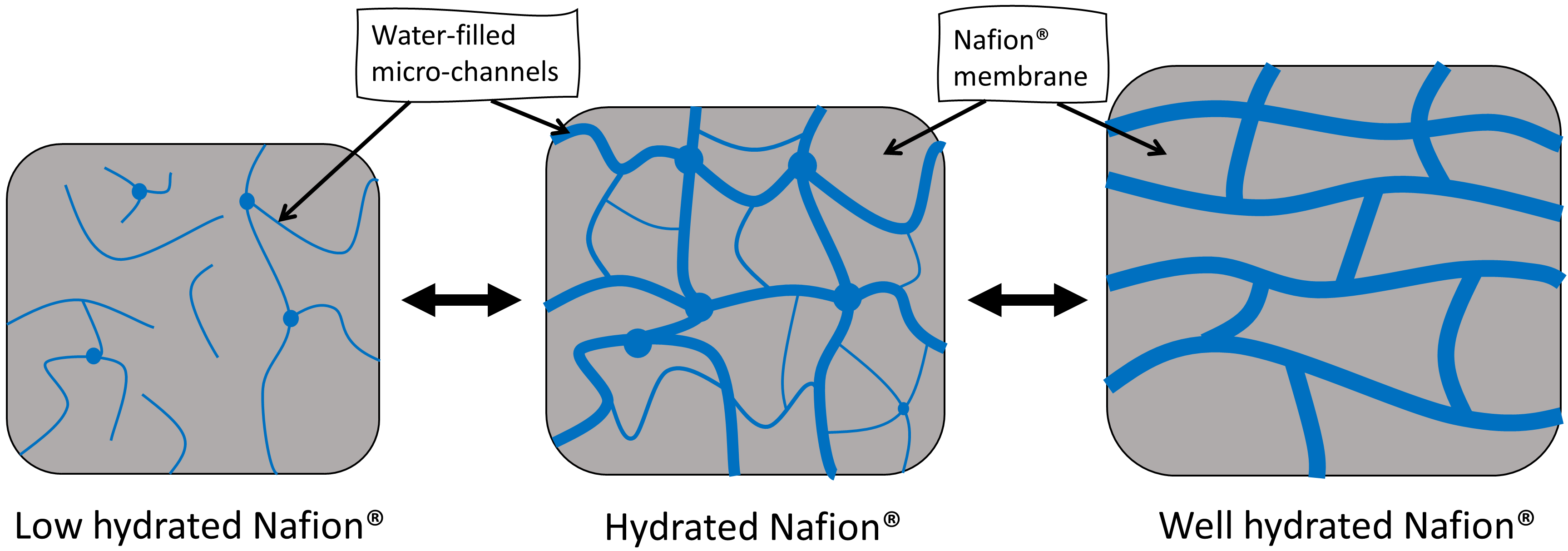}
	\caption{Illustration of PFSA membrane morphology at different levels of hydration.}
	\label{fig:membrane_microchannels}
\end{figure}

The amount of water dissolved in the membrane significantly influences its diffusion coefficient. When the membrane is adequately hydrated, the polymer backbone molecules form water-filled micro-channels, with $SO_{3}^{-}$ groups attached to their walls. Depending on the water content, the membrane exhibits varying numbers of water channels, mean radii, and forms \cite{kulikovskyQuasi3DModelingWater2003}, as shown in Figure \ref{fig:membrane_microchannels}. These structural characteristics directly impact water diffusion, benefiting from higher hydration levels through larger channels, reduced tortuosity, and diminished friction. Consequently, it is essential to account for this dependency when considering the diffusion coefficient, which cannot remain constant but should be expressed as a function of  $\lambda$.

The diffusivity of dissolved water in the electrolyte is typically determined by fitting experimental data. The two commonly used expressions are based on Zawodzinski’s data from 1991 \cite{zawodzinskiDeterminationWaterDiffusion1991}. The first expression, expressed as \eqref{eq:coef_dif_Springer}, was introduced by Springer et al. in 1991 \cite{jiaoWaterTransportPolymer2011,springerPolymerElectrolyteFuel1991}. The second expression, expressed as \eqref{eq:coef_dif_Motupally}, was presented by Motupally et al. in 2000 \cite{jiaoWaterTransportPolymer2011,xuReduceddimensionDynamicModel2021,yangMatchingWaterTemperature2011,motupallyDiffusionWaterNafion2000,wangModelingEffectsCapillary2008,fanCharacteristicsPEMFCOperating2017}. 

\begin{figure}[H]
	\begin{equation}
		D \left( \lambda \right)  = 
		\begin{cases}
			2.692661843 \times 10^{-10}, & \lambda \leq 2 \\ 
			10^{-10} e^{2416\left[\frac{1}{303}-\frac{1}{T_{fc}}\right]} \left[ 0.87\left[ 3-\lambda\right] +2.95\left[ \lambda-2\right] \right] , & 2 < \lambda \leq 3 \\
			10^{-10} e^{2416\left[\frac{1}{303}-\frac{1}{T_{fc}}\right]} \left[ 2.95\left[ 4-\lambda\right] +1.642454\left[ \lambda-3\right] \right] , & 3 < \lambda \leq 4 \\	
			10^{-10} e^{2416\left[\frac{1}{303}-\frac{1}{T_{f c}}\right]} \left[2.563 - 0.33\lambda + 0.0264\lambda^{2} - 0.000671\lambda^{3}\right], & 4 < \lambda < 17					
		\end{cases}	
		\label{eq:coef_dif_Springer}
	\end{equation}
\end{figure}

\begin{equation}
	D \left( \lambda \right)  = 
	\begin{cases}
		3.1 \times 10^{-7}\lambda \left[  e^{0.28\lambda} - 1 \right]  e^{-\frac{2436}{T_{fc}}}, & \lambda < 3 \\
		4.17 \times 10^{-8}\lambda \left[ 161 e^{-\lambda} + 1 \right]  e^{-\frac{2436}{T_{fc}}}, & 3 \leq \lambda < 17
	\end{cases}	
	\label{eq:coef_dif_Motupally}
\end{equation}
\nomenclature[A,1]{$T_{fc}$}{fuel cell temperature $( K )$}
where $T_{fc}$ $( K )$ is the fuel cell temperature. 
Notably, some recent papers have applied an inversion of the coefficient 2436 in the exponential \cite{jiaoWaterTransportPolymer2011,xuReduceddimensionDynamicModel2021,fanCharacteristicsPEMFCOperating2017}. Additionally, both of these expressions do not account for water content values greater than 17. The suitability of these relationships for higher $\lambda$ values is therefore not guaranteed.

\begin{figure}[H]
	\centering
	\includegraphics[width=10cm]{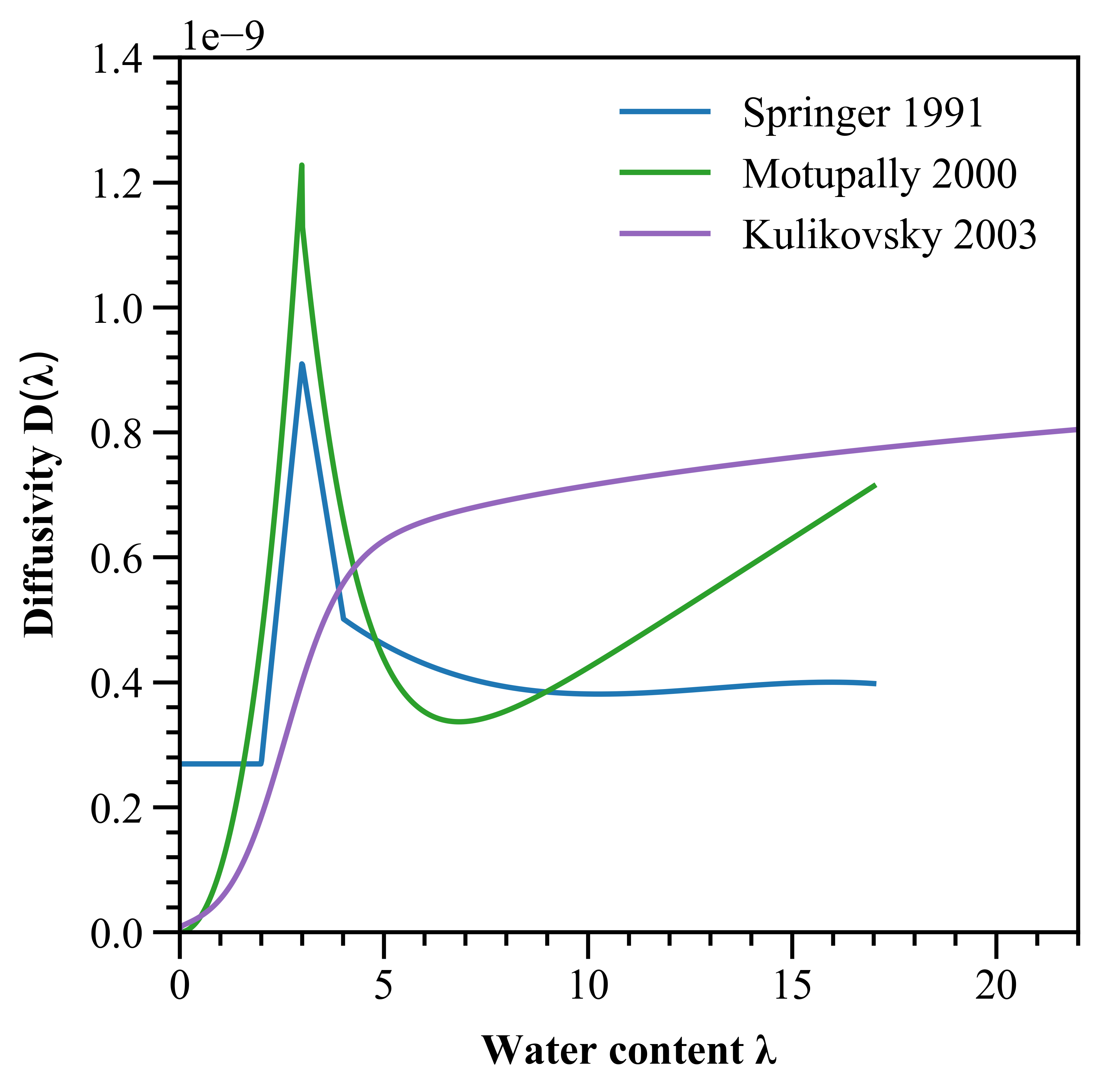}
	\caption{Comparison of the three expressions of the diffusivity coefficient of the membrane at 80°C}
	\label{fig:coef_dif}
\end{figure}

Upon examining figure \ref{fig:coef_dif}, it becomes evident that the difference between the two correlations is non-negligible. However, no conclusive evidence points to one being more accurate than the other. Both correlations have found widespread use in PEMFC modeling \cite{jiaoWaterTransportPolymer2011,springerPolymerElectrolyteFuel1991,motupallyDiffusionWaterNafion2000}. However, the abrupt change in the diffusion coefficient poses challenges for numerical simulations. This peak arises from a correction procedure, which involves differentiating experimental data and may not be consistent with reality. Furthermore, the kinetics of channel formation within \Nafion membranes during water uptake, as well as the geometry of channels, remain insufficiently understood. However, as previously discussed, it is reasonable to assume that lower water content corresponds to a reduced mean pore radius in the membrane, hindering water diffusion \cite{kulikovskyQuasi3DModelingWater2003}. Consequently, there should be no peak, and diffusivity must be a growing function of $\lambda$.

In 1998, Van Bussel et al. \cite{vanbusselDynamicModelSolid1998} conducted measurements that validated these physical considerations. Subsequently, Kulikovsky et al. fitted these values in 2003 and proposed a function expressed as \eqref{eq:coef_dif_Kulikovsky}\cite{kulikovskyQuasi3DModelingWater2003,wangModelingEffectsCapillary2008,fanCharacteristicsPEMFCOperating2017}. This equation appears to be more representative of the underlying physical phenomena. However, it has some drawbacks: the measurements were performed using outdated membranes, similar to the two previous equations \cite{karimiRecentApproachesImprove2019}. Additionally, the temperature dependency was lost in this expression, which was fitted with data at 80°C.

\begin{equation}
	\boxed{
		D \left( \lambda \right) = 4.1 \times 10^{-10} \left[  \frac{\lambda}{25.0} \right] ^{0.15} \left[  1.0 + \tanh\left( \frac{\lambda-2.5}{1.4} \right)\right] 
		\label{eq:coef_dif_Kulikovsky}
	}
\end{equation}

Figure \ref{fig:coef_dif} compares the three proposed equations for the diffusion coefficient at 80°C. 
The authors recommend using the Kulikovsky model \cite{kulikovskyQuasi3DModelingWater2003}, although further improvements could be made by incorporating temperature considerations and utilizing modern membranes in the measurements.

\subsection{Equilibrium water content of the membrane: $\lambda_{eq}$ - an overview}
\label{subsec:equilibrium_water_content}

To calculate the water sorption ($S_{\text{sorp}}$) between the membrane and the catalyst layer, it is essential to first comprehend the equilibrium water content of the membrane ($\lambda_{\text{eq}}$) and the water activity ($a_{\text{w}}$). However, the sorption process is complex, and the existing literature appears incomplete. An overview on $\lambda_{eq}$ in this section, on $a_{w}$ section \ref{subsec:water_activity} and on $S_{sorp}$ section \ref{subsec:Water_flows_at_the_ionomer/CL_interface} are first presented based on the current litterature. Further, section \ref{subsec:another_way_lambda_eq} presents new proposals.

One effective method of quantifying the exchange flow between the membrane and the catalyst layer involves comparing the current water content in the catalyst layer, $\lambda$, with its equilibrium value, $\lambda_{eq}$. The equilibrium water content of the membrane $\lambda_{eq}$ is a variable that is experimentally accessible, and is a function of the water activity $a_{w}$. Subsequently, $\lambda_{v,eq}$ must be differentiated from $\lambda_{l,eq}$. Here, $\lambda_{v,eq}$ refers to an equilibrium of the dissolved water with vapour, which has a certain activity $a_{w}$, whereas $\lambda_{l,eq}$ refers to an equilibrium with pure liquid water. The difference between $\lambda_{l,eq}$ and $\lambda_{v,eq}$ with saturated vapour is noticeable and is referred to as the Schroeder's paradox, as discussed in Section \ref{subsec:Schroeder_paradox}.
\nomenclature[A,1]{$a_{w}$}{water activity in the pores of the CL}

Two experiments widely accepted in the scientific community have been conducted to derive an equation for $\lambda_{eq}$ from experimental data. The first experiment, proposed by Springer et al. \cite{springerPolymerElectrolyteFuel1991}, utilized data provided by Zawodzinski et al. in 1991 \cite{zawodzinskiDeterminationWaterDiffusion1991}. This experiment focused exclusively on a \Nafion 117 membrane, with measurements taken at 30°C for $\lambda_{v,eq}$ and 80°C for $\lambda_{l,eq}$. The equation is expressed as \eqref{eq:lambda_eq_Springer} \cite{jiaoWaterTransportPolymer2011, springerPolymerElectrolyteFuel1991, xingNumericalAnalysisOptimum2015, yangMatchingWaterTemperature2011, zawodzinskiDeterminationWaterDiffusion1991, wangModelingEffectsCapillary2008, fanCharacteristicsPEMFCOperating2017, pasaogullariTwoPhaseModelingFlooding2005, pukrushpanControlOrientedModelingAnalysis2004}.

The second experiment, proposed by Hinatsu et al. in 1994 \cite{hinatsuWaterUptakePerfluorosulfonic1994}, is expressed as \eqref{eq:lambda_eq_Hinatsu} \cite{motupallyDiffusionWaterNafion2000, dickinsonModellingProtonConductiveMembrane2020, geAbsorptionDesorptionTransport2005, hinatsuWaterUptakePerfluorosulfonic1994}. This experiment provided an equation for $\lambda_{v,eq}$ that accurately fits data from various membrane types, including \Nafion 115, \Nafion 117, AC-12, and FL-12. It was conducted at the standard operating temperature of 80°C for PEMFCs. For $\lambda_{l,eq}$, the results varied based on the membrane type. It is the expression for \Nafion 117 which is there considered. Additionally, a temperature dependency was incorporated into the equation. Considering that the experimental conditions proposed by Hinatsu et al. are more realistic than those of Springer et al., their results are preferable.

\begin{figure} [H]
	\centering
	\includegraphics[width=10cm]{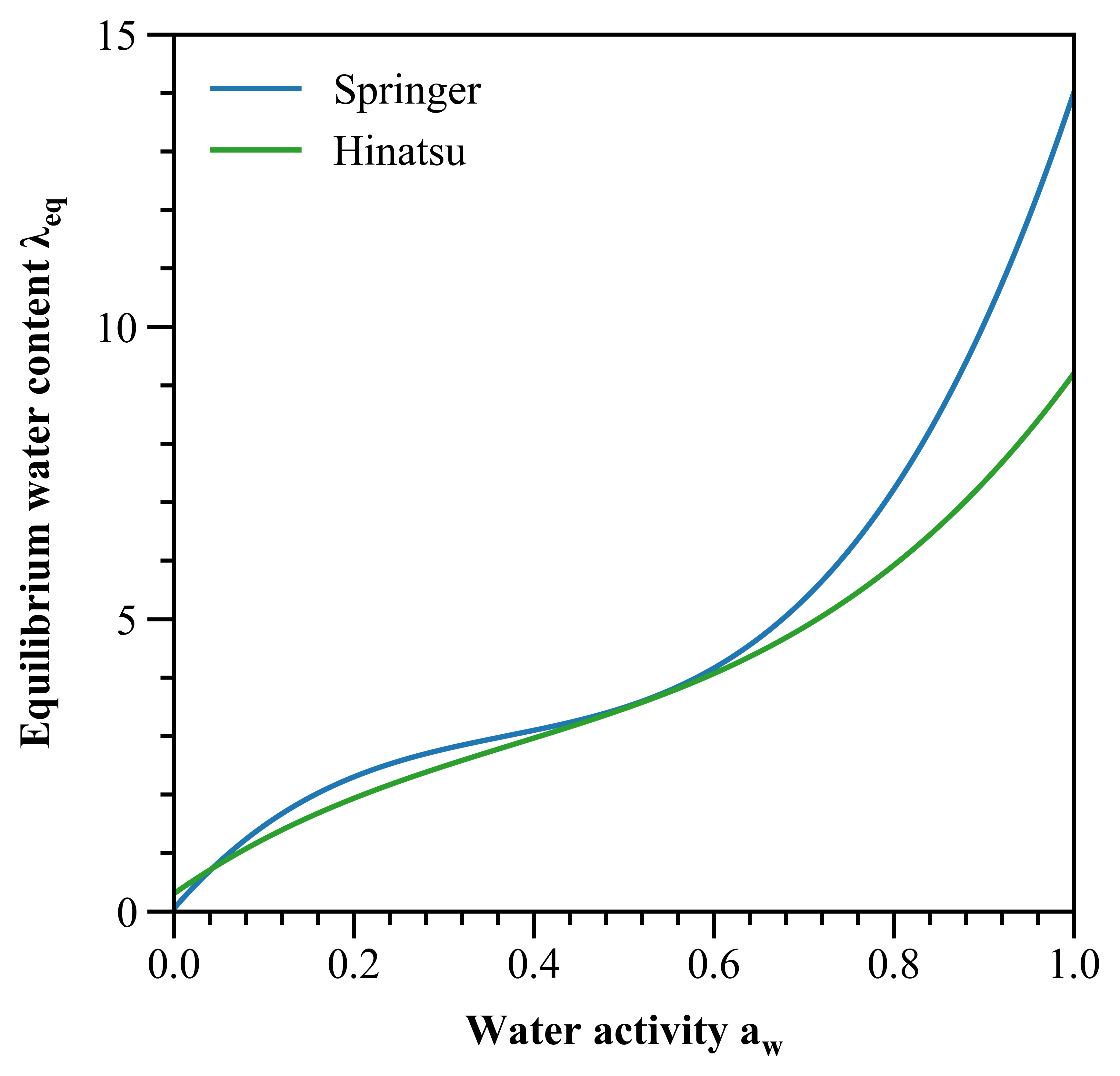}
	\caption{Comparison between Springer’s and Hinatsu's expressions for $\lambda_{v,eq}$.}
	\label{fig:lambda_eq_Hinatsu_Springer}
\end{figure}

Zawodzinski et al. also conducted experiments at 80°C in 1993 \cite{zawodzinskijrCharacterizationPolymerElectrolytes1993}, with the results expressed as an equation by Ye in 2007 \cite{yeThreeDimensionalSimulationLiquid2007}. However, it is the study by Hinatsu et al. which gained the widespread acceptance. Although these measurements are considered outdated for modern models due to advancements in membrane structures and experimental protocols over the last decade \cite{karimiRecentApproachesImprove2019}, they are still widely used. A graphical comparison of these expressions for $\lambda_{v,eq}$ is presented in figure \ref{fig:lambda_eq_Hinatsu_Springer}.

\begin{equation} 
	\lambda_{eq}^{Springer} = 
	\begin{cases}
		\lambda_{v,eq} \left( a_{w} \right) = 0.043 + 17.81 a_{w} - 39.85 a_{w}^{2} + 36.0 a_{w}^{3}, & a_{w} \in \left[ 0,1 \right]\\ 
		\lambda_{l,eq} = 16.8 
	\end{cases}	
	\label{eq:lambda_eq_Springer}
\end{equation}

\begin{equation} 
	\lambda_{eq}^{Hinatsu} = 
	\begin{cases}
		\lambda_{v,eq} \left( a_{w} \right) = 0.300 + 10.8a_{w} - 16.0a_{w}^{2} + 14.1a_{w}^{3}, & a_{w} \in \left[ 0,1 \right]\\ 
		\lambda_{l,eq} = 10.0 + 1.84 \cdot 10^{-2} T_{fc} + 9.90 \cdot 10^{-4} T_{fc}^{2}, & \text{$T_{fc}$ in °C here}
	\end{cases}	
	\label{eq:lambda_eq_Hinatsu}
\end{equation}

Subsequently, numerous researchers have attempted to combine $\lambda_{v,eq}$ and $\lambda_{l,eq}$ into a single quantity, $\lambda_{eq}$. This choice is questioned section \ref{subsec:another_way_lambda_eq}. To establish this connection, Springer et al. initially posited that their formulation for $\lambda_{v,eq}$ remains applicable at 80°C, despite the experimental measurements being conducted at 30°C \cite{springerPolymerElectrolyteFuel1991}. This hypothesis is not necessary for the expression of Hinatsu. Then, Springer et al. correlated the water activity $a_{w}$ with both vapour and liquid phases, although this is uncommon. For $a_{w} \in \left[ 0,1 \right]$, only vapour was present, and for $a_{w} > 1$, liquid water coexisted with saturated vapour. Subsequently, they considered an activity value significantly exceeding 1 and concluded that $a_{w} = 3$ is a suitable value, resulting in the exclusive presence of pure liquid water occupying the entire cavity volume of the triple points zone. The arbitrary selection of $a_{w} = 3$ and the newly formulated expression for $a_{w}$ associated with both vapour and liquid water were left unexplained in their study, posing challenges to the understanding of their model. However, these issues did not impede its prevalence in the literature, and an adapted expression for $a_{w}$ has consequently been introduced, as discussed in section \ref{subsec:water_activity}. 
Finally, a linear expression was arbitrarily employed to connect $\lambda_{v,eq}$ at $a_{w} = 1$ to $\lambda_{l,eq}$ at $a_{w} = 3$, as indicated in \eqref{eq:lambda_eq_Springer_linear} \cite{jiaoWaterTransportPolymer2011,springerPolymerElectrolyteFuel1991,xingNumericalAnalysisOptimum2015,yangMatchingWaterTemperature2011,zawodzinskiDeterminationWaterDiffusion1991,wangModelingEffectsCapillary2008,fanCharacteristicsPEMFCOperating2017,pasaogullariTwoPhaseModelingFlooding2005,pukrushpanControlOrientedModelingAnalysis2004}. In this model, the existence of $a_{w} \geq 3$ is deemed either improbable or impossible; hence, the value of 16.8 is retained for higher $a_{w}$, or higher values should not be considered \cite{pasaogullariTwoPhaseModelingFlooding2005}. Nonetheless, providing precise rules is challenging, given the incomplete and subjective nature of this framework. Nevertheless, this model persists in subsequent studies. The expression by Hinatsu et al. has been adjusted in the same manner at 80°C, resulting in \eqref{eq:lambda_eq_Hinatsu_linear}.

\begin{equation} 
	\lambda_{eq}^{Springer} \left( a_{w} \right) = 
	\begin{cases}
		0.043 + 17.81 a_{w} - 39.85 a_{w}^{2} + 36.0 a_{w}^{3}, & a_{w} \in \left[ 0,1 \right]\\ 
		14 + 1.4 \left[  a_{w} - 1 \right] , & a_{w} \in \left] 1,3 \right]
	\end{cases}	
	\label{eq:lambda_eq_Springer_linear}
\end{equation}

\begin{equation} 
	\lambda_{eq}^{Hinatsu} \left( a_{w} \right) = 
	\begin{cases}
		0.300 + 10.8a_{w} - 16.0a_{w}^{2} + 14.1a_{w}^{3}, & a_{w} \in \left[ 0,1 \right]\\ 
		9.2 + 4.3 \left[  a_{w} - 1 \right] , & a_{w} \in \left] 1,3 \right]
	\end{cases}	
	\label{eq:lambda_eq_Hinatsu_linear}
\end{equation}

As evident, these equations are constructed in two parts, resulting in stiffness when $a_{w} = 1$. Consequently, oscillations occur during the implementation of the models. To enhance the numerical results, Bao et al. replaced the linear expression for $a_{w} \in \left] 1,3 \right]$ by Springer et al., presenting a unique and general equation for all $a_{w}$ values \cite{baoTwodimensionalModelingPolymer2015}. The linear relationship between $\lambda_{eq} \left( a_{w} = 1 \right)$ and $\lambda_{eq} \left( a_{w} = 3 \right)$ being arbitrarily established, there appear to be no immediate obstacles to making this change. However, Bao et al. interpreted Springer's work differently, suggesting a discontinuity at $a_{w} = 3$, where $\lambda_{eq}$ increases from 16.8 to 22 \cite{baoTwodimensionalModelingPolymer2015}. In the authors analysis, the value of 22 is provided for experiments conducted at 100°C, whereas 16.8 is evaluated at 80°C. Only one of them must be chosen depending on the working temperature, which is 80°C in this case. Thus, the authors slightly modified the expression proposed in \cite{baoTwodimensionalModelingPolymer2015} to obtain a more adapted equation, as expressed in \eqref{eq:lambda_eq_Springer_Bao}.

\begin{figure}[H]
	\centering
	\begin{equation}
		\begin{aligned}
			\lambda_{eq} & = 
			\frac{1}{2} \left[  0.043 + 17.81a_{w} - 39.85 a_{w}^{2} + 36.0a_{w}^{3} \right]  \cdot \left[  1 - \tanh\left( 100 \left[  a_{w} - 1 \right]\right)\right] \\
			& + \frac{1}{2} \left[  14 +2.8 \left[  1 - \exp\left( -K_{\text{shape}} \left[  a_{w} - 1        \right]\right)\right]\right] \cdot \left[ 1 + \tanh \left( 100 \left[ a_{w} - 1 \right]\right)\right]
		\end{aligned}
		\label{eq:lambda_eq_Springer_Bao}
	\end{equation}
\end{figure}

\begin{figure}[H]
	\centering
	\includegraphics[width=10cm]{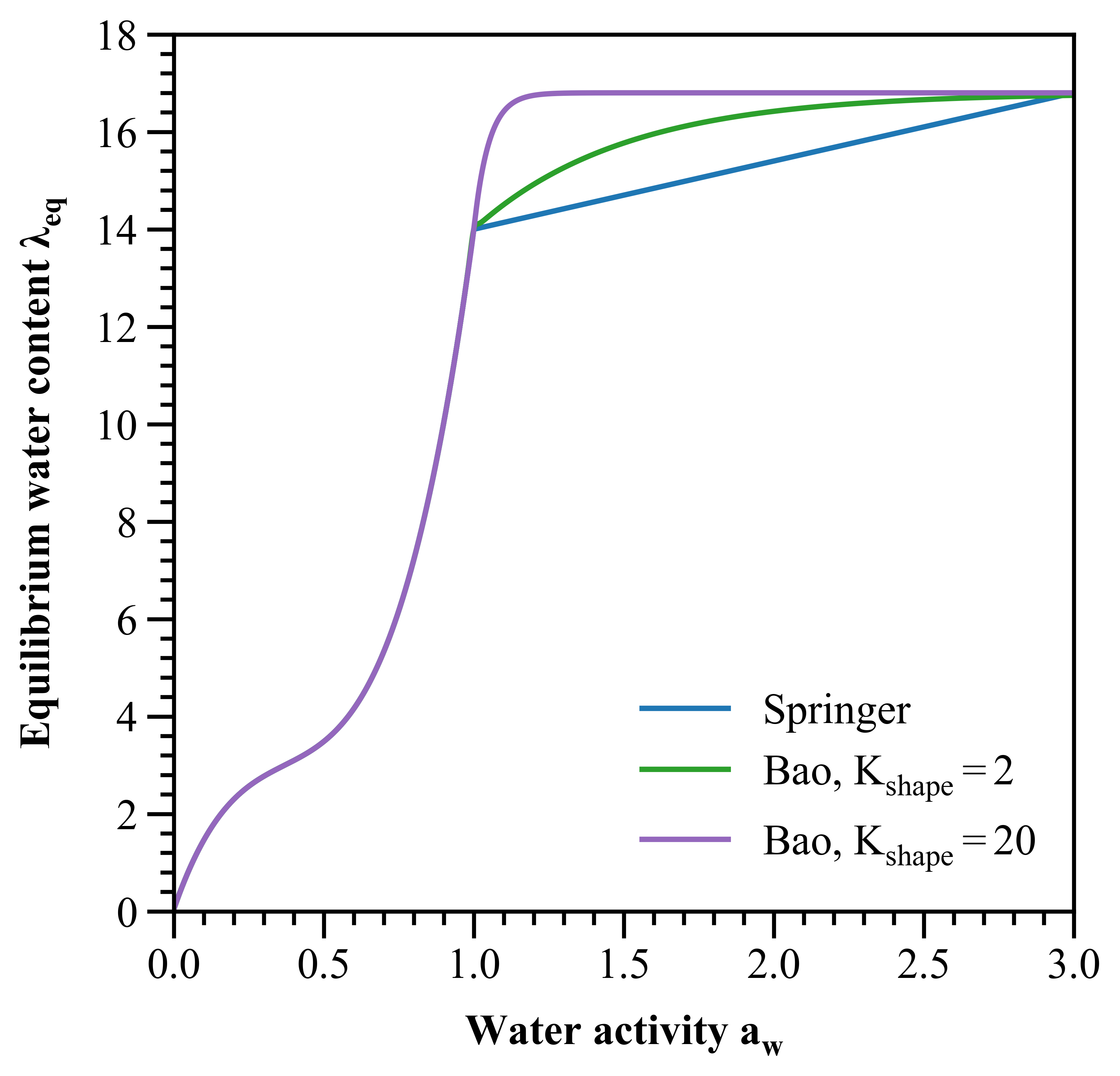}
	\caption{Comparison between Springer's piece wise expression and Bao's function for water content equilibrium with different values of $K_{shape}$}
	\label{fig:lambda_eq_Springer_Bao}
\end{figure}

Various values of the mathematical factor $K_{\text{shape}}$ enable the experimenter to depict either a smooth transition ($K_{\text{shape}} = 2$) or a sharp jump ($K_{\text{shape}} = 20$) between the "two ends of Schroeder's paradox" \cite{baoTwodimensionalModelingPolymer2015}, as illustrated in figure \ref{fig:lambda_eq_Springer_Bao}. It is crucial to emphasize that this is formulated for modeling purposes only, and the physics considerations are disregarded at this point. This study recommends employing a small $K_{\text{shape}}$, such as $K_{\text{shape}} = 2$, to ensure that the model does not deviate excessively from the involved physics. At this stage, it becomes evident that a more robust theory on equilibrium water content needs formulation for enhanced utilization of $\lambda_{v, \text{eq}}$ and $\lambda_{l, \text{eq}}$.
\nomenclature[B,1]{$K_{shape}$}{shape mathematical factor}

\begin{figure} [H]
	\centering
	\includegraphics[width=10cm]{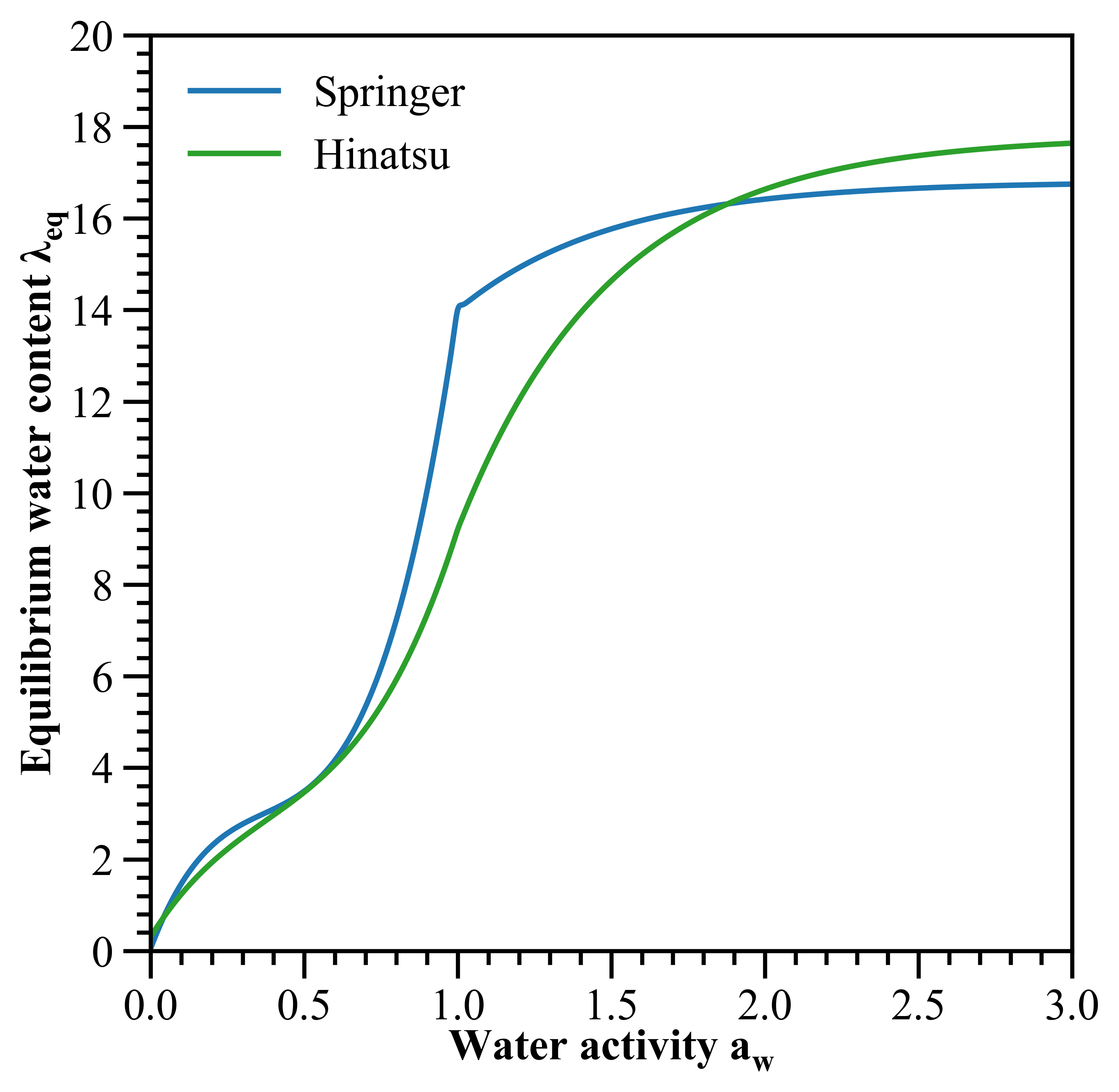}
	\caption{Comparison between Springer's and Hinatsu's expressions for $\lambda_{eq}$ at 80°C, using Bao's form.}
	\label{fig:lambda_eq_Hinatsu_Springer_Bao}
\end{figure}

Finally, the expression for $\lambda_{eq}$ proposed in this study is expressed as \eqref{eq:lambda_eq_Hinatsu_Bao}. It is derived based on Hinatsu's equations at 80°C, following the form suggested by Bao et al. This expression is compared with the one of Springer et al. in figure \ref{fig:lambda_eq_Hinatsu_Springer_Bao}.

\begin{figure}[H]
	\begin{equation}
		\boxed{
			\begin{aligned}
				\lambda_{eq} & = 
				\frac{1}{2} \left[  0.300 + 10.8a_{w} - 16.0a_{w}^{2} + 14.1a_{w}^{3} \right] \cdot \left[  1 - \tanh\left( 100 \left[ a_{w} - 1 \right]\right)\right] \\
				& + \frac{1}{2} \left[ 9.2 +8.6 \left[ 1 - \exp\left( -K_{\text{shape}} \left[ a_{w} - 1        \right]\right)\right]\right] \cdot \left[ 1 + \tanh \left( 100 \left[ a_{w} - 1 \right]\right)\right]
			\end{aligned}
		}
		\label{eq:lambda_eq_Hinatsu_Bao}
	\end{equation}
\end{figure}

\subsection{Water activity: $a_{w}$ - an overview}
\label{subsec:water_activity}

The water activity, denoted as $a_{w}$, quantifies water's ability to humidify the membrane within the catalyst layer. According to Schroeder's paradox, the more condensed the water, the more favorable the humidification. In the literature, two commonly accepted definitions, \eqref{eq:water_activity_Springer} and \eqref{eq:water_activity_simplified}, exist. However, the use of one against the other has never been explained, to the best of the authors knowledge. This section aims to provide an explanation for both of these definitions. Additionally, Section \ref{subsec:another_way_lambda_eq} introduces a novel approach.

The original definition, as expressed in \eqref{eq:water_activity_Springer}, was initially proposed by Springer et al., \cite{springerPolymerElectrolyteFuel1991} and pursued by Ge et al. \cite{geAbsorptionDesorptionTransport2005}. When considering only vapour, $a_{w}$ is equivalent to water humidity, a common consideration across various physics disciplines. However, it's noteworthy that $a_{w}$ can surpass 1 and extend up to 3, indicating a mixture of vapour and liquid water. The value $a_{w}=3$ specifically denotes pure liquid water.

\begin{equation}
	a_{w} \overset{\vartriangle}{=}
	\begin{cases}
		\frac{P}{P_{sat}} \underbrace{=}_{\text{ideal gas law}} \frac{C}{C_{sat}}, &\text{for pure vapour} \\
		3, &\text{for pure liquid water}
	\end{cases}	
	\label{eq:water_activity_Springer}
\end{equation}
\nomenclature[A,1]{$P$}{pressure $(Pa)$}
\nomenclature[A,1]{$C$}{molar concentration $( mol.m^{-3} )$}
where $P$ $( Pa )$ is the vapour pressure, $P_{sat}$ $( Pa )$ is the vapour saturated pressure, $C$ $( mol.m^{-3})$ is the vapour concentration, and $C_{sat}$ $( mol.m^{-3} )$ is the vapour saturated concentration.

The drawback of this model lies in the absence of an explanation regarding the characterization of $a_{w}$ when its values exceed 1. It is implicitly assumed that the expression $\frac{P}{P_{sat}}$ is sufficient, but this assumption is questionable. When vapour is fully saturated at $a_{w} = 1$, condensation initiates, causing the vapour pressure to decrease and giving rise to liquid water. However, vapour continues to be supplied through the chemical reaction of hydrogen and oxygen. As condensation is not instantaneous, the input of vapour can eventually counterbalance condensation, leading to a pressure increase, resulting in $a_{w} > 1$. Nonetheless, the higher the $P > P_{sat}$, the more significant the amount of condensation (as discussed in section \ref{subsec:water_phase_change_rate}). Consequently, it is improbable for $a_{w}$ to increase substantially. Springer et al. considered an arbitrary limit of 3 for $a_{w}$, corresponding to the water activity of pure liquid water. However, achieving this value is unlikely when considering only vapour in water activity $a_{w}$. Thus, while it is acceptable for $\frac{P}{P_{sat}} > 1$, it is unlikely to increase beyond 3.

In the literature, one extrapolation is widely used to incorporate both vapour and liquid water into $a_w$ \cite{jiaoWaterTransportPolymer2011, xingNumericalAnalysisOptimum2015, yangMatchingWaterTemperature2011, fanCharacteristicsPEMFCOperating2017}. This variable is reconstructed following specific rules. Initially, $a_{w} \in \left[ 0,3 \right]$. Subsequently, $a_{w} \in \left[ 0,1 \right]$ represents a vapour phase in the catalyst layer, while $a_{w} \in \left[ 1,3 \right]$ indicates a mixture of vapour and liquid water. Finally, the value of $a_{w}$ increases with the condensation of water. One mathematical expression that fulfills all these criteria is given by \eqref{eq:water_activity_simplified}. However, it is not the sole mathematical approach meeting these conditions, and there is no evidence to suggest that \eqref{eq:water_activity_simplified} is the most accurate model for fuel cell modelling.

\begin{equation}
	a_{w} \overset{\vartriangle}{=} \frac{P}{P_{sat}} + 2 \texttt{s} \underbrace{=}_{\text{ideal gas law}} \frac{C}{C_{sat}} + 2 \texttt{s}
	\label{eq:water_activity_simplified}
\end{equation}
\nomenclature[A,2]{$\texttt{s}$}{liquid water saturation}
where $\texttt{s}$ is the liquid water saturation (explained in \ref{subsec:liquid_water_saturation}).

Finally, there is another equation in the literature \cite{xuReduceddimensionDynamicModel2021,pasaogullariTwoPhaseModelingFlooding2005} that is less commonly employed. This expression, denoted as \eqref{eq:water_activity_bad}, extrapolates the activity definition for water vapour from \eqref{eq:water_activity_Springer}. In this formulation, liquid water is treated as a gas, and its "concentration" in the pore volume is considered. However, the authors strongly discourage the use of this equation for several reasons. Firstly, treating liquid water as a gas represents an overly restrictive assumption. Additionally, this equation is incompatible with Springer's model, even though it relies on it. Indeed, $\lambda_{eq}$ is only applicable for activities between 0 and 3, with the value of 3 established solely for liquid water in the catalyst layer. However, in the case of this equation, for $s \in \left[ 0, 1 \right]$, $a_{w}$ falls within the range $\left[ 0, 5072 \right]$. It is evident that this equation produces inaccurate results.

\begin{equation}
	a_{w} \overset{\vartriangle}{=} \frac{P_{\text{vapour+liquid}}}{P_{\text{sat}}} \underbrace{=}_{\text{ideal gas law}} \frac{C \left[ 1 - \texttt{s} \right] + \frac{\rho_{H_{2}O}}{M_{H_{2}O}} \texttt{s}}{C_{\text{sat}}}
	\label{eq:water_activity_bad}
\end{equation}

\subsection{Water sorption at the ionomer/CL interface: $S_{sorp}$ - an overview}
\label{subsec:Water_flows_at_the_ionomer/CL_interface}

At the interface between the ionomer and the CL, there is an exchange of water. This water exists in a dissolved form within the ionomer and in either vapour or liquid form in the CL. This water conversion, denoted as $S_{sorp}$ in this study, represents either water absorption or desorption and occurs throughout the entire volume of the catalyst layer. It differs from a flow, which involves the transport of matter from one volume to another across a surface. This conversion takes place within a single volume and is therefore volumetric. Due to the discontinuity of matter at the interface, Fick's law does not accurately characterize this flow. More complex phenomena are at play, complicating the applicability of existing laws. Consequently, the caracteristic law for $S_{sorp}$ has undergone refinement over the past decades \cite{jiaoWaterTransportPolymer2011,huAnalyticalCalculationEvaluation2016,geAbsorptionDesorptionTransport2005}. In this section, the authors strive to articulate a clear and precise expression for the sorption term $S_{sorp}$, drawing on the insights from the most recent and original studies \cite{xingNumericalAnalysisOptimum2015,yangMatchingWaterTemperature2011,wuModelingPEMFCTransients2010,geAbsorptionDesorptionTransport2005}.

First and foremost, the following general expression \eqref{eq:j_sorp_classic} is commonly accepted in the community \cite{jiaoWaterTransportPolymer2011,xingNumericalAnalysisOptimum2015,yangMatchingWaterTemperature2011,fanCharacteristicsPEMFCOperating2017,wuModelingPEMFCTransients2010,mengTwophaseNonisothermalMixeddomain2007}.

\begin{equation}
	\boxed{
		S_{sorp} = \gamma_{sorp} \frac{\rho_{\text{mem}}}{M_{\text{eq}}} \left[ \lambda_{\text{eq}} - \lambda \right]
		\label{eq:j_sorp_classic}
	}
\end{equation}

\nomenclature[A,1]{$S$}{matter conversion at the interface of the triple points $( mol.m^{-3}.s^{-1} )$ }
\nomenclature[A,2]{\(\gamma\)}{rate constant $( s^{-1} )$ }
where $\gamma_{sorp}$ $( s^{-1} )$ is the sorption rate.

In this expression, $\lambda_{eq}$ denotes the equilibrium value of $\lambda$. It functions as a virtual variable. Consequently, as long as $\lambda$ deviates from $\lambda_{eq}$, a water flow persists between the membrane and the catalyst layer, endeavoring to achieve equilibrium.

The coefficient $\frac{\rho_{\text{mem}}}{M_{\text{eq}}}$ converts water content to membrane water concentration, enabling the expression of water flow. To eliminate the necessity of tracking membrane swelling in the model, Springer et al. adapted their extensively employed results to a dry membrane model \cite{springerPolymerElectrolyteFuel1991}.

Finally, the sorption rate coefficient reflects the velocity of this sorption. For a given gap between $\lambda_{\text{eq}}$ and $\lambda$, the value of $\gamma$ influences $S_{\text{sorp}}$, and renders the flow as more or less important. Determining  $\gamma$ in an actual PEMFC proved challenging, leading to the adoption of somewhat arbitrary values typically ranging from 0.1 to 100 $s^{-1}$ in the literature \cite{wuModelingPEMFCTransients2010}. The value $\gamma = 1.3 s^{-1}$ is the most commonly encountered one \cite{jiaoWaterTransportPolymer2011, fanCharacteristicsPEMFCOperating2017}.
However, experiments have revealed that the sorption rate of water is not constant; rather, it depends on the volume fraction of water within the membrane, with distinct absorption and desorption rates. Ge et al. \cite{geAbsorptionDesorptionTransport2005} introduced a more precise expression for $\gamma_{sorp}$ in 2005. Subsequently, this expression underwent a slight modification by incorporating the $H_{cl}$ term \cite{wuMathematicalModelingTransient2009}, as illustrated in \eqref{eq:j_sorp_Ge} \cite{xingNumericalAnalysisOptimum2015, yangMatchingWaterTemperature2011, wuModelingPEMFCTransients2010, geAbsorptionDesorptionTransport2005}.

\begin{subequations}
	\begin{equation}
		\gamma_{sorp} = 
		\begin{cases}
			\gamma_{a} = \frac{1.14 \cdot 10^{-5} f_{v}}{H_{cl}} e^{ 2416 \left[ \frac{1}{303} - \frac{1}{T_{f c}} \right] }, & \text{\footnotesize for an absorption flow} \\
			\gamma_{d} = \frac{4.59 \cdot 10^{-5} f_{v}}{H_{cl}} e^{ 2416 \left[ \frac{1}{303} - \frac{1}{T_{f c}} \right] }, & \text{\footnotesize for a desorption flow}
		\end{cases}	
	\end{equation}
	\begin{equation}
		f_{v} = \frac{ \lambda V_{w} }{ V_{\text{mem}} + \lambda V_{w} }
	\end{equation}	
	\label{eq:j_sorp_Ge}
\end{subequations}
\nomenclature[A,1]{$f_{v}$}{water volume fraction of the membrane}
\nomenclature[A,1]{$V$}{molar volume $( m^{3}.mol^{-1} )$}
\nomenclature[A,2]{\(\gamma_{sorp}\)}{sorption rate $( s^{-1} )$ }
where $f_{v}$ is the water volume fraction of the membrane, $V_{w}$ $( m^{3}.mol^{-1} )$ is the molar volume of water, and $V_{mem}$ $( m^{3}.mol^{-1} )$ is the molar volume of dry membrane.

There is one flaw in this equation. The coefficient $\varepsilon_{mc}$, discussed in section \ref{subsec:lambda}, should have been included in this equation to account for the volume fraction of the ionomer in the catalyst layer. Indeed, $\varepsilon_{mc}$ is not universal and varies based on the stack design. Considering this variable is crucial for a more precise model. Including tortuosity considerations with $\varepsilon_{mc}^{\tau}$ would have been even more beneficial (as discussed in section \ref{subsec:effective_diffusion_coefficient}). 

It remains important to know if, in a biphasic state for water in the catalyst layer, $S_{sorp}$ is a vapour, a liquid matter conversion, or both. Ge et al. provided insights into this matter. When liquid water is introduced, it comes into direct contact with the membrane. The water content of the membrane at the membrane/GDL interface is assumed to reach instant equilibrium with liquid water \cite{geAbsorptionDesorptionTransport2005}.
Thus, the expression for the sorption rate, $\gamma_{\text{sorp}}$, is valid only for vapour, and $S_{\text{sorp}}$ necessarily denotes a vapour matter conversion. This observation is significant as it highlights, once again, the common yet inconsistent practice of combining $\lambda_{v,eq}$ and $\lambda_{l,eq}$ in the calculation of $S_{\text{sorp}}$, accomplished by introducing $\texttt{s}$ into $a_w$.

\subsection{New interpretation of $S_{sorp}$, $\lambda_{eq}$ and $a_{w}$}
\label{subsec:another_way_lambda_eq}

Previously, sections \ref{subsec:equilibrium_water_content}, \ref{subsec:water_activity}, and \ref{subsec:Water_flows_at_the_ionomer/CL_interface} provided a comprehensive overview of the current utilization of $\lambda_{eq}$, $a_{w}$, and $S_{sorp}$ in the literature, with their associated limitations highlighted. The present section introduces a novel approach to address these limitations.

Different phenomena give rise to distinct expressions for $\lambda_{eq}$. When vapour is in contact with the membrane, it is denoted as $\lambda_{v,eq}$. Conversely, when liquid water is in contact with the membrane, it is represented by $\lambda_{l,eq}$. Previous studies attempted to amalgamate them as a single entity, treating vapour and liquid water sorption as if they were identical. Consequently, the concepts of $\lambda_{v,eq}$ and $\lambda_{l,eq}$ were subsumed under $\lambda_{eq}$, and water activity was linked to both vapour and liquid water. However, these choices are here reconsidered for the reasons outlined in the previous sections. To address this, the authors propose distinguishing between two water sorption processes. One involves the conversion of vapour into condensed water, denoted as $S_{v,sorp}$. The other entails the conversion of liquid water into condensed water, denoted as $S_{l,sorp}$. As a result, $\lambda_{v,eq}$ and $\lambda_{l,eq}$ are no longer merged, and there is no longer a need to extrapolate the concept of water activity, which can remain equivalent to relative humidity in the CL. These expressions are provided in \eqref{eq:j_sorp_new}.

\begin{subequations}
	\begin{equation}
		S_{v,sorp} = \left( 1 - \varphi_{sp} \right) \gamma_{v,sorp} \frac{\rho_{\text{mem}}}{M_{\text{eq}}} \left[ \lambda_{\text{v,eq}} - \lambda_{cl} \right]
	\end{equation}
	\begin{equation}
		S_{l,sorp} = \varphi_{sp} \gamma_{l,sorp} \left[ \lambda_{\text{l,eq}} - \lambda_{cl} \right]
	\end{equation}	
	\label{eq:j_sorp_new}
\end{subequations}
\nomenclature[B,2]{$\varphi_{sp}$}{Surface proportion function}
where $\gamma_{v,sorp}$ is the sorption rate of vapour to condensed water, equal to the classic expression of $\gamma_{sorp}$, $\gamma_{l,sorp}$ is the sorption rate of liquid water to condensed water, and $\varphi_{sp}$ is the surface proportion of liquid water at the CL ionomer interface.

However, this idea necessitates the identification of two novel quantities in future research. Firstly, $\gamma_{l,sorp}$ is currently unknown and requires experimental measurement. According to Ge et al. \cite{geAbsorptionDesorptionTransport2005}, the value of this sorption rate should be significantly higher than that of $\gamma_{v,sorp}$ to account for their hypothesis of instantaneous equilibrium. Secondly, $\varphi_{sp}$ needs to be determined to properly balance these two water sorptions. These quantities are experimentally evaluated either without liquid water or without vapour. However, in a scenario involving the coexistence of these two quantities, it is essential to balance their access to the membrane. The authors propose considering the surface ratio of liquid water at the CL ionomer interface for $\varphi_{sp}$. Thus, $\varphi_{sp}$ is a function of the volumetric ratio of liquid water $\texttt{s}$. This function does not currently exist and could be the subject of future research. It is noteworthy that the inclusion of $\varphi_{sp}$ might be circumvented if a three-dimensional model with sufficient precision to differentiate regions where liquid water resides from regions with vapour were available. However, constructing such a model would demand computational power to an extent that seems impractical to consider.

Finally, it remains to incorporate $S_{v,sorp}$ and $S_{l,sorp}$ into the matter balances for the water content $\lambda_{cl}$, the vapour concentration $C$, and the liquid water saturation $\texttt{s}$.

\subsection{Water production at the interface of the triple points: $S_{prod}$}
\label{subsec:water_production}

Water is formed through the chemical reaction between hydrogen and oxygen at the triple point interface within the cathode catalyst layer. As redox reactions of oxygen and hydrogen are considered to be infinitely fast, water production is directly associated with $i_{fc}$, representing the current density generated by the fuel cell. Additionally, it is influenced by $i_{sc}$, corresponding to the short circuit current density, as discussed in section \ref{subsec:internal_current_density}. This relationship is expressed in \eqref{eq:S_prod}.

\begin{equation}
	S_{\text{prod}} = 
	\begin{cases}
		\frac{i_{fc} + i_{sc}}{2 F H_{cl}}, &\text{\small{in the CCL}} \\
		0, &\text{\small{elsewhere}}
	\end{cases}
	\label{eq:S_prod}
\end{equation}
\nomenclature[A,1]{$H$}{thickness $( m ) $}
where $S_{prod}$ $(mol.m^{-3}.s^{-1})$ is the water production in the membrane at the triple points zone, $H_{cl}$ $(m)$ is the catalyst layer thickness. It is unclear whether water is initially produced in vapour, liquid, or dissolved form in the membrane. Similar to Jiao et al. \cite{jiaoWaterTransportPolymer2011}, the authors propose to implement it in a dissolved form. Nonetheless, it is straightforward to modify this assumption by relocating the matter conversion term to a different differential equation. 

The stack also generates additional water due to the crossover of hydrogen and oxygen, as discussed in section \ref{subsec:hydrogen_oxygen_consumption}. This water generation is denoted as $S_{co}$. However, it is noteworthy that water can be produced in both the anode and cathode catalyst layers. This water production should be directly associated with the crossover flows, assuming these flows pass through the membrane instantaneously, as if it were of negligible thickness. Additionally, it is assumed that all the matter passing through undergoes instantaneous reaction to form water \cite{namNumericalAnalysisGas2010}. The corresponding equation is expressed in \eqref{eq:S_co}.

\begin{equation}
	S_{co} = 
	\begin{cases}
		2 \cdot S_{O_{2},co}, &\text{\small{in the ACL}} \\
		S_{H_{2},co}, &\text{\small{in the CCL}} \\
		0, &\text{\small{elsewhere}}
	\end{cases}
	\label{eq:S_co}
\end{equation}
where $S_{i,co}$ $(mol.m^{-2}.s^{-1})$ is the crossover flow of molecule i (hydrogen or oxygen), discussed in \ref{subsec:hydrogen_oxygen_consumption}.

Finally, the corrected expression of $S_{prod}$ is expressed as \eqref{eq:S_prod_corrected}.

\begin{equation}
	S_{\text{prod}} = 
	\begin{cases}
		2 k_{O_{2}}\frac{ R T_{fc}}{H_{cl}} \nabla C_{O_{2}}, &\text{\small{in the ACL}} \\
		\frac{i_{fc} + i_{sc}}{2 F H_{cl}} + k_{H_{2}} \frac{R T_{fc}}{H_{cl}} \nabla C_{H_{2}}, &\text{\small{in the CCL}} \\
		0, &\text{\small{elsewhere}}
	\end{cases}
	\label{eq:S_prod_corrected}
\end{equation}

\subsection{Water content dynamic behavior}

In the membrane,  water content is governed by the molar balances presented in equation \ref{subeq:water_content_dynamic_balance} \cite{xuReduceddimensionDynamicModel2021}, along with the boundary condition given in equation \ref{subeq:water_content_dynamic_boundary_ACL}. The system involves two differential equations. Indeed, unlike the ionomer in the membrane, the ionomer in the catalyst layer represents only a fraction of the total volume. This factor affects the governing equation by introducing the CL ionomer volume fraction $\varepsilon_{mc}$. It is important to note that in this context, $\varepsilon_{mc}$ is not associated with tortuosity; this relationship should only be considered in the context of expressing the transport flows.

\begin{subequations}
	\begin{equation}
		\begin{cases}
			\frac{\rho_{\text{mem}}}{M_{\text{eq}}} \frac{\partial \lambda_{\text{mem}}}{\partial t} = - \bm{\nabla} \cdot \bm{J_{mem}}, & \text{\small{in the bulk membrane}} \\
			\frac{\rho_{mem} \varepsilon_{mc}}{M_{eq}} \frac{\partial \lambda_{cl}}{\partial t} =  - \bm{\nabla} \cdot \bm{J_{mem}} + S_{sorp} + S_{prod}, & \text{\small{in the CL}} \\
		\end{cases}
		\label{subeq:water_content_dynamic_balance}
	\end{equation}
	\begin{equation}
		\bm{J_{mem}^{cl,mem}} = \bm{0}, \text{\small{at the ionomer border}}
		\label{subeq:water_content_dynamic_boundary_ACL}
	\end{equation}
\end{subequations}

\section{Liquid water transport in the CL and GDL}
\subsection{Liquid water saturation: $\texttt{s}$}
\label{subsec:liquid_water_saturation}

During the operation of a PEMFC, the water produced through chemical reactions, combined with the moisture in the incoming gas, often reaches a point of vapour saturation, leading to the formation of liquid water within the cell. It is crucial to regulate this quantity, as an excess may submerge the fuel cell, causing a drop in voltage. On one hand, the triple points areas may be submerged in liquid water, introducing additional resistance to fuel transport to the catalysts. On the other hand, the presence of liquid water in the GDLs can impede the gas flow towards the CLs by increasing the material's tortuosity. To quantify the amount of liquid water,  the liquid water saturation variable $\texttt{s}$ is employed. The values of $\texttt{s}$ range from 0 to 1, where 0 signifies the absence of liquid water and 1 indicates the exclusive presence of liquid water within the pore stack. 

\begin{equation}
	\texttt{s} \overset{\vartriangle}{=} \frac{V_{\text{liquid water}}}{V_\text{pore}}
\end{equation}

Three phenomena govern the evolution of liquid water: capillarity, convection, and condensation/evaporation. Each of these phenomena is discussed in the following subsections. Within the CL and the GDL, liquid water is mainly transported by a diffusive force known as capillarity. Referred to in this work as $J_{l,cap}$, this phenomenon is explored in detail in section \ref{subsec:liquid_water_capillary_flow}. Additionally, a secondary flow, denoted as $J_{l,conv}$, arises from gas motions, leading to the hauling of liquid water and thus termed a convective flow. However, as examined in section \ref{subsec:liquid_water_convective_flow}, it is a minor flow compared to the capillary flow and is often disregarded. Darcy's law is employed to characterize both flows. Finally, liquid water at the GDL/GC border is considered equal to 0 due to its rapid expulsion in the GCs and the limited knowledge available on this subject. This is discussed in section \ref{subsec:liquid_water_convective-diffusive_flow}.

\subsection{Liquid water capillary flow in the CL and GDL: $J_{l,cap}$}
\label{subsec:liquid_water_capillary_flow}

The capillary flow, denoted as $J_{l,cap}$ and defined in \eqref{eq:j_cap}, quantifies the capacity of liquid water generated within the electrode through vapour condensation to permeate it \cite{jiaoWaterTransportPolymer2011,xingNumericalAnalysisOptimum2015,wangModelingEffectsCapillary2008,yeThreeDimensionalSimulationLiquid2007,pasaogullariTwoPhaseModelingFlooding2005}. Capillarity represents a specific instance of diffusivity within the liquid phase. The equation is similar to Fick's law, incorporating a matter gradient $\bm{\nabla} \texttt{s}$ and a variable diffusive coefficient $D_{cap}$. This relationship is given by Darcy's law, as demonstrated in \ref{subsec:J_cap}.

It is important to note that \eqref{eq:j_cap} is derived from experiments involving water permeation through beds of sand, representing a significant simplification compared to water permeation through the GDL. Despite this simplification, it currently stands as the most viable model in the literature \cite{jiaoWaterTransportPolymer2011}. Therefore, the development of more pertinent models is imperative. In this context, the porous environment is assumed to be homogeneous, with negligible deformation, and water flow must be slow enough to maintain a small Reynolds number under stationary conditions \cite{whitakerMethodVolumeAveraging1999}. Gravity's impact is typically disregarded in the stack. Additionally, the flow expressed in \eqref{eq:j_cap} is historically in units of $kg.m^{-2}.s^{-1}$, whereas all other flows in PEMFC literature are presented in units of $mol.m^{-2}.s^{-1}$. This formulation is preserved in this review, with adjustments made to the differential equations to accommodate this distinction.

\begin{equation}
	\begin{cases}
		\bm{J_{l,cap}} = - D_{cap}\left( \texttt{s}, \varepsilon \right)  \bm{\nabla} \texttt{s} \\
		D_{cap}\left( \texttt{s}, \varepsilon \right) = \sigma \frac{K_{0}}{\nu_{l}} \left| \cos \left( \theta_{c} \right)\right| \sqrt{\frac{\varepsilon}{K_{0}}} \texttt{s}^{\texttt{e}} \left[ 1.417 - 4.24\texttt{s} + 3.789\texttt{s}^{2} \right]
	\end{cases}
	\label{eq:j_cap}
\end{equation}
\nomenclature[A,1]{$K$}{permeability $(m^{2})$}
\nomenclature[A,2]{\(\sigma\)}{surface tension of liquid water $(N.m^{-1})$}
\nomenclature[A,2]{\(\nu_{l}\)}{liquid water kinematic viscosity $(m^{2}.s^{-1})$}
\nomenclature[A,2]{\(\theta_{c}\)}{contact angle of GDL for liquid water (°)}
\nomenclature[B,3]{\(\bm{\nabla}\)}{gradient notation}
where $\bm{J_{cap}}$ $(kg.m^{-2}.s^{-1})$ is the capillary flow, $\sigma$ $(N.m^{-1})$ is the surface tension of liquid water, $K_{0}$ $(m^{2})$ is the intrinsic permeability, $\nu_{l}$ $(m^{2}.s^{-1})$ is the liquid water kinematic viscosity, $\theta_{c}$ (°) is the contact angle of GDL for liquid water, $\texttt{e}$ is the capillary exponent, and $\bm{\nabla}$ is the gradient notation.
To enhance the clarity of this expression, supplementary information is provided in \eqref{subsec:J_cap}.

It is noticeable that an absolute value was introduced on $\cos \left( \theta_{c} \right)$ in this study, a practice not commonly employed. This modification proves advantageous in ensuring a consistently positive diffusion coefficient $D_{cap}$, thereby preserving the negative sign typically associated with any mass balance. In existing literature, when $\cos \left( \theta_{c} \right)$ is negative, there are instances where the negative sign is occasionally omitted in the overall equation, complicating the proper understanding of the equation and impeding meaningful comparisons between different sources.

Moreover, the capillary exponent $\texttt{e}$ serves as a novel parameter introduced in this study to account for various values of $\texttt{e}$ found in the literature. Notably, the widely utilized cubic correlation assigns a value of 3 to $\texttt{e}$, originating from porous media such as sand/rock with typical porosities ranging from 0.1 to 0.4. Given the similarity between a PEMFC catalyst layer and sand/rock in terms of porosity and morphology, the liquid and gas permeabilities in the catalyst layer are computed using the aforementioned cubic correlations. However, recent research suggests that $\texttt{e}$ should fall within the range of 4.0 to 5.0 for GDL porous materials characterized by high porosities ranging from 0.6 to 0.8. It is noted that the cubic correlation may tend to overestimate liquid permeability, especially at low liquid saturation \cite{jiaoWaterTransportPolymer2011,xingNumericalAnalysisOptimum2015,wangModelingEffectsCapillary2008,yeThreeDimensionalSimulationLiquid2007,mengTwophaseNonisothermalMixeddomain2007}. These considerations are synthesized in \eqref{eq:capillary_exponent}.

\begin{equation}
	\begin{cases}
		\texttt{e} = 3, & \text{ if $\varepsilon \in \left[ 0.1,0.4 \right] $ } \\
		\texttt{e} \in \left[ 4,5 \right], & \text{ if $\varepsilon \in \left[ 0.6,0.8 \right] $ }
	\end{cases}	
	\label{eq:capillary_exponent}
\end{equation}
\nomenclature[A,2]{$\texttt{e}$}{capillary exponent}

Finally, it is important to note that $J_{l,cap}$ (as well as $J_{l,conv}$ in section \ref{subsec:liquid_water_convective_flow}) is based on Darcy's law, which applies only to creeping flow. While this assumption is reasonable within the GDL and the CL, modern stacks, which utilize complex flow-fields in the GC, such as the use of baffles, and operate under high current densities ($> 1 A.cm^{-2}$), may experience convective flows penetrating the GDL. In such critical conditions, Darcy's law becomes inadequate. Instead, Darcy-Forchheimer's law is employed to account for additional inertial effects. For further details, refer to the study by Kim et al. \cite{kimModelingTwophaseFlow2017}.

\subsection{Intrinsic permeability: $K_{0}$}
\label{subsec:intrinsic_permeability}

The intrinsic permeability, denoted as $K_{0}$, measures the porous material's capacity to facilitate the flow of fluids through it. This property is influenced by the material's porosity, the configuration and connectivity of its pores. It stands as an inherent physical characteristic of the material.
The Tomadakis and Sotirchos (T\&S) model is used to calculate intrinsic permeability within random fibrous and porous media \cite{tomadakisViscousPermeabilityRandom2005,fanCharacteristicsPEMFCOperating2017,fishmanHeterogeneousThroughPlaneDistributions2011}. It's worth noting persistent copying errors in the literature concerning this equation. The formulation considered in this study in \eqref{eq:intrinsic_permeability} aligns closely with the original expression \cite{tomadakisViscousPermeabilityRandom2005}.

\begin{equation}
	K_{0} = \frac{\varepsilon}{8 \ln \left( \varepsilon \right)^{2}} \frac{\left[ \varepsilon - \varepsilon_{p} \right]^{\alpha + 2} r_{f}^{2}} {\left[ 1 - \varepsilon_{p} \right]^{\alpha} \left[ \left[ \alpha + 1 \right]  \varepsilon - \varepsilon_{p} \right]^{2}}
	\label{eq:intrinsic_permeability}
\end{equation}
\nomenclature[A,1]{$r_{f}$}{carbon fiber radius $(m)$}
\nomenclature[B,2]{\(\alpha, \beta_{1}, \beta_{2}\)}{fitted values}
where $r_{f}$ $(m)$ is the carbon fibre radius, obtained at $4.6 \cdot 10^{-6}$ $m$ \cite{fishmanHeterogeneousThroughPlaneDistributions2011} or $3.16 \cdot 10^{-6}$ $m$ \cite{fanCharacteristicsPEMFCOperating2017}, $\varepsilon_{p}$ is the percolation threshold porosity, obtained at 0.11 \cite{fanCharacteristicsPEMFCOperating2017,fishmanHeterogeneousThroughPlaneDistributions2011}, and $\alpha$ is a fitted value, obtained at 0.521 for in plane direction and at 0.785 for through plan direction \cite{fanCharacteristicsPEMFCOperating2017,fishmanHeterogeneousThroughPlaneDistributions2011}.

Another element that is often neglected in the literature must be considered in the calculation of the intrinsic permeability. This is the compression of the GDL, described by Bao et al. \cite{baoTransportPropertiesGas2021}. Indeed, when the cells are assembled together, a pressure is applied to them to ensure that the gases between each compartment are sealed. This compression causes deformations in the structure of the GDL and therefore causes changes in the transport properties within it \cite{zamelEffectiveTransportProperties2013}. It is therefore necessary to modify the previously proposed model. The advantage of the proposal by Bao et al. is that it fits any model for calculating the effective diffusivity before compression with the simple addition of an exponential coefficient to account for it. However, this study has a limitation. It can only be used for structures with a porosity of approximately 73$\%$ or 60$\%$. Fortunately, this concerns a large part of the current GDL. Thus, the model of Tomadakis and Sotichos augmented by the work of Bao et al, which can be renamed by the TSB model, yields the following intrinsic permeability given equation \ref{eq:intrinsic_permeability_TSB}.

\begin{equation}
	\boxed{
		K_{0} = \frac{\varepsilon}{8 \ln \left( \varepsilon \right)^{2}} \frac{\left[ \varepsilon - \varepsilon_{p} \right]^{\alpha + 2} r_{f}^{2}} {\left[ 1 - \varepsilon_{p} \right]^{\alpha} \left[ \left[ \alpha + 1 \right]  \varepsilon - \varepsilon_{p} \right]^{2}} e^{ \beta_{1} \varepsilon_{c}}
	}
	\label{eq:intrinsic_permeability_TSB}
\end{equation}
\nomenclature[A,2]{\(\varepsilon_{c}\)}{compression ratio}
where $\beta_{1}$ is a fitted value which varies with the porosity and the matter diffusion direction according to the following table \ref{table:beta1_coefficient} and $\varepsilon_{c}$ is the compression ratio of the GDL, which is defined as the ratio of the thickness reduction to the thickness of uncompressed GDL. According to Yim et al. \cite{yimInfluenceStackClamping2010}, a value of $30\%$ for $\varepsilon_{c}$, which corresponds to high GDL compression, is feasible and exhibits good performances. A minimum value of $15\%$ should be given to $\varepsilon_{c}$ for low GDL compression.

\begin{table*}[htb]
	\centering
	\begin{tabular}{|c|c|c|} \hline
		
		$\beta_{1}$
		& in-plane
		& through-plane \\ \hline
		
		$\varepsilon \approx 0.6$
		& -5.07
		& -3.60  \\ \hline
		
		$\varepsilon \approx 0.73$
		& -3.51
		& -2.60  \\ \hline
		
	\end{tabular}
	\caption{Different values of the fitted parameter $\beta_{1}$ according to the porosity and the diffusion direction of gases.}
	\label{table:beta1_coefficient}
\end{table*}

Table \ref{table:intrinsic_permeability} presents a comparison between the value given by these equations and values found in other works. 

\begin{table*}[htb]
	\centering
	\begin{tabular}{|l|*{8}{c|}} \hline
		
		& TSB
		& T\&S
		& Hu  \cite{huAnalyticalCalculationEvaluation2016,pasaogullariTwoPhaseModelingFlooding2005}
		& Yang \cite{yangMatchingWaterTemperature2011}
		& Wang \cite{wangModelingEffectsCapillary2008}
		& Ye \cite{yeThreeDimensionalSimulationLiquid2007}
		& Meng \cite{mengTwophaseNonisothermalMixeddomain2007} 
		& Wang \cite{wangInvestigationDryIonomer2020}\\ \hline
		
		$K_{0}^{gdl} \left( m^{2} \right) $ 
		& {\normalsize $3.4 \cdot 10^{-13}$} 
		& {\normalsize $10^{-12}$} 
		& {\normalsize $7 \cdot 10^{-13}$} 
		& {\normalsize $3 \cdot 10^{-12}$} 
		& {\normalsize $2 \cdot 10^{-15}$}  
		& {\normalsize $23 \cdot 10^{-12}$}  
		& {\normalsize $10^{-12}$}
		& {\normalsize $10^{-12}$} \\
		
		& {\tiny $\left(\varepsilon=0.6\right)$} 
		& {\tiny $\left(\varepsilon=0.6\right)$} 
		& {\tiny $\left(\varepsilon=0.5-0.6\right)$}
		& {\tiny $\left(\varepsilon=0.7\right)$}
		& {\tiny $\left(\varepsilon=0.5\right)$}
		& {\tiny $\left(\varepsilon=0.7\right)$}
		& {\tiny $\left(\varepsilon=0.6\right)$}
		& {\tiny $\left(\varepsilon=0.7\right)$}  \\ \hline
		
		$K_{0}^{cl} \left( m^{2} \right) $ 
		& {\normalsize $\emptyset$}
		& {\normalsize $1.4 \cdot 10^{-14}$} 
		& {\normalsize $\emptyset$}
		& {\normalsize $3 \cdot 10^{-14}$} 
		& {\normalsize $5 \cdot 10^{-17}$} 
		& {\normalsize $2 \cdot 10^{-15}$} 
		& {\normalsize $10^{-13}$}
		& {\normalsize $10^{-13}$}  \\
		
		&
		& {\tiny $\left(\varepsilon=0.25\right)$} 
		&
		& {\tiny $\left(\varepsilon=0.2\right)$}
		& {\tiny $\left(\varepsilon=0.12\right)$} 
		& {\tiny $\left(\varepsilon=0.2\right)$}
		& {\tiny $\left(\varepsilon=0.12\right)$}
		& {\tiny $\left(\varepsilon=0.3\right)$}  \\ \hline
	\end{tabular}
	\caption{Comparison between the value given by Tamadakis and Sotirchos model and values found in other works}
	\label{table:intrinsic_permeability}
\end{table*}

\subsection{Water surface tension: $\sigma$}

Surface tension is the force that preserves a fluid's specific geometry, corresponding to its minimal surface interface with another fluid, in this case, liquid water with air. This phenomenon enables the two fluids to minimize the energy at their interface. In this case, surface tension is solely a function of temperature, and can be calculated using equation \ref{eq:surface_tension} \cite{vargaftikInternationalTablesSurface1983}.

\begin{equation}
	\sigma = 235.8 \times 10^{-3} \left[ \frac{647.15 - T_{fc}}{647.15} \right]^{1.256} \left[ 1 - 0.625 \frac{647.15 - T_{fc}}{647.15} \right]
	\label{eq:surface_tension}
\end{equation}

The equation produces a result of $\sigma = 0.0627$ $N.m^{-1}$ at 80°C, closely aligning with the prevalent value found in PEMFC literature, which is $0.0625$ $N.m^{-1}$ \cite{huAnalyticalCalculationEvaluation2016,xingNumericalAnalysisOptimum2015,yangMatchingWaterTemperature2011,fanCharacteristicsPEMFCOperating2017,pasaogullariTwoPhaseModelingFlooding2005,wangInvestigationDryIonomer2020}.

\subsection{Liquid water convective flow in the CL and GDL: $J_{l,conv}$}
\label{subsec:liquid_water_convective_flow}

It is noteworthy that the convection flow of liquid water, denoted as $J_{l,conv}$, is not extensively addressed in prior studies. Certain models neglect this flow, such as the unsaturated flow theory ($UFT$) \cite{pasaogullariTwophaseTransportRole2004, wangMultiphaseMixtureModel}, while others rigorously exclude it, like the multi-phase mixture model ($M^{2}$) \cite{wangMultiphaseMixtureModel}. In this study, $J_{l,conv}$ is finally neglected; nevertheless, a discussion is provided on it. This section explores various theories that allow for its omission and outlines resolution methods to be employed when it is deemed significant.

To comprehend $J_{l,conv}$, it is necessary to analyze the impact of gas motion on liquid water. The diffusive transport of gases carries liquid water molecules along with their motion. This transport, quantified by Darcy's law as is $J_{l,cap}$, is expressed as \ref{eq:J_l_conv} \cite{wuMathematicalModelingTransient2009,jiaoWaterTransportPolymer2011} and further demonstrated in \ref{subsec:J_cap}.

\begin{equation}
	\bm{J_{l,conv}} = \frac{\rho_{H_{2}O} \mu_{g}}{\mu_{l}} \frac{\texttt{s}^{\texttt{e}}}{\left[ 1 - \texttt{s} \right]^{\texttt{e}}} \bm{u_{g}}
	\label{eq:J_l_conv}
\end{equation}
\nomenclature[A,1]{$u$}{velocity $(m.s^{-1})$}
\nomenclature[A,2]{\(\mu\)}{dynamic viscosity $(Pa.s)$}
where $\bm{J_{l,conv}}$ $(kg.m^{-2}.s^{-1})$ is the convective flow of liquid water, $\mu_{g}$ $(Pa.s)$ is the gas mixture dynamic viscosity, $\mu_{l}$ $(Pa.s)$ is the liquid water dynamic viscosity, and $\bm{u_{g}}$ $(m.s^{-1})$ is the gas mixture velocity.

The unsaturated flow theory (UFT) \cite{pasaogullariTwophaseTransportRole2004} justify the neglect of this flow by stating that the gas phase pressure remains constant throughout the porous media in a two-phase mixture. This reductionist assumption implies that the pressure variation of the liquid phase is equal to the capillary pressure variation. So, according to Darcy's law, gases are immobile in the porous medium, leading to $\bm{u_{g}} = \bm{0}$ (refer to demonstration \ref{subsec:J_cap} for details). Consequently, the convective flow of liquid water is cancelled out. This theory has found widespread application in the field of fuel cell literature concerning two-phase flow through porous media. 
The UFT theory was excluded from this study due to the restrictive assumption of a constant gas pressure within the stack. Only the convective flow $J_{l,conv}$ was assumed negligible and consequently omitted from further consideration. Although this assumption may seem limiting, it is justified. Firstly, in terms of magnitude, the convective flow is relatively minor compared to the capillary flow. Additionally, liquid water is influenced by both vapour and fuel motions in the electrodes. Given that the vapour and fuel flows are comparable in magnitude and occur in opposite directions, they balance their influence on liquid water. Consequently, $\bm{u_{g}}$ is small, resulting in a minor convective flow.

The conventional approach to consider $\bm{J_{l,conv}}$, also known as the multi-phase approach \cite{wangMultiphaseMixtureModel}, involves utilizing the Cauchy momentum equations to derive the velocity field within the MEA. In the dynamic behavior of liquid water saturation \eqref{subeq:liquid_water_dynamic_behaviour}, $J_{l,conv}$ is typically integrated by applying $\bm{\nabla} \cdot \left( \bm{J_{cap}} + \bm{J_{l,conv}} \right)$ instead of $\bm{\nabla} \cdot \bm{J_{cap}}$, following the methodology proposed by Wu et al. \cite{wuMathematicalModelingTransient2009}. However, adopting this approach leads to the formulation of significantly more intricate equations and longer computational times. Neglecting  $\bm{J_{l,conv}}$ helps circumvent this complexity, relying solely on the continuity equation within the MEA to model the system.

However, an alternative method has been developed to significantly reduce the number of equations inherent in the multi-phase approach, although Cauchy momentum equations are still required. This approach, known as the multi-phase mixture model (M²), treats water vapour and liquid water as a unified multi-phase mixture. As a result, the inter-phase interactions vanish, which eliminates both $J_{l,conv}$ and the terms related to the evaporation and condensation of water, represented by $S_{vl}$ as discussed in section \ref{subsec:water_phase_change_rate}. The M² method maintains results accuracy comparable to the multi-phase approach without necessitating additional reduction assumptions. However, it does require minor adjustments to the present modeling structure by treating water as a singular entity, whether in a liquid or gaseous state. For a more in-depth exploration of the M² model, refer to the study conducted by Wang et al \cite{wangMultiphaseMixtureModel}.

\subsection{Liquid water at the GDL/GC interface - a Dirichlet boundary condition}
\label{subsec:liquid_water_convective-diffusive_flow}

In the literature there are three major methods for modelling liquid water in the GC, and consequently, at the GDL/GC interface. The first, proposed by Pukrushpan et al. \cite{pukrushpanControlOrientedModelingAnalysis2004}, involves considering liquid water as a spray flow, as only a small amount of liquid water is supposed to exist there. These liquid droplets are assumed to be finely dispersed (with zero volume) owing to the strong gaseous motion in the GC and to have transport properties identical to those of vapour. 

Second, several studies have identified similarities between the porous structure of the GDL and the channels of the GC flow-field, although the order of magnitude of the sizes are not the same \cite{kotakaImpactInterfacialWater2014,yangMatchingWaterTemperature2011,haoModelingExperimentalValidation2015,wangModelingTwophaseFlow2008,jiangNumericalModelingLiquid2014,liPorousMediaModeling2015,jieliuPorousMediaModeling2014,imkePorousMediaSimplified2004,wangPorousMediumModel1994,sugumarThermalAnalysisInclined2006,tioThermalAnalysisMicro2000,basuTwophaseFlowMaldistribution2009}. Thus, they proposed to continue the use of the liquid water saturation variable $\texttt{s}$ and to use Darcy's law, with a porosity equal to 1, for modelling liquid water transport, as dicussed previously in Section \ref{subsec:liquid_water_capillary_flow}. This attempt further justified by the fact that liquid water can reach up to $10\%$ of the total mass flow rate in the GC \cite{wangModelingPolymerElectrolyte2005}, which weakens the Pukrushpan hypothesis of a spray flow.

However, as long as 3D complex PEMFC flow-fields are modelled and high current densities are reached ($2 \sim 4$ $A.cm^{-2}$), Darcy's law alone is not sufficient to consider liquid water transport. Thus, Darcy-Forchheimer's law is recommended instead \cite{kimModelingTwophaseFlow2017}. This is an important consideration because the use of advanced GCs, for example baffles, is becoming the norm in modern fuel cells to achieve much higher power densities. 
In this scenario, GC geometry often result in gas flow penetrations into the GDL owing to strong convection. Consequently, extensive modelling is required to consider the 3D geometry of the GC and determine the location of the boundary between the convection-dominated flow within the GC core and the diffusion-dominated flow within the GDL core, which is not a flat boundary anymore. This high-level modelling, partially introduced by Kim et al. in 2017 \cite{kimModelingTwophaseFlow2017}, is not deeply discussed in this article.

To summarize, Pukrushpan's hypothesis is a relatively simplified perspective of reality. Indeed, regarding tiny liquid water droplets as a perfect gas is overly reductionist, despite their small size. Furthermore, it has been demonstrated that liquid water in the GC is not always in the form of a spray but sometimes exists in much condensed forms \cite{wangModelingPolymerElectrolyte2005}.  This hypothesis is therefore not recommended. Then, Darcy-Forchheimer's law requires sophisticated modelling, which is not considered in this study. Darcy's law, although compatible in the current stack model, does not fit well into models that consider all the auxiliaries in addition to the stack. To the best of the authors' knowledge, the auxiliaries are generally still modelled using simple equations wherein only gas flows are considered. Therefore, none of these models were adopted here.

Finally, it has been chosen here to not directly model liquid water in the GC, given the current difficulty in doing so and the potential issues it can pose when integrating stack and auxiliary models. To address this, an assumption is made: the configuration of the bipolar plates as well as the operating conditions allow for a good gas flow within the GC, resulting in excellent removal of liquid water from the stack, preventing it from aggregating within the GC or even on its surface. Thus, liquid water is considered nonexistent in the GC, and a Dirichlet boundary condition is imposed at the GDL/GC interface, setting the liquid water saturation variable \texttt{s} to zero. This boundary condition is presented in equation \eqref{eq:s_boundary_condition_GDL_GC_interface}.

\begin{equation}
	\texttt{s} = 0, \text{at the GDL/GC border}
	\label{eq:s_boundary_condition_GDL_GC_interface}
\end{equation}

\subsection{Water phase change rate: $S_{vl}$}
\label{subsec:water_phase_change_rate}

In the stack, it is imperative to consider the mole variation of liquid water due to its evaporation or formation through vapour condensation. According to kinetic theory, assuming an ideal gas, neglecting interactions between individual molecules and employing constant overall phase change rates \cite{jiaoWaterTransportPolymer2011}, the net matter transfer resulting from evaporation and condensation can be approximated. This is commonly expressed by the following equation \eqref{eq:phase_change_rate} \cite{xingNumericalAnalysisOptimum2015, yangMatchingWaterTemperature2011, wangModelingEffectsCapillary2008, fanCharacteristicsPEMFCOperating2017, yeThreeDimensionalSimulationLiquid2007, mengTwophaseNonisothermalMixeddomain2007, wangInvestigationDryIonomer2020}.

\begin{equation}
	S_{vl} = 
	\begin{cases}
		\gamma_{\text{cond}} \varepsilon \left[ 1 - \texttt{s} \right] x_{v} \left[ C_{v} - C_{\text {v,sat}} \right], & \text{if $C_{v} > C_{\text{v,sat}}$} \\
		-\gamma_{\text{evap}} \varepsilon \texttt{s} \frac{\rho_{H_{2}O}}{M_{H_{2}O}} R T_{fc} \left[ C_{\text{v,sat}} - C_{v} \right], &\text{if $C_{v} \leq C_{\text{v,sat}}$}
	\end{cases}	
	\label{eq:phase_change_rate}
\end{equation}
\nomenclature[A,1]{$S_{vl}$}{phase transfer rate of condensation and evaporation $(mol.m^{-3}.s^{-1})$}
\nomenclature[A,1]{$x_{v}$}{mole fraction of vapour}
where $S_{vl}$ $(mol.m^{-3}.s^{-1})$ is the phase transfer rate of condensation and evaporation, indicating the amount of liquid water converted per units of volume and time, $\gamma_{\text{cond}}$ $(s^{-1})$ is the overall condensation rate constant for water, $\gamma_{\text{evap}}$ $(Pa^{-1}.s^{-1})$ is the overall evaporation rate constant for water, and $x_{v}$ is the mole fraction of vapour. 

Both constants $\gamma_{cond}$ and $\gamma_{evap}$ should be employed carefully, as they are reported in the literature using different units. Consequently, a direct comparison of their values is deemed inappropriate. Table \ref{table:constant_values_1} in the appendix presents various sets of values found in the literature, where the evaporation rate typically surpasses the condensation rate. Notably, Hua Meng's proposed values, $\gamma_{cond} = 5 \cdot 10^{3} s^{-1}$ and $\gamma_{evap} = 10^{-4} s^{-1}Pa^{-1}$, seem to be the most suitable. These values were well justified through numerical studies conducted in the work of Meng \cite{mengTwophaseNonisothermalMixeddomain2007}. 

Finally, given that water phase change rates are significantly influenced by local conditions such as mass and heat transfer, the accuracy of the calculation at the macroscopic level remains debatable \cite{jiaoWaterTransportPolymer2011}.

\subsection{Liquid water saturation dynamic behavior}

Considering all aforementioned phenomena, the dynamic behavior of liquid water can be expressed as given in \eqref{subeq:liquid_water_dynamic_behaviour}, along with its corresponding boundary conditions as indicated in \eqref{subeq:liquid_water_dynamic_behaviour_border}.

\begin{subequations}
	\begin{equation}
		\rho_{H_{2}O} \varepsilon \frac{\partial \texttt{s}}{\partial t} = - \bm{\nabla} \cdot \bm{J_{cap}} + M_{H_{2}O} S_{vl}
		\label{subeq:liquid_water_dynamic_behaviour}
	\end{equation}	
	\begin{equation}
		\begin{cases}
			\bm{J_{l}^{cl,mem}} = \bm{0}, \text{at the ionomer border} \\
			\texttt{s} = 0, \text{at the GDL/GC border}
		\end{cases}
		\label{subeq:liquid_water_dynamic_behaviour_border}
	\end{equation}
\end{subequations}

\section{Vapour transport in the CL and GDL}
\subsection{Vapour diffusive flow in the CL and GDL: $J_{dif}$}
\label{subsec:vapour_diffusive_flow_GDL}

Concentration gradients govern matter transport within the electrodes, with convection being neglected. This is due to the diminishing velocity of the gas stream near the GDL/GC boundary caused by frictional effects. In the absence of convective mixing, concentration gradients can arise within the stagnant gas of the electrode \cite{ohayreFuelCellFundamentals2016}. To describe this flow, a simple Fick equation is used, as depicted in \eqref{eq:J_diff_vapour}.

\begin{equation}
	\bm{J_{v,dif}} = - D_{v}^{eff} \bm{\nabla} C_{v}
	\label{eq:J_diff_vapour}
\end{equation}
where $J_{v,dif}$ is the vapour diffusive flow and $D_{v}^{eff}$ is the vapour diffusion coefficient.

\subsection{Effective diffusion coefficient of two species i and j: $D_{i/j}^{eff}$}
\label{subsec:effective_diffusion_coefficient}

The conventional binary diffusivity, denoted as $D_{i/j}$ and discussed in section \ref{subsec:binary_diffusion_coefficient}, is typically calculated in an open space, which corresponds to an environment different from the matter transport within GDLs or CLs, both of which are porous solids. Consequently, it is necessary to adjust $D_{i/j}$ to incorporate the influence of its surroundings on matter transport. Therefore, the effective diffusivity of species i/j serves as a method to consider both the porosity and tortuosity of the material in which the species evolve, in addition to the space occupied by liquid water that constrains their movement. Tortuosity is introduced to characterize the additional hindrance to diffusion arising from a convoluted or tortuous flow path. 
Notably, the GDLs and the CLs exhibit disparate structures, leading to distinct flow behaviors within them. Furthermore, these structures are anisotropic, implying that their evolution is direction-dependent. Consequently, it becomes imperative to account for these variations in the mathematical expressions used to describe the effective diffusivity.

The Bruggeman model stands out as the most widely employed expression in the field \cite{xuReduceddimensionDynamicModel2021,xingNumericalAnalysisOptimum2015,yangMatchingWaterTemperature2011,fanCharacteristicsPEMFCOperating2017,fishmanHeterogeneousThroughPlaneDistributions2011,wangInvestigationDryIonomer2020,namMicroporousLayerWater2009,ohayreFuelCellFundamentals2016}. It introduces a pore structure coefficient denoted as $\tau$, which characterizes the material's tortuosity. This coefficient exhibits a range of variability from 1.5 to 4.0 \cite{ohayreFuelCellFundamentals2016,xieValidationMethodologyPEM2022}, depending on the configuration of the pore structure. For example, highly 'maze-like' or meandering pore structures result in elevated tortuosity values \cite{ohayreFuelCellFundamentals2016}. Nevertheless, the Bruggeman model tends to overestimate the effective diffusion coefficient of GDLs \cite{zamelCorrelationEffectiveGas2009}, as it is based on the porosity of packed spherical particles rather than the cylindrical fibers constituting the GDLs.
In contrast, Tomadakis and Sotirchos proposed an alternative model designed for randomly oriented fibrous porous media to characterise vapour infiltration through them \cite{fishmanHeterogeneousThroughPlaneDistributions2011}. Based on this, Nam et al. \cite{namMicroporousLayerWater2009} proposed the adoption of the Bruggeman model for the CLs and the Tomadakis and Sotirchos model for the GDLs. They further adjusted these models to incorporate liquid water saturation considerations. Consequently, the effective diffusion coefficient $D_{i/j}^{eff}$ is expressed in \eqref{eq:effective_diffusion_coefficient}.

\begin{equation}
	D_{i/j}^{eff} = 
	\begin{cases}
		\varepsilon^{\tau}\left[ 1 - \texttt{s}\right] ^{\tau} D_{i/j}, &\text {at the CL (Bruggeman model)} \\
		\varepsilon \left[ \frac{\varepsilon - \varepsilon_{p}} {1 - \varepsilon_{p}} \right]^{\alpha} \left[ 1 - \texttt{s} \right]^{2} D_{i/j}, &\text {at the GDL (Tomadakis and Sotirchos model)}
	\end{cases}	
	\label{eq:effective_diffusion_coefficient}
\end{equation}
\nomenclature[A,1]{$D_{i/j}$}{binary diffusivity of two species i and j in open space $(m^{2}.s^{-1})$}
\nomenclature[A,2]{\(\tau\)}{pore structure coefficient}
where $\tau$ is the pore structure coefficient, commonly obtained at $\tau = 1.5$ for PEMFC \cite{fishmanHeterogeneousThroughPlaneDistributions2011}, $D_{i/j}$ $(m^{2}.s^{-1})$ is the binary diffusivity of two species in open space, $\varepsilon_{p}$ is the percolation threshold porosity, obtained at 0.11 \cite{fanCharacteristicsPEMFCOperating2017,fishmanHeterogeneousThroughPlaneDistributions2011}, and $\alpha$ is a fitted value, obtained at 0.521 for in plane direction and at 0.785 for through plan direction \cite{fanCharacteristicsPEMFCOperating2017,fishmanHeterogeneousThroughPlaneDistributions2011}.

$\varepsilon_{p}$ represents the minimum porosity necessary within a porous material to allow for diffusion or permeation. Tomadakis and Sotirchos determined the percolation threshold porosity for a random, two-dimensional (2D) fibrous structure, determining it to be $\varepsilon_{p} = 0.11$. This conclusion was drawn by extrapolating the outcomes of their model \cite{tomadakisViscousPermeabilityRandom2005,fishmanHeterogeneousThroughPlaneDistributions2011}.

A distinct correlation for effective diffusivity also exists; nonetheless, it has been proven to be more accurate exclusively for fuel cells operating at elevated temperatures. PEMFCs are consequently unconcerned with this. This correlation is expressed as $D_{i/j}^{eff} = D_{i/j} \frac{\varepsilon_{gdl}}{\tau}$ \cite{ohayreFuelCellFundamentals2016}.

Another aspect often overlooked in the literature is the compression of GDLs, as detailed by Bao et al. \cite{baoTransportPropertiesGas2021} and discussed in section \ref{subsec:intrinsic_permeability}. Consequently, a modification to the previously proposed model becomes necessary. It is important to clarify that this adjustment concerns only GDLs and not CLs. Due to the elastic properties of GDLs, they deform to a greater extent than CLs, where deformation can be disregarded. Thus, the Tomadakis and Sotichos model, enhanced by the contributions of Bao et al., renamed as the TSB model, provides the effective diffusion coefficient for the GDL, as given by \ref{eq:effective_diffusion_coefficient_TSB}.

\begin{equation}
	\boxed{
		D_{i/j}^{eff} = 
		\begin{cases}
			\varepsilon^{\tau}\left[ 1 - \texttt{s}\right] ^{\tau} D_{i/j}, &\text {at the CL (Bruggeman model)} \\
			\varepsilon \left[ \frac{\varepsilon - \varepsilon_{p}} {1 - \varepsilon_{p}} \right]^{\alpha} \left[ 1 - \texttt{s} \right]^{2} e^{ \beta_{2} \varepsilon_{c}} D_{i/j}, &\text {at the GDL (TSB model)}
		\end{cases}
	}
	\label{eq:effective_diffusion_coefficient_TSB}
\end{equation}
where $\beta_{2}$ represents a fitted parameter that fluctuates based on porosity and the diffusion direction of gases, as outlined in table \ref{table:beta_coefficient}.

\begin{table*}[htb]
	\centering
	\begin{tabular}{|c|c|c|} \hline
		
		$\beta_{2}$
		& in-plane
		& through-plane \\ \hline
		
		$\varepsilon \approx 0.6$
		& -2.05
		& -1.59  \\ \hline
		
		$\varepsilon \approx 0.73$
		& -1.04
		& -0.90  \\ \hline
		
	\end{tabular}
	\caption{Different values of the fitted parameter $\beta_{2}$ according to the porosity and the diffusion direction of gases.}
	\label{table:beta_coefficient}
\end{table*}

Finally, to ensure a comprehensive study, it is assumed that the electrolyte in the CLs exhibits tortuosity characteristics identical to those of the carbon structure.

\subsection{Binary diffusion coefficient: $D_{i/j}$}
\label{subsec:binary_diffusion_coefficient}

As previously stated, diffusion coefficients are typically measured in open space. However, in PEMFCs, gas species are not alone during their transport in CLs and GDLs. They diffuse alongside other species, and this coexistence has an impact on their diffusion behavior. For the sake of simplicity, analyses often focus on only two gases simultaneously, with nitrogen assumed to have no interfering effect. As a result, when two species are transported jointly, they share the same diffusion coefficient. This is why $D_{i/j}$ is referred to as a binary diffusion coefficient.

For a binary system comprising gases i and j, $D_{i/j}$ is a function dependent on temperature, pressure, and the molecular weights of both species \cite{ohayreFuelCellFundamentals2016}. Two close expressions, derived from experimental data, can be found in the literature and are represented as \eqref{eq:binary_diffusion_coef_Ohayre} \cite{ohayreFuelCellFundamentals2016} and \eqref{eq:binary_diffusion_coef_classic} \cite{yangMatchingWaterTemperature2011,fanCharacteristicsPEMFCOperating2017,fishmanHeterogeneousThroughPlaneDistributions2011,wangInvestigationDryIonomer2020}.

\begin{equation}
	\boxed{
		\begin{cases}
			& D_{H_{2}O/H_{2}} = 1.644 \cdot 10^{-4} \left[ \frac{T_{fc}}{333} \right]^{2.334} \left[ \frac{101325}{P} \right] \\
			& D_{H_{2}O/O_{2}} = 3.242 \cdot 10^{-5} \left[ \frac{T_{fc}}{333} \right]^{2.334} \left[ \frac{101325}{P} \right]
		\end{cases}
	} 
	\label{eq:binary_diffusion_coef_Ohayre}
\end{equation}

The expressions \eqref{eq:binary_diffusion_coef_classic}, commonly employed in the literature, originate from a single source that does not provide a clear explanation of its calculation. Additionally, there is a disparity between $D_{vc}$, representing the vapour diffusion coefficient at the cathode, and $D_{O2}$, the dioxygen diffusion coefficient, in \eqref{eq:binary_diffusion_coef_classic}, despite both denoting the binary diffusivity of vapour and dioxygen in the GDLs, implying they should be equal. This disparity may be attributed to the presence of nitrogen in the fuel cell; however, these studies do not furnish explanations on this matter.

\begin{subequations}
	\begin{equation}
		D_{H_{2}O/H_{2}} = 1.005 \cdot 10^{-4} \left[ 	\frac{T_{fc}}{333} \right]^{1.75} \left[ \frac{101325}{P} \right] 
	\end{equation}
	\begin{equation}
		\begin{cases}
			& D_{vc} = 2.982 \cdot 10^{-5} \left[ 	\frac{T_{fc}}{333} \right]^{1.75} \left[ \frac{101325}{P} \right] \\
			& D_{O_{2}} = 2.652 \cdot 10^{-5} \left[ 	\frac{T_{fc}}{333} \right]^{1.75} \left[ \frac{101325}{P} \right]
		\end{cases}
	\end{equation}
	\label{eq:binary_diffusion_coef_classic}
\end{subequations}

Table \ref{table:binary_diffusion_coef} provides a comparison between the two equations and data sourced from alternative references.

\begin{table*}[htb]
	\centering
	\begin{tabular}{|l|*{5}{c|}} \hline
		
		& O'Hayre \cite{ohayreFuelCellFundamentals2016}
		& Yang \cite{yangMatchingWaterTemperature2011}
		& Hu, Pasaogullari \cite{huAnalyticalCalculationEvaluation2016,pasaogullariTwoPhaseModelingFlooding2005}
		& Jiao \cite{jiaoWaterTransportPolymer2011}
		& Nam, Bultel \cite{namMicroporousLayerWater2009,bultelInvestigationMassTransport2005} \\
		
		& {\scriptsize  (at 353 K and 1.5 atm) } 
		& {\scriptsize  (at 353 K and 1.5 atm) } 
		& {\scriptsize  (at 353 K and 1.5 atm) } 
		& 
		& 								 \\ \hline
		
		$D_{va} \left( m^{2} \cdot s^{-1} \right) $ 
		& {\normalsize $1.256 \cdot 10^{-4}$} 
		& {\normalsize $7.420 \cdot 10^{-5}$}  
		& {\normalsize $5.457 \cdot 10^{-5}$} 
		& {\normalsize $1 \cdot 10^{-5}$} 
		& {\normalsize $\emptyset$} \\ \hline
		
		$D_{vc} \left( m^{2} \cdot s^{-1} \right) $ 
		& {\normalsize $2.477 \cdot 10^{-5}$} 
		& {\normalsize $2.202 \cdot 10^{-5}$}  
		& {\normalsize $2.236 \cdot 10^{-5}$} 
		& {\normalsize $1 \cdot 10^{-5}$} 
		& {\normalsize $\emptyset$} \\ \hline
		
		$D_{O_{2}} \left( m^{2} \cdot s^{-1} \right) $ 
		& {\normalsize $2.477 \cdot 10^{-5}$} 
		& {\normalsize $1.958 \cdot 10^{-5}$}  
		& {\normalsize $1.806 \cdot 10^{-5}$} 
		& {\normalsize $\emptyset$} 
		& {\normalsize $2.9 \cdot 10^{-5}$}  \\ \hline
		
	\end{tabular}
	\caption{Comparison between the values given by the mentioned expressions for the binary diffusion coefficients and values found in other works}
	\label{table:binary_diffusion_coef}
\end{table*}

\subsection{Vapour convective-diffusive flow at the GDL/GC interface: $J_{v,codi}$}
\label{subsec:vapour_convective-diffusive_flow}

To achieve a complete model of matter transports within the cell, it is imperative to account for the sorption flow between the GDL and GC. In this study, this flow is approximated by an alternative, easier to calculate, denoted as the vapour convective-diffusive flow at the GDL/GC interface and represented by $J_{v,codi}$. A subtle distinction exists between these two flows: $J_{v,codi}$ exclusively occurs within the GC, characterizing vapour flow from the GDL/GC interface on the GC side to the core of the GC. Conversely, the sorption flow characterizes vapour flow between two distinct layers, the GDL and the GC. It is reasonable to assume that concentrations at the two interface sides instantaneously balance. Consequently, in the absence of matter accumulation, the sorption flow at the GDL/GC interface equates to the convective-diffusive flow $J_{v,codi}$.

In the literature, $J_{v,codi}$ is commonly denoted as a convective flow \cite{ohayreFuelCellFundamentals2016}. However, the authors find this term to be potentially confusing and have opted to rename it as a convective-diffusive flow. Indeed, this flow arises from the coupling of convective mass transfer driven by the pressure difference between the GCs inlet and outlet, and diffusive mass transfer between the GCs interface and its core. These two flows mainly evolve orthogonally to each other. The mathematical expression for this phenomenon is provided in \eqref{eq:vapour_convective-diffusive_flow}, derived from diffusive theory while incorporating characteristics of external convective flow \cite{taineTransfertsThermiquesIntroduction2014,ohayreFuelCellFundamentals2016}. Further details on the derivation of this equation and the determination of the convective-diffusive mass transfer coefficient $h_{v}$ can be found in \ref{subsec:additional_information_concerning_Jv_codi}.

\begin{equation}
	\bm{J_{v,codi}} = 
	\begin{cases}
		& h_{v} \left[C_{v,gc} - C_{v,gc}^{\text{inter}} \right] \bm{\imath}, \text{at the anode} \\
		& h_{v} \left[ C_{v,gc}^{\text{inter}} - C_{v,gc} \right] \bm{\imath}, \text{at the cathode}
	\end{cases}
	\label{eq:vapour_convective-diffusive_flow}
\end{equation}
\nomenclature[A,1]{$h$}{convective-conductive mass transfer coefficient $(m.s^{-1})$}
where $h_{v}$ $(m.s^{-1})$ is the convective-diffusive mass transfer coefficient of vapour, $C_{v,gc}^{\text{inter}}$ $(mol.m^{-3})$ is the vapour concentration in the GC at its interface with the GDL, and $\bm{\imath}$ is a unit vector along the x-axis. Notably, $h_{v}$ is not an "effective" coefficient, as convective-diffusive flow happens in the GC where vapour moves into an empty space.

To use equation \eqref{eq:vapour_convective-diffusive_flow}, establishing a correlation between $C_{v,gc}^{\text{inter}}$, representing vapour concentration in the GC at its interface with the GDL, and $C_{v,gdl}^{\text{inter}}$, denoting vapour concentration in the GDL at its interface with the GC, is imperative. While $C_{v,gc}^{\text{inter}}$ remains unknown, $C_{v,gdl}^{\text{inter}}$ is accessible through the diffusion theory outlined in \ref{subsec:vapour_diffusive_flow_GDL}. This requirement is similar to the relationship between $\lambda_{eq}$ and $a_{w}$, albeit without matter conversion in this context. To date, a correlation of the form $ C_{v,gc}^{\text{inter}} = f\left( C_{v,gdl}^{\text{inter}} \right) $ is not present in the existing literature. The authors encourage the scientific community to undertake experiments to determine this correlation. Meanwhile, the following simplification is suggested: $C_{v,gc}^{\text{inter}} = C_{v,gdl}^{\text{inter}}$, leading to equation \eqref{eq:vapour_convective-conductive_flow_final}.

\begin{equation}
	\boxed{
		\bm{J_{v,codi}} = 
		\begin{cases}
			& h_{v} \left[C_{v,gc} - C_{v,gdl}^{\text{inter}} \right] \bm{\imath}, \text{at the anode} \\
			& h_{v} \left[ C_{v,gdl}^{\text{inter}} - C_{v,gc} \right] \bm{\imath}, \text{at the cathode}
	\end{cases}}
	\label{eq:vapour_convective-conductive_flow_final}
\end{equation}
where $C_{v,gdl}^{\text{inter}}$ $(mol.m^{-3})$ is the vapour concentration in the GDL at its interface with the GC. 

Finally, an implicit assumption is considered when referring to a convective-diffusive flow. The dividing line, or boundary delineating the convective-dominated flow within the core of the GC and the diffusive-dominated flow within the core of the electrodes, is presumed to occur at the interface between the GC and the GDL. This assumption is reasonable for medium current density operation (approximately $1$–$1.5$ $A.cm^{-2}$). However, its accuracy is contingent upon various factors, such as flow conditions, flow channel geometry, or electrode structure. For instance, under very low gas velocities in the GC, the diffusion layer may extend into the middle of the gas channels. Conversely, at extremely high gas velocities, convective mixing may infiltrate the electrode itself, causing the diffusion layer to recede. Nevertheless, precisely defining its location proves challenging, and determining the true diffusion layer thickness under such conditions necessitates sophisticated models \cite{ohayreFuelCellFundamentals2016}, such as the one presented by Kim et al. \cite{kimModelingTwophaseFlow2017}. These models are not incorporated into the scope of this study.

\subsection{Water effective convective-diffusive mass transfer coefficient: $h_{v}$}
\label{subsec:water_effective_convective-diffusive_mass_transfer_coefficient}

To calculate $h_{v}$, it is common to use the Sherwood number $S_{h}$, establishing a connection between $h_{v}$ and $D_{v}$ as depicted in \eqref{eq:Sherwood_number_def} \cite{ohayreFuelCellFundamentals2016}. The Sherwood number, a dimensionless parameter, is employed in mass-transfer operations to compare convective-diffusion with classical diffusion.

\begin{equation}
	h_{v} = S_{h} \frac{D_{v}}{H_{gc}}
	\label{eq:Sherwood_number_def}
\end{equation}
\nomenclature[A,1]{$S_{h}$}{Sherwood number}

Then, by fitting the data provided by O'Hayre \cite{ohayreFuelCellFundamentals2016} with a correlation coefficient of $R^{2} = 0.9869$, the authors derived the following expression for $S_{h}$, which exclusively depends on channel geometry. Nevertheless, equation \eqref{eq:Sherwood_number_expression} is applicable only under the assumption of uniform gas density along the channel.

\begin{equation}
	S_{h} = 0.9247 \cdot \ln \left( \frac{W_{gc}}{H_{gc}} \right) + 2.3787, \text{for} \frac{W_{gc}}{H_{gc}} \in \left[ 0.2,10.0 \right] 
	\label{eq:Sherwood_number_expression}
\end{equation}
\nomenclature[A,1]{$W_{gc}$}{width of the gas channel $(m)$}
where $W_{gc}$ $(m)$ is the width of the gas channel.

\subsection{Vapour concentration dynamic behavior in the CL and GDL}

After examining the aforementioned phenomena, the dynamic behavior of vapour concentration can be given. Equation \eqref{subeq:vapour_dynamic_behaviour_balance} represents a molar balance of vapour in the CL or the GDL, while \eqref{subeq:vapour_dynamic_behaviour_border} aligns with the boundary conditions at the CL/membrane and the GDL/GC interfaces.

\begin{subequations}
	\begin{equation}
		\varepsilon \frac{\partial}{\partial t} \left( \left[ 1 - \texttt{s} \right] C_{v} \right) = - \bm{\nabla} \cdot \bm{J_{v,dif}} - S_{sorp} - S_{vl}
		\label{subeq:vapour_dynamic_behaviour_balance}
	\end{equation}	
	\begin{equation}
		\begin{cases}
			\bm{J_{v}^{cl,mem}} = \bm{0}, \text{\small{at the ionomer border}} \\
			\bm{J_{v}^{gdl,gc}} = \bm{J_{v,codi}}, \text{\small{at the GDL/GC border}}
		\end{cases}
		\label{subeq:vapour_dynamic_behaviour_border}
	\end{equation}
\end{subequations}

\section{Vapour transport in the GC}
\label{sec:vapour_GC}

To complete water evolution in the stack, water concentrations in the gas channels must be considered. 

\subsection{Vapour convective flow in the GC : $J_{v,conv}$}

The flow in the gas channels is convection dominated, and the driving force is the pressure at the flow channel inlets \cite{jiaoWaterTransportPolymer2011}. Within the GC, convection ensures that the gas streams are well mixed, preventing the occurrence of concentration gradients \cite{ohayreFuelCellFundamentals2016}. Then, GC being similar to a classical pipe, $J_{v,conv}$ is simply expressed as \eqref{eq:vapour_convective_flow_GC}. 

\begin{equation}
	\bm{J_{v,conv}} = C_{v} \bm{u_{g}}
	\label{eq:vapour_convective_flow_GC}
\end{equation}

where $\bm{u_{g}}$ $(m.s^{-1})$ is the gas mixture velocity. 
It is assumed, as a hypothesis, that all gases evolve collectively at the same velocity and are not independent. However,  $\bm{u_{g}}$ is not necessarily fixed. The calculation of the gas mixture velocity is not explicitly detailed in this study, as gas transport in flow-fields adheres to classical fluid mechanic equations and is highly dependent on the chosen geometry for the GC.
Various GC configurations, such as interdigitated, serpentine, baffle, or porous flow fields, exist and continue to evolve over time. Each configuration has a significant and distinct impact on stack performance. However, the specifics of the gas mixture velocity calculation are beyond the scope of this study which does not delve into the detailed analysis of each configuration. However, it is essential to keep in mind that the choice of GC geometry plays a crucial role in gas transport within flow-fields. 

\subsection{Simplified vapour flows at the inlet and outlet of the GC: $J_{v,in/out}^{gc}$ }
\label{subsec:simplified_vapour_flow}

In real conditions, gas flows at the inlet and outlet of the GC are dependent on the auxiliary system: the nozzles, manifolds, humidifiers, pump and compressor. Other components can also be added depending on the installation that is simulated. Pukrushpan et al. proposed a very simple model involving these components \cite{pukrushpanControlOrientedModelingAnalysis2004}, which has been refined by Xu et al. \cite{xuRobustControlInternal2017} and Shao and al. \cite{shaoComparisonSelfHumidificationEffect2020} while remaining simple. However, this significantly complicates the calculation of the inlet and outlet gas flows. Additionally, modeling the auxiliaries serves solely to determine the error and delay they introduce in the matter supply to the stack, in comparison to the operator's instructions. They are not necessary for modeling the stack's operation. Thus, during the building of a simulation, it is a better to have first a simplified model for these flows to check the accuracy of the matter transport simulated in the stack. The equations mentioned in this study are already numerous, complex, and dependent of one another. Thus, being able to verify the algorithm before using a more complex model is desirable. To this end, simplified equations inspired by Pukrushpan's work \cite{pukrushpanControlOrientedModelingAnalysis2004} for the inlet and outlet water flows at the GC are proposed here.

At the stack inlets, the flows are directly equal to the setpoints normally imposed on the reducer (for the anode) and the compressor (for the cathode). A demonstration of these expressions is provided in \ref{subsec:simplified_flows_inlet_outlet_AGC} and \ref{subsec:simplified_flows_inlet_outlet_CGC}. To achieve this, it is simply assumed that the flows must be proportional to the consumed current, with a proportionality coefficient named stoichiometry $S_a/S_c$. It is considered that the gases are ideal, the channels of the GC are cubic, the incoming gases already have a humidity equal to the desired one $\Phi_{des}$, and the pressure of the incoming gases is equal to the pressure at the GC inlet, with pressure losses being neglected. The equations are shown in \eqref{eq:vapour_inlet_and_outlet_flows}.

At the stack outlets, the flows are not directly controlled by a machine, as it is the case at the inlets. They naturally evacuate the stack due to the pressure difference with the external environment. The outlet flows are therefore expressed based on this pressure difference. For simplicity, they are considered proportional to the latter, with $k_{em,\text{in}}$ the proportional constant. This is valid for compressible, adiabatic, and steady flows. The gases must also be ideal, and pressure losses neglected \cite{pukrushpanControlOrientedModelingAnalysis2004}. These are however strong assumptions. The back-pressure valve is indirectly modeled here by assuming that the outlet pressures of the gas channels directly match the desired pressures $P_{des}$. Finally, balances similar to those explained in \ref{subsec:simplified_flows_inlet_outlet_AGC} and \ref{subsec:simplified_flows_inlet_outlet_CGC} lead to equations \eqref{eq:vapour_inlet_and_outlet_flows}.

\begin{equation}
	\begin{cases}
		J_{v,\text{in}}^{agc} = \frac{\Phi_{a,\text{des}} P_{sat}}{P_{\text{agc,in}} - \Phi_{a,\text{des}} P_{sat}} \frac{A_{act}}{H_{gc} W_{gc}} \frac{S_{a} \left[i_{fc} + i_{n}\right]}{2F} \\
		
		J_{v,\text{out}}^{agc} = \frac{\Phi_{agc,out} P_{sat}}{P_{agc,out}} \frac{k_{em,\text{in}}}{H_{gc} W_{gc} M_{agc,out}} \left[P_{agc,out} - P_{a,\text{des}}\right]   \\
		
		J_{v,\text{in}}^{cgc} = \frac{\Phi_{c,\text{des}} P_{sat}}{P_{\text{cgc,in}} - \Phi_{c,\text{des}} P_{sat}} \frac{1}{y_{O_{2},ext}} \frac{A_{act}}{H_{gc} W_{gc}} \frac{S_{c} \left[i_{fc} + i_{n}\right]}{4F} \\
		
		J_{v,\text{out}}^{cgc} = \frac{\Phi_{cgc,out} P_{sat}}{P_{cgc,out}} \frac{k_{em,\text{in}}}{H_{gc} W_{gc} M_{cgc,out}} \left[P_{cgc,out} - P_{c,\text{des}}\right]
	\end{cases}	
	\label{eq:vapour_inlet_and_outlet_flows}
\end{equation}
\nomenclature[A,1]{$A_{act}$}{active area $(m^{2})$}
\nomenclature[A,1]{$y_{O_{2}}$}{molar fraction of $O_{2}$ in dry air}
\nomenclature[A,1]{$S_{a}/S_{c}$}{stoichiometric ratio at the anode/cathode}
\nomenclature[A,2]{\(\Phi\)}{relative humidity}
where $A_{act}$ ($m^{2}$) is the active area, $\Phi_{des}$ and $P_{des}$ ($Pa$) are the desired humidity and pressure fixed by the user, $P_{gc}$ is the GC total pressure, $y_{O_{2},ext}$ is the molar fraction of $O_{2}$ in dry air, $M_{gc,out}$ is the molar mass of the gas mixture at the GC exit, and $k_{em,\text{in}}$ ($kg.s^{-1}.Pa^{-1}$) is the exhaust manifold inlet orifice constant, usually taken between $\left[3.5, 8.0\right] \times 10^{-6}$ $kg.s^{-1}.Pa^{-1}$ \cite{pukrushpanControlOrientedModelingAnalysis2004,xuRobustControlInternal2017}.

\subsection{Vapour dynamic behavior in the GC}

Finally, assuming that no phase change occurs in the GC, the following dynamic behaviour of vapour concentration can be obtained. \eqref{eq:vapour_dynamic_behaviour_GC} corresponds to a molar balance of vapour in the GC and \eqref{eq:vapour_dynamic_behaviour_GC_boundary_conditions} matches the boundary conditions at the GDL/GC interface, inlet, and outlet of the GC.

\begin{subequations}
	\begin{equation}
		\frac{\partial C_{v}}{\partial t} = - \bm{\nabla} \cdot \bm{J_{v,conv}}
		\label{eq:vapour_dynamic_behaviour_GC}
	\end{equation}
	\begin{equation}
		\begin{cases}
			\bm{J_{v}^{gdl,gc}} = \bm{J_{v,codi}}, \text{at the GDL/GC border} \\
			J_{v}^{in/out,gc} = J_{v,\text{in/out}}^{gc}, \text{at the inlet/outlet of the GC} 
			\label{eq:vapour_dynamic_behaviour_GC_boundary_conditions}
		\end{cases}
	\end{equation}
\end{subequations}

\section{Hydrogen and oxygen transports}

The behaviors of hydrogen and oxygen closely resemble vapor transport in the cell. Consequently, the following governing equations are presented without further explanations.

\subsection{Hydrogen and oxygen flows: $J_{H_{2},dif}$, $J_{H_{2},codi}$, $J_{H_{2},conv}$, $J_{O_{2},dif}$, $J_{O_{2},codi}$, $J_{O_{2},conv}$}
\label{subsec:hydrogen_oxygen_flows}

Hydrogen diffusive, convective-diffusive, convective flows, and inlet and outlet flows at the AGC are respectively expressed in \eqref{eq:hydrogen_flows}. 

\begin{equation}
	\begin{cases}
		\bm{J_{H_{2}, dif}} = - D_{H_{2}}^{eff} \bm{\nabla} C_{H_{2}} \\
		
		\bm{J_{H_{2},codi}} = h_{H_{2}} \left[ C_{H_{2},\text{agc}} - C_{H_{2},\text{agdl}}^{\text{inter}} \right] \bm{\imath} \\
		
		\bm{J_{H_{2},conv}} = C_{H_{2}} \bm{u_{g}} \\
		
		J_{H_{2},\text{in}} = \frac{A_{act}}{H_{gc} W_{gc}} \frac{S_{a} \left[i_{fc} + i_{n}\right]}{2F} \\
		
		J_{H_{2},\text{out}} = \frac{P_{\text{agc,out}} - \Phi_{\text{agc,out}} P_{sat}}{P_{\text{agc,out}}} \frac{k_{em,\text{in}}}{H_{gc} W_{gc} M_{agc,out}} \left[P_{agc,out} - P_{a,\text{des}}\right]
	\end{cases}
	\label{eq:hydrogen_flows}
\end{equation}

Oxygen diffusive, convective-diffusive, convective flows, and inlet and outlet flows at the CGC are respectively expressed in \eqref{eq:oxygen_flows}. 

\begin{equation}
	\begin{cases}
		\bm{J_{O_{2}, dif}} = - D_{O_{2}}^{eff} \bm{\nabla}C_{O_{2}} \\
		
		\bm{J_{O_{2},codi}} = h_{O_{2}} \left[ C_{H_{2},\text{cgdl}}^{\text{inter}} - C_{O_{2},\text{cgc}} \right] \bm{\imath} \\
		
		\bm{J_{O_{2},conv}} = C_{O_{2}} \bm{u_{g}} \\
		
		J_{O_{2},\text{in}} = \frac{A_{act}}{H_{gc} W_{gc}} \frac{S_{c} \left[i_{fc} + i_{n}\right]}{4F}  \\
		
		J_{O_{2},\text{out}} = y_{\text{cgc,out}}  \frac{P_{\text{cgc,out}} - \Phi_{\text{cgc,out}} P_{sat}}{P_{\text{cgc,out}}} \frac{k_{em,\text{in}}}{H_{gc} W_{gc} M_{cgc,out}} \left[P_{cgc,out} - P_{c,\text{des}}\right]  \\
	\end{cases}
	\label{eq:oxygen_flows}
\end{equation}

\subsection{Hydrogen and oxygen consumption at the interface of the triple points: $S_{H_{2},cons}$}
\label{subsec:hydrogen_oxygen_consumption}

Hydrogen and oxygen consumption are respectively expressed as \eqref{eq:hydrogen_consumption} and \eqref{eq:oxygen_consumption}.

\begin{equation}
	S_{H_{2},cons} = 
	\begin{cases}
		- \frac{i_{fc}}{2 F H_{cl}}, & \text{in the ACL} \\
		0, & \text{elsewhere} 
	\end{cases}
	\label{eq:hydrogen_consumption}
\end{equation}

\begin{equation}
	S_{O_{2},cons} = 
	\begin{cases}
		- \frac{i_{fc}}{4 F H_{cl}}, & \text{in the CCL} \\
		0, & \text{elsewhere} 
	\end{cases}
	\label{eq:oxygen_consumption}
\end{equation}

It is possible to incorporate mathematical terms into the equations \eqref{eq:hydrogen_consumption} and \eqref{eq:oxygen_consumption} for accounting crossover effects. The membrane, chosen for its impermeability to gases, aims to prevent direct mixing of hydrogen and oxygen, which is crucial for electricity generation and safety. However, it is not perfectly impermeable, allowing a small gas transfer in both directions. Consequently, a portion of the hydrogen and oxygen, initially designated for participation in the fuel cell mechanism, instead traverse the membrane and react directly, forming water. Thus, a portion of the fuels is lost. This phenomenon is commonly known as crossover, and the equation describing the additional water production due to it is delineated in section \ref{subsec:water_production}.

The crossover flows, denoted as $S_{H_{2},co}$ and $S_{O_{2},co}$, are expressed in units of $mol.m^{-3}.s^{-1}$. It is more relevant to conceptualize them as volume flows since they traverse the membrane via the dispersed CL ionomer within the CL volume. The calculation of these flows classically involves the application of Fick's law across the two interfaces of the membrane \cite{weberTransportPolymerElectrolyteMembranes2004,giner-sanzHydrogenCrossoverInternal2014,namNumericalAnalysisGas2010,trucNumericalExperimentalInvestigation2018,ahluwaliaBuildupNitrogenDirect2007,kochaCharacterizationGasCrossover2006}, as illustrated in equations \ref{eq:crossover_flow_H2} and \ref{eq:crossover_flow_O2}.

\begin{equation}
	S_{H_{2},co} = 
	\begin{cases}
		k_{H_{2}} \frac{R T_{fc}}{H_{cl}} \nabla C_{H_{2}}, & \text{in the ACL}\\
		0, & \text{elsewhere}
	\end{cases}
	\label{eq:crossover_flow_H2}
\end{equation}
\begin{equation}
	S_{O_{2},co} = 
	\begin{cases}
		k_{O_{2}}\frac{ R T_{fc}}{H_{cl}} \nabla C_{O_{2}}, & \text{in the CCL}\\
		0, & \text{elsewhere}
	\end{cases}
	\label{eq:crossover_flow_O2}
\end{equation}
\nomenclature[A,1]{$k$}{permeability coefficient in the membrane $(mol.m^{-1}.s^{-1}.Pa^{-1})$}
where $k_{i}$ $(mol.m^{-1}.s^{-1}.Pa^{-1})$ is the permeability coefficient of molecule i (hydrogen or oxygen) in the membrane.

An experimental expression for these permeability coefficients was proposed by Weber et al. in 2004 \cite{weberTransportPolymerElectrolyteMembranes2004, ahluwaliaBuildupNitrogenDirect2007,trucNumericalExperimentalInvestigation2018}, providing the most accurate prediction to date with a coefficient function of both $\lambda$ and $T_{fc}$. Kocha et al. conducted another experiment in 2006 \cite{kochaCharacterizationGasCrossover2006}, but did not account for the variation of $\lambda$ in $k_{i}$. Gas permeability in PEM fuel cells strongly depends on membrane hydration level and temperature, leading to fluctuations in $k_{i}$ with changes in operating conditions \cite{kochaCharacterizationGasCrossover2006}. The Weber proposal is expressed by equations \eqref{eq:permeability_coefficients_crossover_H2} and \eqref{eq:permeability_coefficients_crossover_O2}.

\begin{equation}
	k_{H_{2}} = 
	\begin{cases}
		\left[ 0.29 + 2.2 f_{v}\left( \lambda \right) \right] 10^{-14} \exp \left( \frac{E_{act,H_{2},v}}{R} \left[ \frac{1}{T_{ref}} - \frac{1}{T_{fc}} \right]\right) & if \lambda < \lambda_{l,eq}  \\
		1.8 \cdot 10^{-14} \exp \left( \frac{E_{act,H_{2},l}}{R} \left[ \frac{1}{T_{ref}} - \frac{1}{T_{fc}} \right]\right) & if \lambda = \lambda_{l,eq} \\
	\end{cases}
	\label{eq:permeability_coefficients_crossover_H2}
\end{equation}
\begin{equation}
	k_{O_{2}} = 
	\begin{cases}
		\left[ 0.11 + 1.9 f_{v}\left( \lambda \right) \right] 10^{-14} \exp \left( \frac{E_{act,O_{2},v}}{R} \left[ \frac{1}{T_{ref}} - \frac{1}{T_{fc}} \right]\right) & if \lambda < \lambda_{l,eq}  \\
		1.2 \cdot 10^{-14} \exp \left( \frac{E_{act,O_{2},l}}{R} \left[ \frac{1}{T_{ref}} - \frac{1}{T_{fc}} \right]\right) & if \lambda = \lambda_{l,eq}
	\end{cases}
	\label{eq:permeability_coefficients_crossover_O2}
\end{equation}
where $E_{act,H_{2},v} = 2.1 \cdot 10^{4} J.mol^{-1}$ and $E_{act,O_{2},v} = 2.2 \cdot 10^{4} J.mol^{-1}$ are the crossover activation energies of hydrogen and oxygen for an under saturated membrane, $E_{act,H_{2},l} = 1.8 \cdot 10^{4} J.mol^{-1}$ and $E_{act,O_{2},l} = 2.0 \cdot 10^{4} J.mol^{-1}$ are the crossover activation energies of hydrogen and oxygen for a liquid-equilibrated membrane, $T_{ref} = 303.15 K$ is a referenced temperature, and $f_{v}$ is the water volume fraction of the membrane described in \ref{subsec:Water_flows_at_the_ionomer/CL_interface}.

After hydrogen and oxygen molecules traverse the membrane, their consumption needs to be considered, denoted as $S_{i,wasted}$. Existing equations in the literature simplify this process by assuming instantaneous passage of matter through the membrane, disregarding its thickness, and immediate reaction with its complementary molecule to form water \cite{namNumericalAnalysisGas2010}. Under this assumption, the equations can be expressed as \eqref{eq:wasted_flow_H2} and \eqref{eq:wasted_flow_O2}.

\begin{equation}
	S_{H_{2},wasted} = 
	\begin{cases}
		- 2 \cdot S_{O_{2},co}, & \text{in the ACL}\\
		0, & \text{elsewhere}
	\end{cases}
	\label{eq:wasted_flow_H2}
\end{equation}
\begin{equation}
	S_{O_{2},wasted} = 
	\begin{cases}
		- 0.5 \cdot S_{H_{2},co}, & \text{in the CCL}\\
		0, & \text{elsewhere}
	\end{cases}
	\label{eq:wasted_flow_O2}
\end{equation}

Finally, the corrected formulations of $S_{i,cons}$, incorporating the short-circuited current density $i_{sc}$, are expressed in \eqref{eq:hydrogen_consumption_corrected} and \eqref{eq:oxygen_consumption_corrected}.

\begin{equation}
	S_{H_{2},cons} = 
	\begin{cases}
		- \frac{i_{fc} + i_{sc}}{2 F H_{cl}} - \frac{R T_{fc}}{H_{cl}} \left[ k_{H_{2}} \nabla C_{H_{2}} + 2 k_{O_{2}} \nabla C_{O_{2}} \right], & \text{in the ACL} \\
		0, & \text{elsewhere} 
	\end{cases}
	\label{eq:hydrogen_consumption_corrected}
\end{equation}

\begin{equation}
	S_{O_{2},cons} = 
	\begin{cases}
		- \frac{i_{fc} + i_{sc}}{4 F H_{cl}} - \frac{ R T_{fc}}{H_{cl}} \left[ k_{O_{2}} \nabla C_{O_{2}} + \frac{k_{H_{2}}}{2} \nabla C_{H_{2}} \right], & \text{in the CCL} \\
		0, & \text{elsewhere} 
	\end{cases}
	\label{eq:oxygen_consumption_corrected}
\end{equation}

\subsection{Hydrogen and oxygen concentration dynamic behavior in the CL and GDL}

The hydrogen dynamic behaviour is given by the molar balance of $H_{2}$ in \eqref{subeq:hydrogen_dynamic_behaviour_balance} and the boundary conditions at the CL/membrane, GDL/GC interfaces, and inlet/outlet of the GC in \eqref{subeq:hydrogen_dynamic_behaviour_border}.

\begin{subequations}
	\begin{equation}
		\begin{cases}
			\varepsilon \frac{\partial}{\partial t} \left( \left[  1 - \texttt{s} \right] C_{H_{2}} \right)  = - \bm{\nabla} \cdot \bm{J_{H_{2},dif}} + S_{H_{2},cons}, & \text{in the anode} \\
			\frac{\partial C_{H_{2}}}{\partial t} = - \bm{\nabla} \cdot \bm{J_{H_{2},conv}}, & \text{in the AGC} 
		\end{cases}
		\label{subeq:hydrogen_dynamic_behaviour_balance}
	\end{equation}	
	\begin{equation}
		\begin{cases}
			\bm{J_{H_{2}}^{cl,mem}} = \bm{0}, &\text{\small{at the CL/membrane border}} \\
			\bm{J_{H_{2}}^{gdl,gc}} = \bm{J_{H_{2},codi}}, &\text{\small{at the GDL/GC border}} \\
			J_{H_{2}}^{in/out,gc} = J_{H_{2}, \text{in/out}}, &\text{at the inlet/outlet of the GC}
		\end{cases}
		\label{subeq:hydrogen_dynamic_behaviour_border}
	\end{equation}
\end{subequations}

The oxygen dynamic behaviour is given by the molar balance of $O_{2}$ in \eqref{subeq:oxygen_dynamic_behaviour_balance} and the boundary conditions at the CL/membrane, GDL/GC interfaces, and inlet/outlet of the GC in \eqref{subeq:oxygen_dynamic_behaviour_border}.

\begin{subequations}
	\begin{equation}
		\begin{cases}
			\varepsilon \frac{\partial}{\partial t} \left( \left[  1 - \texttt{s} \right] C_{O_{2}} \right)  = - \bm{\nabla} \cdot \bm{J_{O_{2},dif}} + S_{O_{2},cons}, & \text{in the cathode} \\
			\frac{\partial C_{O_{2}}}{\partial t} = - \bm{\nabla} \cdot \bm{J_{O_{2},conv}}, & \text{in the CGC} 
		\end{cases}
		\label{subeq:oxygen_dynamic_behaviour_balance}
	\end{equation}	
	\begin{equation}
		\begin{cases}
			\bm{J_{O_{2}}^{cl,mem}} = \bm{0}, &\text{at the CL/membrane border} \\
			\bm{J_{O_{2}}^{gdl,gc}} = \bm{J_{O_{2},codi}}, &\text{at the GDL/GC border} \\
			J_{O_{2}}^{in/out,gc} = J_{O_{2}, \text{in/out}}, &\text{at the inlet/outlet of the GC} 
		\end{cases}
		\label{subeq:oxygen_dynamic_behaviour_border}
	\end{equation}
\end{subequations}

\section{Nitrogen transport}

For the modeling of nitrogen transport, it is crucial to assume homogeneity of $N_{2}$ throughout the stack (CCL, CGDL and CGC). This assumption facilitates the use of binary coefficients for the calculation of oxygen and water flows at the cathode. Consequently, in the following differential equation as depicted in \eqref{eq:nitrogen_dynamic_behaviour_agc}, the control volume encompasses the cathode electrode and CGC volume. The internal nitrogen flow is disregarded, which is a reasonable assumption as no nitrogen is consumed in this process. Moreover, this study does not address $N_{2}$ crossover, as its relevance is confined to specific modeling tasks. For further details, please refer to \cite{baikCharacterizationNitrogenGas2011}.

Thus, nitrogen evolution fully depends on the inlet and outlet flows at the CGC. Similarly to vapor discussed in section \ref{sec:vapour_GC}, simplifications of these flows are suggested to obtain preliminary results before incorporating auxiliaries.

\subsection{Simplified nitrogen concentration flows at the inlet and outlet of the CGC}

The nitrogen inlet concentration flow in the stack is represented by \eqref{eq:nitrogen_agc_in}, while the outlet concentration flow is denoted by \eqref{eq:nitrogen_agc_out}. Demonstrations are provided in the appendix.

\begin{equation}
	W_{N_{2},\text{in}} = \frac{1 - y_{O_{2},ext}}{y_{O_{2},ext}} \frac{A_{act}}{H_{gc} W_{gc}} \frac{S_{c} \left[i_{fc} + i_{n}\right]}{4F}
	\label{eq:nitrogen_agc_in}
\end{equation}

\begin{equation}
	W_{N_{2},\text{out}} = \left[ 1 - y_{O_{2},\text{cgc,out}} \right] \frac{P_{\text{cgc,out}} - \Phi_{\text{cgc,out}} P_{sat}}{P_{\text{cgc,out}}} \frac{k_{em,\text{in}}}{H_{gc} W_{gc} M_{cgc,out}} \left[P_{cgc,out} - P_{c,\text{des}}\right]
	\label{eq:nitrogen_agc_out}
\end{equation}
\nomenclature[A,1]{$L_{gc}$}{cumulated length of the gas channel $(m)$}
where $L_{gc}$ is the cumulated length of the gas channel.

\subsection{Nitrogen concentration dynamic behavior in the cathode}

The nitrogen dynamic behaviour in the cathode is expressed as \eqref{eq:nitrogen_dynamic_behaviour_agc}.

\begin{equation}
	\frac{\mathrm{dC}_{\mathrm{N_{2}}}}{\mathrm{dt}} = W_{N_{2},\text{in}} - W_{N_{2},\text{out}} 
	\label{eq:nitrogen_dynamic_behaviour_agc}
\end{equation}

\section{Voltage polarisation}

In the literature, the current density $i_{fc}$ is typically imposed independently of other variables, with the resulting voltage then being calculated or measured. However, it is important to acknowledge that under extreme conditions, such as severe fuel starvation or intense membrane drying, it becomes impractical to maintain a fixed current density, leading to the stack ceasing operation. Utilizing a model that consistently imposes a fixed current density may yield inaccurate results, including negative voltages. Since this study does not consider such extreme scenarios, it is crucial to work with acceptable values for both operating conditions and current density.

\subsection{The apparent voltage: $U_{cell}$}
\label{subsec:apparent_voltage}

To determine the apparent voltage $U_{cell}$ in a fuel cell, various phenomena must be taken into account. 
Initially, the equilibrium voltage $U_{eq}$ defines the maximum energy available in the reaction $H_{2}(g) + \frac{1}{2}O_{2}(g) \rightarrow H_{2}O$ $(l)$ through thermodynamics. This equation also implicitly contributes to concentration losses, arising from insufficient fuel stored in the CLs due to gas diffusion limitations, which cannot counterbalance excessive fuels demand at high current density. This aspect is elaborated upon in section \ref{subsec:equilibrium_potential}.

Consequently, multiple voltage losses need consideration. The overpotential $\eta$ encompasses kinetic losses from the redox reactions, fuel crossover, internal short circuit, and contributes to concentration losses. Kinetic losses serve to accelerate the rate-limiting step from redox reactions, ensuring molecules in the CLs are appropriately directed to triple point areas and decomposed into ions. Voltage losses from fuel crossover and internal short circuits result from membrane imperfections, allowing a portion of fuels and electrons to pass through, translating to an energy loss. This discussion is presented in section \ref{subsec:overpotential}.

Finally, both the electrical resistances of protons $R_p$ and electrons $R_e$ counterbalance the equilibrium voltage. These voltage losses originate from the transport of electric charges which experience resistances from the materials in which they move. They result from microscopic collisions between electric charges and materials. The resistances associated with protons, denoted as $R_p$, are distinguished from those associated with electrons, denoted as $R_e$, because the charges and materials are different. Indeed, the transport of protons through the membrane and CLs to the triple point areas is much more resistant than the transport of electrons through the GDLs and bipolar plates, which are good electrical conductors. An expression for $R_p$ is provided in section \ref{subsec:proton_conductive_resistance}. However, to the best of the authors' knowledge, there is no widely disseminated expression for $R_e$ in the literature. This quantity is generally either neglected or considered as an undetermined parameter that must be calibrated with experimental data from the specific fuel cell under study.

Based on this theory, the following relation \eqref{eq:apparent_voltage} is obtained for calculating the apparent cell voltage \cite{dicksFuelCellSystems2018,jiaoWaterTransportPolymer2011,xuReduceddimensionDynamicModel2021,ohayreFuelCellFundamentals2016}: 

\begin{equation}
	U_{cell} = U_{eq} - \eta_{c} - i_{fc} \left[ R_{p} + R_{e} \right] 
	\label{eq:apparent_voltage}
\end{equation}
\nomenclature[A,1]{$U$}{voltage $(V)$}
\nomenclature[A,1]{$R_{e}/R_{p}$}{electron/proton conduction resistance $(\Omega.m^{2})$}
\nomenclature[A,2]{\(\eta\)}{overpotential $(V)$}
where $U_{cell}$ $(V)$ is the cell voltage, $U_{eq}$ $(V)$ is the equilibrium voltage, $\eta_{c}$ $(V)$ is the cathode overpotential, $R_{p}$ $(\Omega.m^{2})$ is the area specific resistance of the protons, and $R_{e}$ $(\Omega.m^{2})$ is the area specific resistance of the electrons.

\subsection{Equilibrium potential at the cathode $U_{eq}$}
\label{subsec:equilibrium_potential}

The theoretical maximum energy extractable from chemical reactions is represented by the Gibbs free energy, grounded in thermodynamics. Subsequently, it is reformulated to adopt the expression of a potential, denoted as the equilibrium potential $U_{eq}$ or the Nernst equation, as shown in \ref{eq:equilibrium_voltage}. In literature, the convention is to designate the anode potential as zero. As a result, the equilibrium voltage is equivalent to the cathode equilibrium potential \cite{jiaoWaterTransportPolymer2011,fanCharacteristicsPEMFCOperating2017,pukrushpanControlOrientedModelingAnalysis2004,baoTwodimensionalModelingPolymer2015,wangInvestigationDryIonomer2020,ohayreFuelCellFundamentals2016,yangEffectsOperatingConditions2019}.

\begin{figure}[H]
	\begin{equation}
		U_{eq} = V_{eq}^{c} = E^{0} - 8.5 \cdot 10^{-4} \left[ T_{fc} - 298.15 \right] + \frac{R T_{fc}}{2 F} \left[ \ln \left( \frac{R T_{fc} C_{H_{2},\text{acl}}}{P_{\text{ref}}} \right) + \frac{1}{2} \ln \left( \frac{R T_{fc} C_{O_{2},\text{ccl}}}{P_{\text{ref}}} \right)\right]
		\label{eq:equilibrium_voltage}
	\end{equation} 
\end{figure} 
\nomenclature[A,1]{$E^{0}$}{standard-state reversible voltage $(V)$}
where $E^{0}$ $(V)$ is the standard-state reversible voltage taken at $E^{0} = 1.229 V$, $P_{ref}$ $(Pa)$ is the reference pressure taken at $10^{5}$ $Pa$, $C_{H_{2},\text{acl}}$ and $C_{O_{2},\text{ccl}}$ $(mol.m^{-3})$ are the $H_{2}$ concentration at the anode catalyst layer and the $O_{2}$ concentration at the cathode catalyst layer, respectively. 
These concentrations should be taken into account at the CLs, where reactions occur and chemical energy is converted \cite{zihrulVoltageCyclingInduced2016}. By doing so, $U_{eq}$ contributes to calculating the concentration drop, as any reduction in fuel concentrations in the CLs will consequently decrease the apparent voltage. Further details are discussed in section \ref{subsec:concentration_loss}.

\subsection{The overpotential at the cathode $\eta_{c}$}
\label{subsec:overpotential}

The overpotential is a complex quantity that encompasses various voltage drops. All equations describing overpotential currently rely on the Butler-Volmer theory, derived from transition state theory applied to single-electron transfer reactions \cite{bardElectrochemicalMethodsFundamentals2001}. However, redox reactions are more complex, and this assumption is simplifying. Consequently, employing the Butler-Volmer theory to model overpotential entails limitations, such as the inability to account for potential effects of electrode flooding or membrane drying on the overpotential. Furthermore, current knowledge about overpotential is restricted, without the existence of widely accepted alternative theories. Consequently, several researchers have proposed modifications to the Butler-Volmer equation to incorporate other phenomena \cite{fanCharacteristicsPEMFCOperating2017, baoTwodimensionalModelingPolymer2015, xuReduceddimensionDynamicModel2021}. However, these ideas have been criticized by others \cite{dickinsonButlerVolmerEquationPolymer2019}. Section \ref{subsubsec:fair_use_Butler-Volmer_equation} explores the Tafel equation, based on the conventional Butler-Volmer theory, while section \ref{subsubsec:manipulation_Butler-Volmer_equation} discusses the proposed modifications to the equation.

\subsubsection{A fair use of the Butler-Volmer theory}
\label{subsubsec:fair_use_Butler-Volmer_equation}

The oxidation reaction of hydrogen at the anode is significantly faster than the reduction reaction of oxygen at the cathode. Therefore, it is common to neglect the overpotential resulting from this chemical reaction at the anode compared to the overpotential at the cathode. By doing so, the Butler-Volmer theory leads to a simplified equation called the Tafel equation, expressed in \eqref{eq:overpotential_fair_Tafel} \cite{dickinsonButlerVolmerEquationPolymer2019}. It alludes to the pioneering research of Swiss chemist Julius Tafel, who derived this equation from empirical data. 

Given the omnipresence of Tafel equation, the original Butler-Volmer equation is not discussed here. It is however important to note that Dickinson et al. \cite{dickinsonButlerVolmerEquationPolymer2019} have identified a major misuse of the Butler-Volmer equation, which has been widely propagated in the literature, and renamed the Bernardi-Verbrugge formulation. This formulation contains a single concentration-dependence term, utilizing oxygen concentration, which is applied equally between the two processes of oxidation and reduction (each modeled by an exponential), contradicting the essence of the Butler-Volmer equation. This error is concealed in practice by the simplifying assumption of neglecting hydrogen overpotential, but it is crucial not to build a model on the wrong foundation. The authors therefore urge the community to exercise caution in choosing litterature equations.

\begin{figure}[H]
	\begin{equation}
		\boxed{
			i_{fc} + i_{n} =  i_{0,c}^{ref} \left[ \frac{ C_{O_{2},ccl}}{C_{O_{2}}^{ref}} \right]^{\kappa_{c}} \exp \left(\frac{F \alpha_{c}}{R T_{fc}} \eta_{c} \right)
		}
		\label{eq:overpotential_fair_Tafel}
	\end{equation}
\end{figure}

where $i_{n}$ $(A.m^{-2})$ is the internal current density, $i_{0,c}^{ref}$ is the referenced exchange current density at the cathode for a given oxygen concentration $C_{O_{2}}^{ref}$, $C_{O_{2}}^{ref}$ $(mol.m^{-3})$ is the reference concentration of oxygen, $\kappa_{c}$ is the overpotential correction exponent, $\alpha_{c}$ is the charge-transfer coefficient of the cathode, $R$ $(J.mol^{-1}.K^{-1})$ is the universal gas constant.

In equation \eqref{eq:overpotential_fair_Tafel}, the internal current density $i_{n}$ is used to consider the fuel crossover in the membrane and the electronic short circuit \cite{kimDegradationModelingOperational2014, dicksFuelCellSystems2018}. It is deeply explained in section \ref{subsec:internal_current_density}. 

The exchange current density, denoted as $i_{0,c}$, serves to measure the number of chemical reactions occurring at equilibrium within the triple point regions of the CCL, when the fuel cell is not generating current. This quantity is normalized to a current density, as the difference between this value and the imposed current density on the fuel cell provides a measure of the requirement to accelerate the redox reaction and, consequently, the associated kinetic losses.
$i_{0,c}$ is at least a function of the oxygen concentration in the CCL. This dependency is explicitly expressed in the Tafel equation provided in \eqref{eq:overpotential_fair_Tafel}, and further detailed in \eqref{eq:i_0,c_ref}. Therefore, $i_{0,c}^{ref}$ corresponds to the exchange current density measured at a reference oxygen concentration $C_{O_{2}}^{ref}$.
In this study, the overpotential correction exponent denoted as $\kappa_{c}$ is introduced. Indeed, numerous exponent values can be found in the literature, typically ranging between $[0.25, 4.0]$ \cite{xieValidationMethodologyPEM2022,fanCharacteristicsPEMFCOperating2017,baoTwodimensionalModelingPolymer2015,xuReduceddimensionDynamicModel2021,yangEffectsOperatingConditions2019,dickinsonButlerVolmerEquationPolymer2019}, but none of these values appears to be dominant. Then, the authors propose considering $\kappa_{c}$ as an undetermined parameter which should be estimated for each specific fuel cell stack.

\begin{figure}[H]
	\begin{equation}
		i_{0,c} = i_{0,c}^{ref} \left[ \frac{ C_{O_{2},ccl}}{C_{O_{2}}^{ref}} \right]^{\kappa_{c}}
		\label{eq:i_0,c_ref}
	\end{equation}
\end{figure}

The term $exp \left(\frac{F \alpha_{c}}{R T_{fc}} \eta_{c} \right)$ serves as the final component connecting the current density $i_{fc}$ to the overpotential $\eta_{c}$. The parameter $\alpha_{c}$, commonly known as the 'charge-transfer coefficient,' represents the proportion of the electrical  energy utilized to modify the rate of an electrochemical reaction by altering the activation barrier. Its value is contingent upon the specific reaction and electrode material, but it must fall within the 0–1.0 range. Typically, for most electrochemical reactions, $\alpha_{c}$ falls within the approximate range of 0.2–0.5. At the oxygen electrode, there is greater variability in the charge-transfer coefficient, ranging from about 0.1–0.5 in most scenarios. In the case of 'symmetric' reactions, $\alpha_{c}$ is generally considered as 0.5 \cite{dicksFuelCellSystems2018,ohayreFuelCellFundamentals2016}.

Furthermore, it is crucial to incorporate $C_{O_{2},ccl}$ in the calculation of $\eta_{c}$ since the overpotential manifests in the triple point region \cite{zihrulVoltageCyclingInduced2016}. This indirectly contributes to the modeling of concentration losses, as discussed in \ref{subsec:concentration_loss}.

Finally, for improved precision in calculating the model voltage or addressing specific issues such as electrode flooding, it is essential to move away from the use of the Butler-Volmer equation. Instead, adopting a rigorous multi-step mechanism-based model is crucial. However, further extensive research is needed before effectively implementing such a model \cite{dickinsonButlerVolmerEquationPolymer2019}. The addition of terms to the Tafel equation is a temporary proposition embraced by many researchers while awaiting more advanced models. These propositions are detailed in Section \ref{subsubsec:manipulation_Butler-Volmer_equation}. Nonetheless, Dickinson et al. \cite{dickinsonButlerVolmerEquationPolymer2019} caution against the uselessness of such an approach. Since the Butler-Volmer theory is inherently simple and reductionist, there is no reason to believe that such manipulations can be effective. There is a risk of rendering the model more unstable and complex without achieving any significant gains.

\subsubsection{Manipulations on the Butler-Volmer equation}
\label{subsubsec:manipulation_Butler-Volmer_equation}

There is a need to consider temperature variations in the CCL \cite{baoTwodimensionalModelingPolymer2015, fanCharacteristicsPEMFCOperating2017, wangQuasi2DTransientModel2018}, membrane drying \cite{baoTwodimensionalModelingPolymer2015}, CCL flooding \cite{fanCharacteristicsPEMFCOperating2017, wangQuasi2DTransientModel2018, shaoComparisonSelfHumidificationEffect2020, xuReduceddimensionDynamicModel2021} and electrochemical surface area reduction over time \cite{khajeh-hosseini-dalasmParametricStudyCathode2010, hamnettComponentsElectrochemicalCell2010, zihrulVoltageCyclingInduced2016, zhaoReviewPhysicsbasedDatadriven2021} when calculating overpotential. Since these requirements are not addressed by a robust theory, empirical coefficients have been suggested by the community to modify the Tafel equation derived from the simple Butler-Volmer theory. Several of these proposals have been synthesized in \eqref{eq:overpotential_Fan_Bao}. Each of these coefficients can be used independently, depending on the desired application.

\begin{figure}[H]
	\begin{equation}
		\begin{cases}
			i_{fc} + i_{n} = i_{0,c} \exp \left(\frac{F \alpha_{c}}{R T_{fc}} \eta_{c} \right) \\
			
			i_{0,c} = i_{0,353}^{ref} a_{+}^{1 - 2 \alpha_{c}} \left[ 1-s_{ccl} \right]^{1.5} r_{f} \left( t \right) \exp \left( \frac{E_{act}}{R} \left[ \frac{1}{T_{ref}} - \frac{1}{T_{fc}} \right]\right) \left[ \frac{C_{O_{2},ccl}}{C_{O_{2}}^{ref}} \right]^{\kappa_{c}} \\
			
			r_{f} \left( t \right) = ECSA \left( t \right) \cdot L_{Pt} \\
			
			a_{+} = \frac{\left[ \lambda_{ccl} + 1 \right] - \sqrt{\left[ \lambda_{ccl} + 1 \right]^{2} - 4 \lambda_{ccl} \left[ 1 - \frac{1}{K_{e}} \right]}}{2 \left[ 1 - \frac{1}{K_{e}} \right]} \\
			
			K_{e} = K_{e}^{0} \exp \left(- \frac{\Delta H^{0}}{R} \left[ \frac{1}{T_{fc}} - \frac{1}{298} \right]\right)
		\end{cases}
		\label{eq:overpotential_Fan_Bao}
	\end{equation}
\end{figure}
\nomenclature[A,1]{$R$}{universal gas constant $(J.mol^{-1}.K^{-1})$}
\nomenclature[A,1]{$E_{act}$}{activation energy $(J.mol^{-1})$}
\nomenclature[A,1]{$K_{e}$}{acid-base equilibrium constant}
\nomenclature[A,1]{$r_{f}$}{electrode roughness factor $(m_{Pt}^{2}.m^{-2})$}
\nomenclature[A,1]{$ECSA$}{electrochemical surface area $(cm_{Pt}^{2}.mg_{Pt}^{-1})$}
\nomenclature[A,1]{$L_{Pt}$}{initial Pt loading of the electrode $(mg_{Pt}.cm^{-2})$}
\nomenclature[A,2]{\(\kappa\)}{overpotential correction exponent}
\nomenclature[A,2]{\(\alpha_{c}\)}{charge-transfer coefficient of the cathode}
\nomenclature[A,2]{\(\Delta H^{0}\)}{standard enthalpy of reaction $(J.mol^{-1})$}
where $i_{0,353}^{ref}$ $(A.m^{-2})$ is the referenced exchange current density at the cathode for a given oxygen concentration $C_{O_{2}}^{ref}$, a humidified membrane, a dry electrode, an initial electrode roughness factor and at $353.15 K$. $r_{f}$ $(m_{Pt}^{2}.m^{-2})$ is the electrode roughness factor, representing the ratio of the active platinum surface area to the flat surface area of the electrode. This active platinum surface is non-planar, defined in three dimensions as the sum of surfaces where redox reactions are accelerated by the catalyst within the electrode's pores. $r_{f}$ is on the order of a hundred \cite{hamnettComponentsElectrochemicalCell2010} and is commonly decomposed to reveal the electrochemical surface area ($ECSA$ in $cm_{Pt}^{2}.mg_{Pt}^{-1}$), a more practical quantity widely used in the literature. ECSA accounts for the active surface area surrounding a mass of platinum, facilitating the development of models for the evolution of this active surface. Its value is on the order of a hundred $cm_{Pt}^{2}.mg_{Pt}^{-1}$ for a PEMFC \cite{hamnettComponentsElectrochemicalCell2010}. $L_{Pt}$ $(mg_{Pt}.cm^{-2})$ corresponds to the initial platinum loading of the electrode, a fixed quantity over time, typically around $0.5$ $mg_{Pt}.cm^{-2}$ \cite{hamnettComponentsElectrochemicalCell2010}. Then, $a_{+}$ is the activity of solvated protons, $E_{act}$ $(J.mol^{-1})$ is the activation energy term, $T_{ref}$ $(K)$ is the referenced temperature taken at $353.15$ $K$, $K_{e}$ is the acid-base equilibrium constant, $K_{e}^{0}$ is the standard acid-base equilibrium constant, and $\Delta H^{0}$ $(J.mol^{-1})$ is the standard enthalpy of reaction. 

\begin{figure}[H]
	\centering
	\includegraphics[width=8cm]{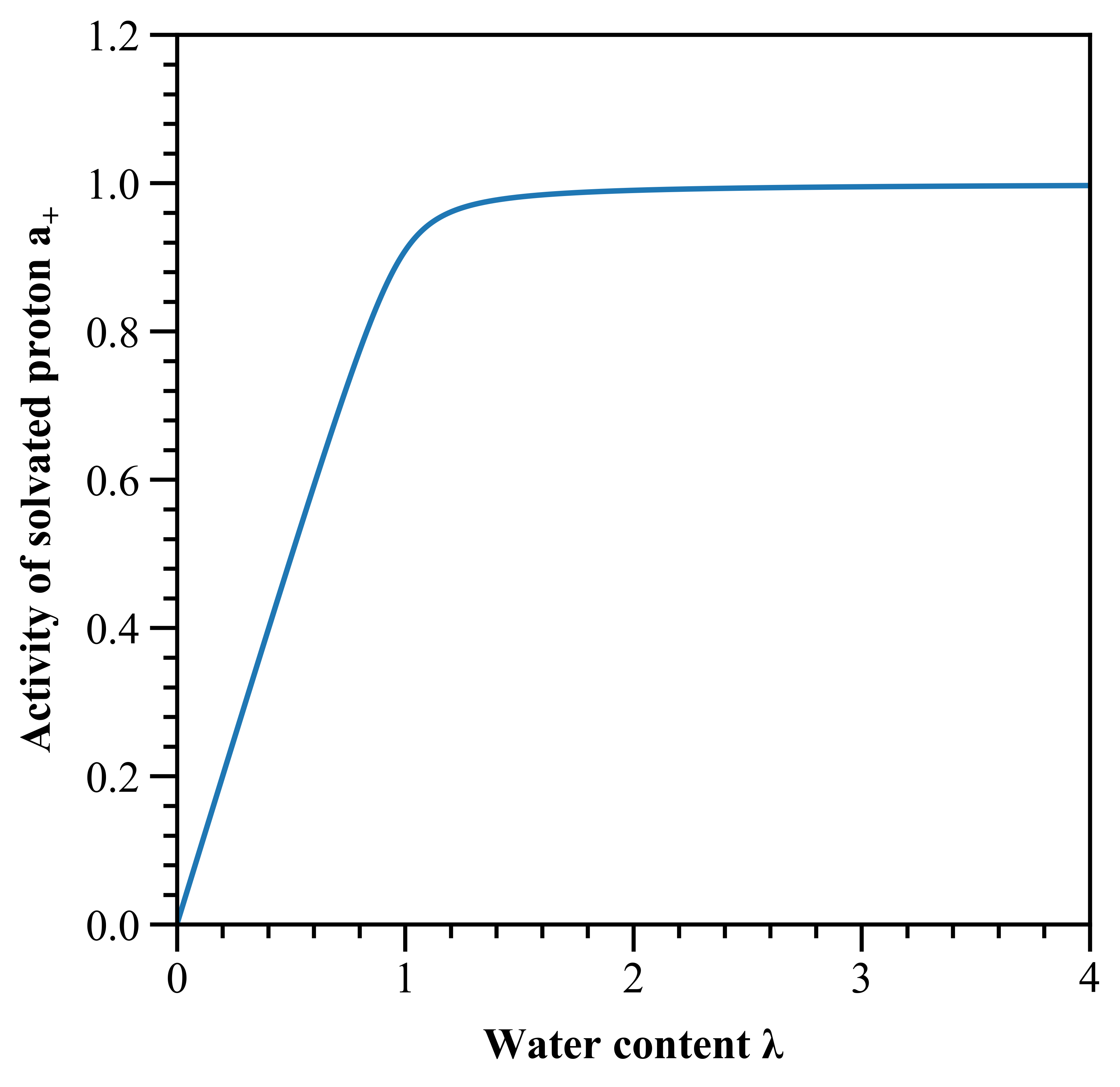}
	\caption{Plot of the activity of the solvated protons function of the water content}
	\label{fig:activity_solvated_protons}
\end{figure}

The solvated protons' activity, denoted as $a_{+}$, serves as a metric to examine the impact of a notably dry membrane on the exchange current density. This analysis is carried out for analysing start operation and current ignition \cite{baoTwodimensionalModelingPolymer2015}. Additionally, as illustrated in figure \ref{fig:activity_solvated_protons}, when $T = 353 K$ and $\lambda > 1$, the value of $a_{+}$ is approximately 1. Consequently, it exerts negligible influence on the equation for a hydrated membrane.

The purpose of $\left( 1-s_{ccl} \right)^{1.5}$ is to account for the impact of flooding on the cell voltage by examining the covering effect of liquid water on the active area of the catalyst. When the liquid saturation at the CCL increases, the active area of the catalyst becomes covered by liquid water, leading to a drop in cell voltage. Several formulations have been proposed for this coefficient \cite{fanCharacteristicsPEMFCOperating2017, wangQuasi2DTransientModel2018, shaoComparisonSelfHumidificationEffect2020, xuReduceddimensionDynamicModel2021}, yet none has been experimentally validated. 

The purpose of the electrode roughness factor $r_f$ is to account for the diminishing active surface area of platinum over time within the electrode on the cell voltage \cite{khajeh-hosseini-dalasmParametricStudyCathode2010, hamnettComponentsElectrochemicalCell2010, zihrulVoltageCyclingInduced2016, zhaoReviewPhysicsbasedDatadriven2021}. Several degradation phenomena contribute to this decrease in the platinum's active surface, such as the oxidation of the platinum surface. Various degradation models available in the literature could be employed to quantify the evolution of ECSA over time \cite{darlingKineticModelPlatinum2003,rinaldoCatalystDegradationNanoparticle2013,baroodyPredictingPlatinumDissolution2021}. The details of these models, however, are not covered in this work and merit a specific discussion.

Another dependency of $i_{0,c}$ is highlighted here: temperature \cite{baoTwodimensionalModelingPolymer2015, fanCharacteristicsPEMFCOperating2017, wangQuasi2DTransientModel2018}. Similar to the approach taken for the concentration dependency, a referenced constant $i_{0,353}^{ref}$ is identified, representing the exchange current density measured at a reference oxygen concentration $C_{O_{2}}^{ref}$ and at a specified temperature, typically set at $353$ K. The temperature dependence is then expressed through the coefficient $\exp \left( \frac{E_{act}}{R} \left[ \frac{1}{T_{ref}} - \frac{1}{T_{fc}} \right]\right)$.

Finally, it is crucial to highlight that $\eta_{c}$ is associated with indeterminate parameters, $i_{0,353}^{ref}$ and $\kappa_{c}$, due to the limitations of current knowledge. These parameters have to be calibrated based on experimental data. The utilization of predefined values for them is not feasible, given their dependence on the particular fuel cell being employed.

\subsubsection{Internal current density: crossover and short circuit}
\label{subsec:internal_current_density}

In the fuel cell, a small amount of matter naturally permeates the membrane, even though it is designed to be impermeable. This dissipation of chemical energy within the stack results in a voltage drop. This matter can be either oxygen or hydrogen, termed as crossover, or in the case of electrons, it is identified as an electronic short circuit. Together, these two phenomena contribute to the internal current density, as formulated in \ref{eq:internal_current_density}.

\begin{equation}
	i_{n} = i_{co,H_{2}} + i_{co,O_{2}} + i_{sc}
	\label{eq:internal_current_density}
\end{equation}
where $i_{co,i}$ $(A.m^{-2})$ is the internal crossover current density of the molecule i (hydrogen or oxygen) and $i_{sc}$ $(A.m^{-2})$ is the internal short circuit current density.

During a crossover, the matter which was expected to react following the fuel cell mechanism instead permeates the membrane and interacts with its complementary molecule by direct contact. Consequently, both the matter and the electrons it carries are lost. The volume flow of matter through the membrane during this crossover, denoted as $S_{i,co}$ and discussed in section \ref{subsec:hydrogen_oxygen_consumption}, can be correlated with a flow of lost electrons for the calculation of $i_{co}$. This relationship is expressed in \eqref{eq:internal_crossover_current_density}.

\begin{equation}
	\begin{cases}
		i_{co,H_{2}} = 2 F H_{cl} S_{H_{2},co} = 2 F k_{H_{2}} R T_{fc} \nabla C_{H_{2}} \\
		i_{co,O_{2}} = 4 F H_{cl} S_{O_{2},co} = 4 F k_{O_{2}} R T_{fc} \nabla C_{O_{2}}
	\end{cases}
	\label{eq:internal_crossover_current_density}
\end{equation}

The way $i_{co,i}$ is introduced into the equation for $\eta_{c}$ in \eqref{eq:overpotential_fair_Tafel} needs clarification. As crossover corresponds to a matter loss in the cell, it is expressed in the gas transport equations. Thus, it is already indirectly included in $C_{O_{2},ccl}$. However, the calculation of the equilibrium potential $U_{eq}$ also utilizes the value of $C_{O_{2},ccl}$. Therefore, the theoretically maximum energy extractable by redox reactions is biased by the crossover. To account for this energy loss, it should be necessary to add the value of $C_{O_{2},ccl}$ to $U_{eq}$ along with the oxygen concentration lost due to crossover. Another method widely employed in the literature and proposed in \eqref{eq:overpotential_fair_Tafel}, considered equivalent and more practical, involves adding $i_{co,i}$ to $i_{fc}$ in the calculation of $\eta_{c}$. This means that additional fictitious current is required to compensate for the crossover loss, thereby increasing the overpotential value. However, the equivalence of these two methods is questionable. Transferring information that should be contained in $U_{eq}$ to $\eta_{c}$ is not straightforward. Future research should be conducted to better incorporate crossover effects in voltage calculations.

During an electronic short circuit across the membrane, the reaction between oxygen and hydrogen occurs as expected on both sides of the membrane. However, the electrons released by hydrogen do not traverse the external circuit. Consequently, they do not contribute to $i_{fc}$ and manage to pass through the membrane, even though it is designed to resist their passage. 
There is limited literature on this topic. The equation presented here, standardized since the work of Giner-Sanz et al. \cite{giner-sanzHydrogenCrossoverInternal2014} for broader applicability and expressed as \eqref{eq:i_sc}, includes several assumptions that significantly constrain its utility. This experimental study was conducted using a single commercial \Nafion 117 membrane and employed linear voltammetry. The measurements were carried out under constant temperature and relative humidity of the incoming gases. The pressures at the anode and cathode were the only variables and were adjusted independently, without the necessity for their equality. Consequently, it was assumed that pressure is the crucial variable for calculating $i_{sc}$, although, in reality, temperature should also be considered. Additionally, pressure variations are relatively small, ranging between 1.12–1.45 bar at the cathode and only between 1.01–1.06 bar at the anode. Given these significant limitations, further extensive experimental tests are necessary to refine these results \cite{giner-sanzHydrogenCrossoverInternal2014}.

\begin{equation}
	\begin{cases}
		i_{sc} =  \frac{U_{cell}}{r_{sc}}\\
		r_{sc} = 1.79 \cdot 10^{-2} \left[ \frac{P_{agc}}{101325} \right]^{-9.63} \left[ \frac{P_{cgc}}{101325} \right]^{0.38} 
	\end{cases}
	\label{eq:i_sc}
\end{equation}
where $r_{sc}$ $(\Omega.m^{2})$ is the area specific short circuit resistance.

Giner-Sanz et al. proposed a physical explanation for the correlation between the inlet pressures of the stack and the internal electronic short circuit. Elevating the gas pressure in a PEMFC can induce two opposing effects on the short-circuit resistance. On one hand, it may increase the effective interfacial contact area between layers, thereby reducing resistance. On the other hand, it can lead to porosity and morphological changes, which may increase or decrease resistance depending on the specific characteristics of the PEMFC. Whether an increase in pressure will result in heightened or diminished resistance depends on the relative significance of these two effects. Nonetheless, further research is needed to validate this hypothesis \cite{giner-sanzHydrogenCrossoverInternal2014}.

Next, the authors provide an explanation of how electronic short circuits affect voltage. As the redox reactions proceed as normal, oxygen molecules that receive electrons traversing the membrane are still attracted to the triple point regions to form ions. Thus, the electronic short circuit phenomenon leads to overpotential, yet does not contribute to $i_{fc}$. Therefore, it is acceptable to calculate overpotential by adding $i_{sc}$ to $i_{fc}$.

It is noteworthy, as mentioned by O'Hayre et al. \cite{ohayreFuelCellFundamentals2016}, and in contrast to the proposition by Dicks et al. \cite{dicksFuelCellSystems2018}, that adding $i_{n}$ to $i_{fc}$ should not be done when calculating electronic and proton resistances. These resistances are only associated with the external current density and do not encompass internal losses.

Finally, the internal current density is weak, approximately $0.01$–$0.05$ $A\cdot cm^{-2}$. Under normal working conditions, it is highly negligible. However, at low current density, its impact becomes significant. It is responsible for the reduced value of the open circuit voltage of fuel cells, which is approximately $0.95$ $V$, whereas the Nernst potential is around $1.2$ $V$. Therefore, accurate modeling of these losses is crucial for developing PEMFC models that faithfully replicate the experimental behavior of PEMFCs operating at low current densities.

\subsection{Concentration losses}
\label{subsec:concentration_loss}

Under high loads or suboptimal operating conditions, the concentrations of $O_{2}$ and $H_{2}$ within the CLs may significantly decrease, resulting in a voltage loss. Indeed, these are the fuel concentrations within the CLs that mainly influence the voltage in the cell, as reflected in the mathematical expression for $U_{cell}$. Various factors could contribute to this phenomenon, commonly known as concentration loss.

Diffusion between the GC and the CL may result in concentration losses. While effective control of the auxiliary system can stabilize fuel concentrations in the GC, an increase in fuel consumption at the CL inherent to the rise in current density, diminishes fuel concentration at the CL. Indeed, achieving matter equilibrium requires fuel supply to equal consumption. As the main mechanism of matter transport in the MEA is diffusion, and with concentrations in the GCs stabilized, enhancing the supply flow necessitates reducing fuel concentrations in the CLs. The extent of this concentration loss depends on the diffusion characteristics of the stack, load, and operating conditions \cite{ohayreFuelCellFundamentals2016}.

Then, across the width of the stack, inhomogeneities can lead to a partial concentration drop near the stack outlet. While diffusion alone may effectively fill the initial active sites near the GC inlets, it may be insufficient for those located in proximity to the outlets. This is because, as the gas mixture traverses the channels of the bipolar plates, it becomes depleted in fuel, consequently diminishing the diffusion capabilities of fuels within the stack. Given that the dimensions across the width are significantly larger than those along the depth, lateral diffusion alone may prove insufficient to compensate for this shortfall \cite{ohayreFuelCellFundamentals2016}.

Additionally, the crossover of $N_{2}$ through the membrane may result in a loss of $H_{2}$ concentration, as its accumulation leads to a reduction in the partial pressure of $H_{2}$ in the gas mixture, at a constant total pressure in the GC. However, this challenge can be addressed by strategically employing an ejector at the anode outlet, which expels gases when the presence of $N_{2}$ is substantial.

Furthermore, in the absence of an effective drainage system, the accumulation of liquid water can impede oxygen flow to the catalyst sites, leading to a loss of concentration. Under fixed operating conditions, it is common for this concentration loss to occur at lower current densities than those required to saturate the diffusion capabilities of the fuel cell in the absence of liquid water. Hence, the occurrence of liquid water is a critical point for concentration loss.

To account for all these phenomena, the following expression \eqref{eq:concentration_loss} \cite{dicksFuelCellSystems2018, ohayreFuelCellFundamentals2016,pukrushpanControlOrientedModelingAnalysis2004,yangEffectsOperatingConditions2019,santarelliParametersEstimationPEM2006,williamsAnalysisPolarizationCurves} is mainly used in the literature as another voltage loss to add in \eqref{eq:apparent_voltage}. 

\begin{figure}[htb]
	\begin{equation}
		\Delta V_{conc} = \frac{RT}{2F} ln \left( \frac{i_{lim}}{i_{lim}-i_{fc}} \right) 
		\label{eq:concentration_loss}
	\end{equation}
\end{figure}
where $i_{lim}$ $(A.m^{-2})$ is the limiting current density. 
However, \eqref{eq:concentration_loss} represents a simplified approach to concentration losses. It proves valuable solely for black box and equivalent electrical models where concentrations at the CLs remain inaccessible. Additionally, a significant drawback is its dependence on experimental data. The value of $i_{lim}$ varies with stack technology and operating conditions, and no universal expressions exist to define it. Consequently, it lacks the precision needed to predict concentration losses accurately under changing operating conditions, imposing limitations on its applicability.

Nevertheless, when PEMFC is spatially modeled, some of the information regarding concentration losses already exists. Indeed, it is embedded in the fuel concentrations at the CLs. This is why $C_{O_2,ccl}$ is used in both the equilibrium potential $U_{eq}$ (see \ref{subsec:equilibrium_potential}) and the overpotential $\eta_{c}$ (see \ref{subsec:overpotential}) \cite{zihrulVoltageCyclingInduced2016}. These two equations contribute to representing the concentration drop in the voltage calculation, albeit indirectly. The modeling of mass transport directly represents this phenomenon. However, current models do not satisfactorily simulate the impact of liquid water on fuel transport to triple point areas. Therefore, it is still necessary to use \eqref{eq:concentration_loss} to fully consider the concentration drop, although improvements in models should lead to its elimination.

\subsection{Proton conductive resistance}
\label{subsec:proton_conductive_resistance}
\subsubsection{Proton conductivity of the membrane: $\sigma_{m}$}

The proton conductivity, denoted by $\sigma$, is commonly defined as shown in \eqref{eq:proton_conductivity_global}.

\begin{figure}[H]
	\begin{equation}
		\frac{1}{R_{p}} \overset{\vartriangle}{=} \sigma \frac{d S}{d x}
		\label{eq:proton_conductivity_global}
	\end{equation}
\end{figure}
\nomenclature[A,2]{\(\sigma_{m}\)}{conductivity of the membrane $(\Omega^{-1}.m^{-1})$}

A confusing convention exists in the literature on  PEMFC. The term "resistance" ($R_{p}$) is intended to represent the area-specific resistance ($r_{p}$) in units of $\Omega.m^{2}$. Theoretically, we should have $r_{p}=R_{p} d S$. However, in the litterature, the resistance is referred to as the area-specific resistance, and the symbol $R_{p}$ is still used, leading to the statement "$r_{p}=R_{p}$". Consequently, due to this ambiguous convention, the authors adopt the following definition for the local proton conductivity (see equation \eqref{eq:proton_conductivity_local}).

\begin{figure}[htb]
	\begin{equation}
		\frac{1}{R_{p}} = \frac{\sigma}{d x}
		\label{eq:proton_conductivity_local}
	\end{equation}
\end{figure}

Then, to determine the resistance $R_{p}$, Springer et al. experimentally derived an expression for proton conductivity in the membrane $\sigma_{m}$ in 1991 \cite{springerPolymerElectrolyteFuel1991}. This expression has since become widely adopted in the literature \cite{jiaoWaterTransportPolymer2011,xuReduceddimensionDynamicModel2021,xingNumericalAnalysisOptimum2015,yangMatchingWaterTemperature2011,fanCharacteristicsPEMFCOperating2017,yeThreeDimensionalSimulationLiquid2007,pasaogullariTwoPhaseModelingFlooding2005,baoTwodimensionalModelingPolymer2015,wangInvestigationDryIonomer2020}, and is represented by \eqref{eq:proton_conductivity_Springer}.

\begin{equation}
	\sigma_{m} = 
	\begin{cases}
		\left[ 0.5139 \lambda - 0.326 \right] \exp \left( 1268 \left[ \frac{1}{303.15} - \frac{1}{T_{fc}} \right]\right), & \text{for $\lambda \geq 1$} \\
		
		0.1879 \exp \left( 1268 \left[ \frac{1}{303.15} - \frac{1}{T_{fc}} \right]\right), & \text{for $\lambda < 1$} \\
	\end{cases}
	\label{eq:proton_conductivity_Springer}
\end{equation}

\begin{figure}[htb]
	\centering
	\includegraphics[width=10cm]{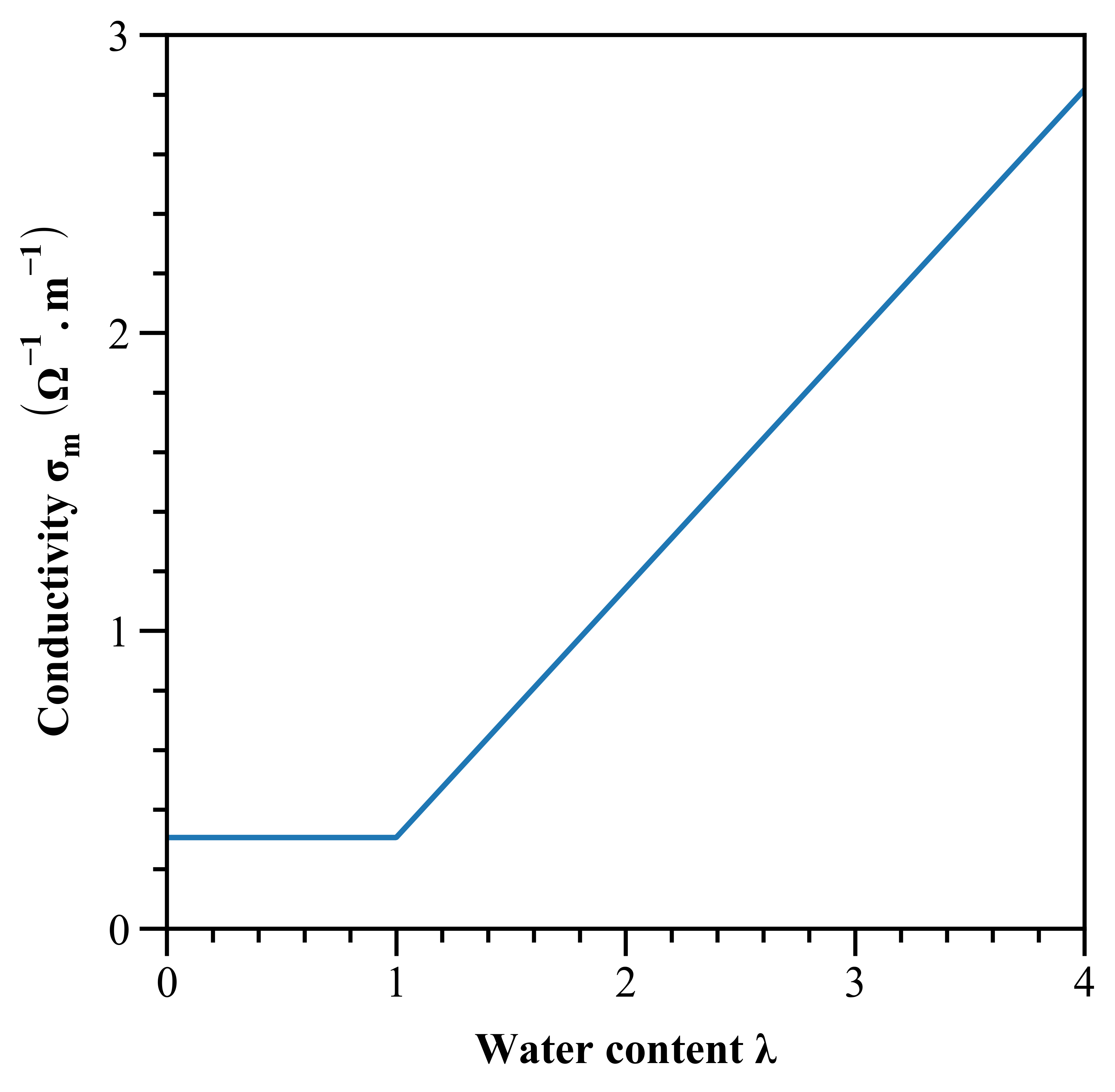}
	\caption{Shape of the conductivity $\sigma_{m}$ function of the water content $\lambda$ at $T_{fc} = 343.15 K$}
	\label{fig:conductivity}
\end{figure}

The linear term $0.5139 \lambda - 0.326$ is derived from measurements at 30°C, while the exponential component enables the extension to other temperature ranges. An activation energy, $E_{\text{act}} = 10542$ $J.mol^{-1}$, was measured and assumed to be independent of $\lambda$. Subsequently, the coefficient $1268$ was computed using the equation $1268 = \frac{E_{act}}{R} = \frac{10542}{8.314}$. Additionally, when the quantity of water molecules per charge site is below one $\left( \lambda < 1 \right)$, the conductivity is presumed to remain constant. The graphical representation of this function is illustrated in figure \ref{fig:conductivity}.

In certain papers, there was an innacurate modification of this expression when $\lambda < 1$, employing a linear decrease with $\lambda$ as illustrated in \eqref{eq:proton_conductivity_bad} \cite{yeThreeDimensionalSimulationLiquid2007}. The aim was to "avoid negative conductivity". While the intention was to rectify the expression for $\sigma_{m}$, which exhibits negative values when the expression for $\lambda \geq 1$ is applied to $\lambda < 1$, this approach is flawed. The constant term in \eqref{eq:proton_conductivity_Springer} is mentioned in Springer's original work \cite{springerPolymerElectrolyteFuel1991}, albeit only in the text and not in the equation, potentially causing confusion. Moreover, while it is reasonable to expect conductivity to decrease with $\lambda$ for a 117 \Nafion membrane, attributing $\sigma_{m} = 0$ $\Omega^{-1}.m^{-1}$ when $\lambda = 0$ seems excessive, suggesting a perfect insulator under feasible conditions.

\begin{equation}
	\sigma_{m} = 0.1879 \lambda \exp \left( 1268 \left[ \frac{1}{303.15} - \frac{1}{T_{fc}} \right]\right), \text{for $\lambda < 1$}
	\label{eq:proton_conductivity_bad}
\end{equation} 

The expression \eqref{eq:proton_conductivity_Springer} is limited in its applicability to modern models due to its creation with outdated membranes \cite{karimiRecentApproachesImprove2019}. While recent models do exist \cite{dickinsonModellingProtonConductiveMembrane2020}, they have not gained widespread acceptance in the literature due to various shortcomings. These include reliance on outdated data, lack of accessibility to the data used, or restriction of the equation's applicability to single-phase systems without liquid water. Consequently, the community expects a robust, well-documented study for $\sigma_{m}$ with minimal limitations across a broad range of membranes.

\begin{figure}[htb]
	\centering
	\includegraphics[width=10cm]{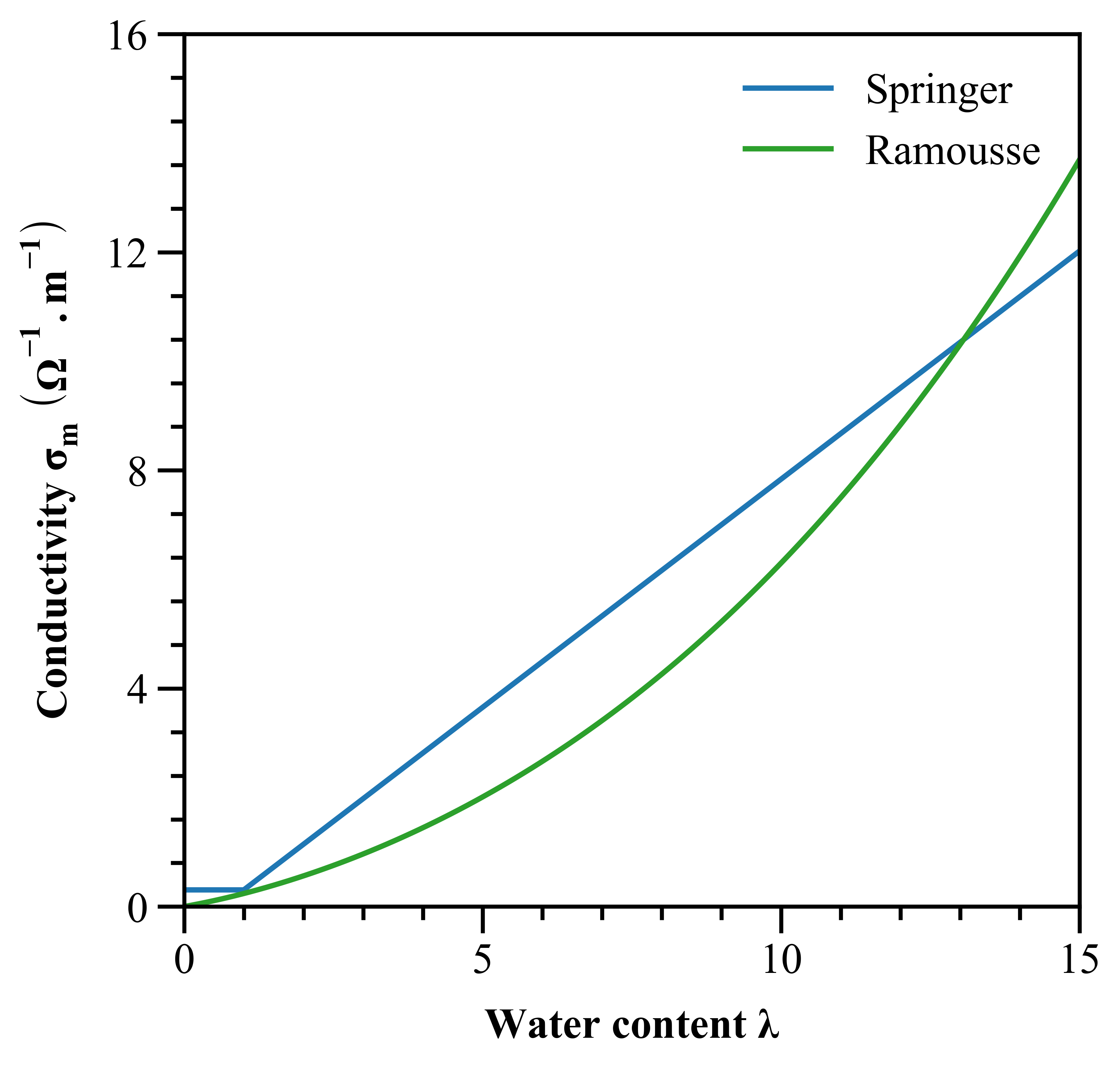}
	\caption{Comparison between Springer and Ramousse expressions for the conductivity at $T_{fc} = 343.15 K$}
	\label{fig:conductivity_Springer_Ramousse}
\end{figure}

Moreover, \eqref{eq:proton_conductivity_Springer} has the disadvantage of being constructed in two parts, resulting in a discontinuous derivative. This characteristic could potentially introduce parasitic oscillations in models, especially when the discontinuity occurs around $\lambda \approx 1$. To circumvent this issue, it is possible to opt for the expression proposed by Ramousse et al. \cite{ramousseModellingHeatMass2005}, as given by \eqref{eq:proton_conductivity_Ramousse}. However, it is worth noting that this expression relies on outdated and hardly accessible data. Additionally, it yields $\sigma_{m} = 0$ $\Omega^{-1}.m^{-1}$ when $\lambda = 0$.
A comparative analysis of the Springer and Ramousse expressions is illustrated in Figure \ref{fig:conductivity_Springer_Ramousse}.

\begin{equation}
	\begin{cases}
		\sigma_{m} = \left[ 0.0013\lambda^{3} + 0.0298\lambda^{2} + 0.2658\lambda \right] \exp \left( E_{A} \left[ \frac{1}{353} - \frac{1}{T_{fc}} \right]\right) \\
		E_{A} = 2640 \exp \left( -0.6\lambda \right) + 1183
	\end{cases}
	\label{eq:proton_conductivity_Ramousse}
\end{equation} 

\subsubsection{Proton conductivity resistance: $R_{p}$}

The proton conductivity resistance $R_{p}$ encompasses both the proton resistance within the membrane and the proton resistance within the ionomer of the CCL. The proton resistance within the ionomer of the ACL is neglected because the hydrogen oxidation reaction is fast, and gas diffusion resistances for pure H2 are minor. As a result, the reaction current occurs near the membrane, establishing a short route for proton transport and minimizing voltage loss at the anode \cite{neyerlinCathodeCatalystUtilization2007}. 

Springer's relationship characterizes the conductivity in the membrane but does not address the CCL ionomer's conductivity. For this purpose, it is assumed that the ionomer within the CCL exhibits a conductivity equivalent to that of the membrane, but adjusted by a factor $\frac{\varepsilon_{mc}}{\tau}$ to account for both the dispersion of the ionomer within the CCL (represented by $\varepsilon_{mc}$) and its tortuosity (denoted by $\tau$) \cite{xuReduceddimensionDynamicModel2021,makhariaMeasurementCatalystLayer2005, neyerlinCathodeCatalystUtilization2007}. In this context, tortuosity is represented using a different form than the one introduced in section \ref{subsec:effective_diffusion_coefficient}: $\varepsilon_{mc}^{\tau}$. The authors did not find reasons for this choice, and it remains unclear whether substituting $\frac{\varepsilon_{mc}}{\tau}$ with $\varepsilon_{mc}^{\tau}$ produces acceptable results. This aspect will require further clarification in future research. Furthermore, several studies employing different theories \cite{makhariaMeasurementCatalystLayer2005, neyerlinCathodeCatalystUtilization2007}, such as the transmission line model, emphasize the importance of an additional correction. This involves incorporating a coefficient of $\frac{1}{3}$ into $R_{ccl}$ to achieve satisfactory results, and has been implemented in the present work. Hence, the following equation \eqref{eq:proton_conductivity_resistance} for $R_{p}$ is proposed.

\begin{equation}
	R_{p} = R_{mem} + R_{ccl} = \int_{mem} \frac{d x}{\sigma_{m}} + \frac{1}{3} \int_{ccl} \frac{d x}{\frac{\varepsilon_{mc}}{\tau} \sigma_{m}}
	\label{eq:proton_conductivity_resistance}
\end{equation}

\section{Summary and outlook}

This study attempted to synthesise and document the matter transport and voltage polarisation governing laws proposed in the literature. New laws, being the combination of several ideas presented in the literature, have also been presented. Certain expressions were discussed in detail, as they were more representative of the physics phenomena at stake. A synthesis is presented in tables \ref{table:synthesis_spotlighted_transport_expressions_1},\ref{table:synthesis_spotlighted_transport_expressions_2}, \ref{table:synthesis_flow_coef}, and \ref{table:synthesis_spotlighted_voltage_expressions}. 

Finally, certain perspectives are discussed in this last section.

\subsection{Model development}
From this review, it is noted that more investigations are needed to model more clearly and precisely the processes at stake. In particular, explorations should be undertaken to improve the water sorption at the triple points in a biphasic state, the matter sorption at the GDL/GC interface, and the flooding impact on the voltage. Better modelling of these processes requires more targeted experimental investigation at both the material and cell levels.

\subsection{Model parameter identification}
The values of involved parameters are determinative for model performance. As shown in Tables \ref{table:constant_values_1} and \ref{table:constant_values_2}, different parameters are selected and used in both mater transport and voltage polarisation models. The unclarified selection and use conditions of these parameters could lead to poor model performance, and even model invalidity. In addition, most of the available experimental data dedicated to model identification in the literature are outdated. The data were mainly extracted from experiments at the beginning of the 1990s. When more recent expressions are given, they are often based on outdated experimental or hardly accessible data, or usable only with strong limitations. Thus, a strong well documented study with few limitations over a large brand of fuel cells must be conducted to update the electro-osmotic drag, equilibrium water content, capillary pressure, and protonic conductivity expressions. The components have evolved during the last years and the modern measurement protocols have become more precise \cite{karimiRecentApproachesImprove2019}. It is therefore highly necessary to update the database dedicated to model parameter identification. Overall, the limitations of the identified model parameters should be well noted in the model development and use stages.

However, the authors recognise that doing these precise experimentations is very challenging. This explains why the equations in the literature are not unified, as they were done by different teams on different stacks and under different experimental conditions. It is therefore difficult to separate equations that model the same physical phenomenon. It also remains to be demonstrated that these equations from different experimental conditions remain valid in combination in a global model. 

\subsection{Targeted multiscale experiments}
The development of both the matter and voltage polarisation models involves multiscale physicochemical phenomena. The global operating conditions can only be controlled and assigned at the macroscopic level. However, the matter transport and electrochemical processes are concerned at microscale to mesoscale. Most existing models were developed without considering the link between the different scales. For instance, processes at the microscale to mesoscale are often concerned with modelling the mass transport along the MEA and GDL. The corresponding models were thus often built using the data from ex-situ characterisations without considering the impacts of dynamic macroscopic operating parameters. It is questionable that the developed model can still conserve the performance when the multiscale interactions must be considered, which is the condition in practice. Thus, to achieve more reliable models, experiments and model development must be undertaken with multiscale characterisations and analysis. In these experiments, it is often necessary to combine macroscopic in-situ characterisations at the stack/system level and ex-situ microscopic characterisations at the component/cell level. Moreover, the in-situ and operando characterisation techniques are promising tools to gather the relevant microscopic data during operation \cite{meyerSituOperandoCharacterization2019}. 

\subsection{Model resolution}
As discussed in this review, the matter transport models, developed based on different theories, are governed by partial differential equations (PDEs) based on the Navier-Stokes equations. It is mainly the conservation equations which were used, as most flows are Fick-like ones \cite{weberCriticalReviewModeling2014}. However, for more complex models that consider multidimensional space (from 2D to 3D) or that consider convective flows within the GDL and CL, it is necessary to add the Navier-Stokes momentum balance equations to obtain a solvable model. These PDEs, in most cases, can be solved only by numerical simulations \cite{zhangMultiphaseModelsWater2018}. The high computation complexity renders it difficult to upscale the developed models in terms of space and time. In addition, the PDEs governed models are naturally not able to satisfy the requirements of certain model applications. For instance, inferring material properties must solve inverse problems, that is, calculating model parameters from online measured data. The inverse problems of PDEs and molecular simulations are prohibitively expensive and require complex formulations, and new algorithms \cite{karniadakisPhysicsinformedMachineLearning2021}. Moreover, the models represented by PDEs cannot handle the noisy boundary data \cite{shuklaPhysicsinformedNeuralNetwork2020}. This results in the development of reduced-order modelling (ROM) that seeks to build low-dimensional models for efficient solutions with noisy boundary data \cite{chenPhysicsinformedMachineLearning2021}. Particularly, recent studies have shown that machine learning can be adopted as an efficient ROM tool and provide robust and efficient model resolutions \cite{dingApplicationMachineLearning2022}.

\subsection{Model use}
The reviewed matter transport and voltage polarisation models are essential for optimisation of cell design, materials preparation, and operating conditions. It should be noted that different uses of the models recall different requirements for model order reduction, simplification, and formulation \cite{mayurLifetimePredictionPolymer2018}. Nowadays, the analysis and optimisation of high-power fuel cell stacks/systems and the prediction of performance degradation has become increasingly important for fuel cell large deployment. In these large spatial-tempo scale applications, how to maintain the high-fidelity model performance without losing the model efficiency remains a challenging issue \cite{jahnkePerformanceDegradationProton2016}.

To proceed further, the authors have also built a one-dimensional dynamic two-phases model in a complementary work and the numerical results are discussed there. It is an interesting application to observe the deeper simplifications that were made to adapt the following equations for a control-command use.
In addition, a review of the preponderant degradation phenomena is scheduled. In combination with these reviews and the one-dimensional two-phases model, these works are expected to allow a more accurate control on running devices, such as buses, to extend their lifetime and help to reach the European goal of $25,000$ operating hours by 2023 \cite{FuelCellsHydrogen2018}.

\section{Acknowledgments}

This work has been supported by French National Research Agency via project DEAL (Grant no. ANR-20-CE05-0016-01), the Region Provence-Alpes-Côte d'Azur, the EIPHI Graduate School (contract ANR-17-EURE-0002) and the Region Bourgogne Franche-Comté.

\begin{landscape}
	\begin{table}[H]
		\centering
		\linespread{1.2}
		\begin{tabularx}{\linewidth}{|Y|Y|} \hline
			\bf{Dynamical models} & \bf{Matter flow expressions} \\ \hline \hline
			
			\multicolumn{2}{|c|}{\bf{Dissolved water in the membrane}} \\ \hline
			
			\multirow{2}{*}{
				\normalsize{
					$\begin{cases}
						\frac{\rho_{\text{mem}}}{M_{\text{eq}}} \frac{\partial \lambda_{\text{mem}}}{\partial t} = - \bm{\nabla} \cdot \bm{J_{mem}}, \text {\small{in the bulk membrane}} \\ 
						\frac{\rho_{mem} \varepsilon_{mc}}{M_{eq}} \frac{\partial \lambda_{cl}}{\partial t} =  - \bm{\nabla} \cdot \bm{J_{mem}} + S_{sorp} + S_{prod}, \text{\small{in the CL}}
					\end{cases}$
					\eqref{subeq:water_content_dynamic_balance}}}  &
			\normalsize{
				$S_{\text{prod}} = 
				\begin{cases}
					2 k_{O_{2}}\frac{ R T_{fc}}{H_{cl}} \nabla C_{O_{2}}, &\text{\small{in the ACL}} \\
					\frac{i_{fc} + i_{sc}}{2 F H_{cl}} + k_{H_{2}} \frac{R T_{fc}}{H_{cl}} \nabla C_{H_{2}}, &\text{\small{in the CCL}} \\
					0, &\text{\small{elsewhere}}
				\end{cases}$ \eqref{eq:S_prod_corrected}} \\ &  \\

			&
			$S_{sorp} = \gamma_{sorp} \frac{\rho_{\text{mem}}}{M_{\text{eq}}} \left[ \lambda_{\text{eq}} - \lambda \right]$ \eqref{eq:j_sorp_Ge}  \\ 
			
			$\bm{J_{mem}^{cl,mem}} = \bm{0}, \text{\small{at the ionomer border}}$
			\eqref{subeq:water_content_dynamic_boundary_ACL} & \\
			
			&
			$\bm{J_{mem}} = \frac{2.5}{22} \frac{i_{fc}}{F} \lambda \text{ } \bm{\imath} - \frac{\rho_{mem}}{M_{eq}} D\left( \lambda\right)  \bm{\nabla \lambda}$
			\eqref{eq:water_flow_membrane} \\ \hline

			\multicolumn{2}{|c|}{\bf{Liquid water in the GDL and the CL}} \\ \hline
			
			\normalsize{
				$\rho_{H_{2}O} \varepsilon \frac{\partial \texttt{s}}{\partial t} = - \bm{\nabla} \cdot \bm{J_{l,cap}} + M_{H_{2}O} S_{vl}$
				\eqref{subeq:liquid_water_dynamic_behaviour}} &
			\normalsize{
				$S_{vl} = 
				\begin{cases}
					\gamma_{\text{cond}} \varepsilon \left[ 1 - \texttt{s} \right] x_{v} \left[ C_{v} - C_{\text {v,sat}} \right], & \text{if $C_{v} > C_{\text{v,sat}}$} \\
					-\gamma_{\text{evap}} \varepsilon \texttt{s} \frac{\rho_{H_{2}O}}{M_{H_{2}O}} R T_{fc} \left[ C_{\text{v,sat}} - C_{v} \right], &\text{if $C_{v} \leq C_{\text{v,sat}}$}
				\end{cases}$
				\eqref{eq:phase_change_rate}} \\ 
			
			\normalsize{
				$\begin{cases}
					\bm{J_{l}^{cl,mem}} = \bm{0}, \text{\small{at the ionomer border}} \\
					\texttt{s} = 0, \text{\small{at the GDL/GC border}}
				\end{cases}$
				\eqref{subeq:liquid_water_dynamic_behaviour_border}} &
			\normalsize{
				$\bm{J_{l,cap}} = -\sigma \frac{K_{0}}{\nu_{l}} \left| \cos \left( \theta_{c} \right)\right| \sqrt{\frac{\varepsilon}{K_{0}}} \texttt{s}^{\texttt{e}} \left[ 1.417 - 4.24\texttt{s} + 3.789\texttt{s}^{2} \right] \bm{\nabla \texttt{s}}$ \eqref{eq:j_cap}}  \\ \hline

			\multicolumn{2}{|c|}{\bf{Vapour in the GDL and the CL}} \\ \hline
			
			\multirow{2}{*}{
				\normalsize{
					$ \varepsilon \frac{\partial}{\partial t} \left( \left[ 1 - \texttt{s} \right] C_{v} \right) = - \bm{\nabla} \cdot \bm{J_{v,dif}} - S_{sorp} - S_{vl}$
					\eqref{subeq:vapour_dynamic_behaviour_balance}}} &
			\multirow{2}{*}{
				\normalsize{
					$\bm{J_{v,dif}} = - D_{v}^{eff} \bm{\nabla C_{v}}$ \eqref{eq:J_diff_vapour}}} \\ & \\
			
			\normalsize{
				$\begin{cases}
					\bm{J_{v}^{cl,mem}} = \bm{0}, \text{\small{at the ionomer border}} \\
					\bm{J_{v}^{gdl,gc}} = \bm{J_{v,codi}}, \text{\small{at the GDL/GC border}}
				\end{cases}$
				\eqref{subeq:vapour_dynamic_behaviour_border}} &
			\multirow{1}{*}{
				\normalsize{
					$\bm{J_{v,codi}} = \pm h_{v} \left[C_{v,gc} - C_{v,gdl}^{\text{inter}} \right] \bm{\imath}$
					\eqref{eq:vapour_convective-conductive_flow_final}}} \\  \hline

			\multicolumn{2}{|c|}{\bf{Vapour in the GC}} \\ \hline
			
			&
			$\bm{J_{v,conv}} = C_{v} \bm{u_{g}} $
			\eqref{eq:vapour_convective_flow_GC}  \\
			
			$\frac{\partial C_{v}}{\partial t} = - \bm{\nabla} \cdot \bm{J_{v,conv}}$
			\eqref{eq:vapour_dynamic_behaviour_GC}
			&
			$J_{v,\text{in}}^{agc} = \frac{\Phi_{a,\text{des}} P_{sat}}{P_{\text{agc,in}} - \Phi_{a,\text{des}} P_{sat}} \frac{A_{act}}{H_{gc} W_{gc}} \frac{S_{a} \left[i_{fc} + i_{n}\right]}{2F}$ \eqref{eq:vapour_inlet_and_outlet_flows} \\
			
			\multirow{4}{*}{
				$\begin{cases}
					\bm{J_{v}^{gdl,gc}} = \bm{J_{v,codi}}, \text{at the GDL/GC border} \\
					J_{v}^{in/out,gc} = J_{v,\text{in/out}}^{gc}, \text{at the inlet/outlet of the GC} 
				\end{cases}$
				\eqref{eq:vapour_dynamic_behaviour_GC_boundary_conditions} }
			& 
			$J_{v,\text{out}}^{agc} = \frac{\Phi_{agc,out} P_{sat}}{P_{agc,out}} \frac{k_{em,\text{in}}}{H_{gc} W_{gc} M_{agc,out}} \left[P_{agc,out} - P_{a,\text{des}}\right]$ \eqref{eq:vapour_inlet_and_outlet_flows} \\
			
			& $J_{v,\text{in}}^{cgc} = \frac{\Phi_{c,\text{des}} P_{sat}}{P_{\text{cgc,in}} - \Phi_{c,\text{des}} P_{sat}} \frac{1}{y_{O_{2},ext}} \frac{A_{act}}{H_{gc} W_{gc}} \frac{S_{c} \left[i_{fc} + i_{n}\right]}{4F}$ \eqref{eq:vapour_inlet_and_outlet_flows} \\
			
			&
			$J_{v,\text{out}}^{cgc} = \frac{\Phi_{cgc,out} P_{sat}}{P_{cgc,out}} \frac{k_{em,\text{in}}}{H_{gc} W_{gc} M_{cgc,out}} \left[P_{cgc,out} - P_{c,\text{des}}\right]$ \eqref{eq:vapour_inlet_and_outlet_flows} \\ \hline

		\end{tabularx}
		\caption{Synthesis of the partial differential equations and the spotlighted matter transport expressions (1/2)}
		\label{table:synthesis_spotlighted_transport_expressions_1}
	\end{table}
\end{landscape}

\begin{landscape}
	\begin{table}[H]
		\centering
		\linespread{1.2}
		\begin{tabularx}{\linewidth}{|Y|Y|} \hline
			
			\bf{Dynamical models} & \bf{Matter flow expressions} \\ \hline \hline	
			
			\multicolumn{2}{|c|}{\bf{Hydrogen in the GDL and the CL}} \\ \hline
			
			\multirow{2}{*}{
				\normalsize{
					$ \varepsilon \frac{\partial}{\partial t} \left( \left[  1 - \texttt{s} \right] C_{H_{2}} \right)  = - \bm{\nabla} \cdot \bm{J_{H_{2},dif}} + S_{H_{2},cons}$
					\eqref{subeq:hydrogen_dynamic_behaviour_balance}}} &
			\normalsize{
				$S_{H_{2},cons} = 
				\begin{cases}
					- \frac{i_{fc} + i_{sc}}{2 F H_{cl}} - \frac{R T_{fc}}{H_{cl}} \left[ k_{H_{2}} \nabla C_{H_{2}} + 2 k_{O_{2}} \nabla C_{O_{2}} \right], \text{in the ACL} \\
					0, \text{elsewhere} 
				\end{cases}$
				\eqref{eq:hydrogen_consumption_corrected}}  \\ & \\
			
			\multirow{2}{*}{
				\normalsize{
					$\begin{cases}
						\bm{J_{H_{2}}^{cl,mem}} = \bm{0}, \text{\small{at the CL/membrane border}} \\
						\bm{J_{H_{2}}^{gdl,gc}} = \bm{J_{H_{2},codi}}, \text{\small{at the GDL/GC border}}
					\end{cases}$
					\eqref{subeq:hydrogen_dynamic_behaviour_border}}} &
			\normalsize{
				$\bm{J_{H_{2}, dif}} = - D_{H_{2}}^{eff} \bm{\nabla} C_{H_{2}}$ \eqref{eq:hydrogen_flows}} \\ & \\
			
			&
			$\bm{J_{H_{2},codi}} = h_{H_{2}} \left[ C_{H_{2},\text{agc}} - C_{H_{2},\text{cgdl}}^{\text{inter}} \right] \bm{\imath}$ \eqref{eq:hydrogen_flows} \\ \hline

			\multicolumn{2}{|c|}{\bf{Hydrogen in the GC}} \\ \hline
			
			\multirow{2}{*}{
				\normalsize{
					$\frac{\partial C_{H_{2}}}{\partial t} = - \bm{\nabla} \cdot \bm{J_{H_{2},conv}}$
					\eqref{subeq:hydrogen_dynamic_behaviour_balance}}} &
			\normalsize{
				$\bm{J_{H_{2},conv}} = C_{H_{2}} \bm{u_{g}}$
				\eqref{eq:hydrogen_flows}} \\ & \\
			
			\multirow{2}{*}{
				\normalsize{
					$\begin{cases}
						\bm{J_{H_{2}}^{gdl,gc}} = \bm{J_{H_{2},codi}}, \text{at the GDL/GC border} \\
						J_{H_{2}}^{in/out,gc} = J_{H_{2}, \text{in/out}}, \text{at the inlet/outlet of the GC} 
					\end{cases}$
					\eqref{subeq:hydrogen_dynamic_behaviour_border}}} &
			\normalsize{
				$J_{H_{2},\text{in}} = \frac{A_{act}}{H_{gc} W_{gc}} \frac{S_{a} \left[i_{fc} + i_{n}\right]}{2F}$ \eqref{eq:hydrogen_flows}} \\ & \\
			
			&
			$J_{H_{2},\text{out}} = \frac{P_{\text{agc,out}} - \Phi_{\text{agc,out}} P_{sat}}{P_{\text{agc,out}}} \frac{k_{em,\text{in}}}{H_{gc} W_{gc} M_{agc,out}} \left[P_{agc,out} - P_{a,\text{des}}\right]$ \eqref{eq:hydrogen_flows} \\ \hline

			\multicolumn{2}{|c|}{\bf{Oxygen in the GDL and the CL}} \\ \hline
			
			\multirow{2}{*}{
				\normalsize{
					$ \varepsilon \frac{\partial}{\partial t} \left( \left[  1 - \texttt{s} \right] C_{O_{2}} \right)  = - \bm{\nabla} \cdot \bm{J_{O_{2},dif}} + S_{O_{2},cons}$
					\eqref{subeq:oxygen_dynamic_behaviour_balance}}} &
			\normalsize{
				$S_{O_{2},cons} = 
				\begin{cases}
					- \frac{i_{fc} + i_{sc}}{4 F H_{cl}} - \frac{ R T_{fc}}{H_{cl}} \left[ k_{O_{2}} \nabla C_{O_{2}} + \frac{k_{H_{2}}}{2} \nabla C_{H_{2}} \right], \text{in the CCL} \\
					0, \text{elsewhere} 
				\end{cases}$
				\eqref{eq:oxygen_consumption_corrected}}  \\ & \\
			
			\multirow{2}{*}{
				\normalsize{
					$\begin{cases}
						\bm{J_{O_{2}}^{cl,mem}} = \bm{0}, \text{at the CL/membrane border} \\
						\bm{J_{O_{2}}^{gdl,gc}} = \bm{J_{O_{2},codi}}, \text{at the GDL/GC border}
					\end{cases}$
					\eqref{subeq:oxygen_dynamic_behaviour_border}}} &
			\normalsize{
				$\bm{J_{O_{2}, dif}} = - D_{O_{2}}^{eff} \bm{\nabla}C_{O_{2}}$ \eqref{eq:oxygen_flows}} \\ & \\
			
			&
			$\bm{J_{O_{2},codi}} = h_{O_{2}} \left[ C_{O_{2},\text{cgdl}}^{\text{inter}} - C_{O_{2},\text{cgc}} \right] \bm{\imath}$ \eqref{eq:oxygen_flows}  \\ \hline

			\multicolumn{2}{|c|}{\bf{Oxygen in the GC}} \\ \hline
			
			\multirow{2}{*}{
				\normalsize{
					$\frac{\partial C_{O_{2}}}{\partial t} = - \bm{\nabla} \cdot \bm{J_{O_{2},conv}}$
					\eqref{subeq:oxygen_dynamic_behaviour_balance}}} &
			\normalsize{
				$\bm{J_{O_{2},conv}} = C_{O_{2}} \bm{u_{g}}$
				\eqref{eq:oxygen_flows}} \\ & \\
			
			\multirow{2}{*}{
				\normalsize{
					$\begin{cases}
						\bm{J_{O_{2}}^{gdl,gc}} = \bm{J_{O_{2},codi}}, \text{at the GDL/GC border} \\
						J_{O_{2}}^{in/out,gc} = J_{O_{2}, \text{in/out}}, \text{at the inlet/outlet of the GC} 
					\end{cases}$
					\eqref{subeq:oxygen_dynamic_behaviour_border}}} &
			\normalsize{
				$J_{O_{2},\text{in}} = \frac{A_{act}}{H_{gc} W_{gc}} \frac{S_{c} \left[i_{fc} + i_{n}\right]}{4F}$ \eqref{eq:oxygen_flows}} \\ & \\
			
			&
			$J_{O_{2},\text{out}} = y_{O_2,\text{cgc,out}}  \frac{P_{\text{cgc,out}} - \Phi_{\text{cgc,out}} P_{sat}}{P_{\text{cgc,out}}} \frac{k_{em,\text{in}}}{H_{gc} W_{gc} M_{cgc,out}} \left[P_{cgc,out} - P_{c,\text{des}}\right]$ \eqref{eq:oxygen_flows} \\ \hline

			\multicolumn{2}{|c|}{\bf{Nitrogen}} \\ \hline
			
			\multirow{3}{*}{
				\normalsize{
					$\frac{\mathrm{dC}_{\mathrm{N_{2}}}}{\mathrm{dt}} = W_{N_{2},\text{in}} - W_{N_{2},\text{out}} $
					\eqref{eq:nitrogen_dynamic_behaviour_agc}}} &
			\multirow{2}{*}{
				\normalsize{
					$W_{N_{2},\text{in}} = \frac{1 - y_{O_{2},ext}}{y_{O_{2},ext}} \frac{A_{act}}{H_{gc} W_{gc}} \frac{S_{c} \left[i_{fc} + i_{n}\right]}{4F}$ \eqref{eq:nitrogen_agc_in}}} \\ & \\
			
			&
			$W_{N_{2},\text{out}} = \left[ 1 - y_{O_{2},\text{cgc,out}} \right] \frac{P_{\text{cgc,out}} - \Phi_{\text{cgc,out}} P_{sat}}{P_{\text{cgc,out}}} \frac{k_{em,\text{in}}}{H_{gc} W_{gc} M_{cgc,out}} \left[P_{cgc,out} - P_{c,\text{des}}\right]$
			\eqref{eq:nitrogen_agc_out} \\ \hline

		\end{tabularx}
		\caption{Synthesis of the partial differential equations and the spotlighted matter transport expressions (2/2)}
		\label{table:synthesis_spotlighted_transport_expressions_2}
	\end{table}
\end{landscape}

\begin{landscape}
	\begin{table}[H]
		\centering
		\linespread{1.2}
		\begin{tabularx}{\linewidth}{|YYY|} \hline
			
			\multicolumn{3}{|>{\hsize=\dimexpr3\hsize+4\tabcolsep+2\arrayrulewidth\relax}Y|}{
				\bf{Coefficients associated to the dissolved water in the membrane}} \\ \hline
			
			$a_{w} (C,\texttt{s}) = \frac{C}{C_{sat}} + 2 \texttt{s}$ \eqref{eq:water_activity_simplified}
			& \multicolumn{2}{>{\hsize=\dimexpr2\hsize+2\tabcolsep+\arrayrulewidth\relax}Y|}{
				$D (\lambda) = 4.1 \times 10^{-10} \left[ \frac{\lambda}{25.0} \right]^{0.15} \left[ 1.0 + \tanh\left( \frac{\lambda-2.5}{1.4} \right)\right]$
				\eqref{eq:coef_dif_Kulikovsky}} \\ & & \\
			
			\multicolumn{3}{|>{\hsize=\dimexpr3\hsize+4\tabcolsep+2\arrayrulewidth\relax}Y|}{
				$ \begin{aligned}
					\lambda_{eq} & = 
					\frac{1}{2} \left[  0.300 + 10.8a_{w} - 16.0a_{w}^{2} + 14.1a_{w}^{3} \right] \cdot \left[  1 - \tanh\left( 100 \left[ a_{w} - 1 \right]\right)\right] \\
					& + \frac{1}{2} \left[ 9.2 +8.6 \left[ 1 - \exp\left( -K_{\text{shape}} \left[ a_{w} - 1        \right]\right)\right]\right] \cdot \left[ 1 + \tanh \left( 100 \left[ a_{w} - 1 \right]\right)\right]
				\end{aligned} $ \eqref{eq:lambda_eq_Hinatsu_Bao} } \\  & & \\
			
			$f_{v}(\lambda) = \frac{ \lambda V_{w} }{ V_{\text{mem}} + \lambda V_{w} }$ \eqref{eq:j_sorp_Ge}
			& \multicolumn{2}{>{\hsize=\dimexpr2\hsize+2\tabcolsep+\arrayrulewidth\relax}Y|}{
				$\gamma_{sorp} (\lambda,T) = 
				\begin{cases}
					\frac{1.14 \cdot 10^{-5} f_{v}(\lambda)}{H_{cl}} e^{ 2416 \left[ \frac{1}{303} - \frac{1}{T_{f c}} \right] }, & \text{\footnotesize absorption flow} \\
					\frac{4.59 \cdot 10^{-5} f_{v}(\lambda)}{H_{cl}} e^{ 2416 \left[ \frac{1}{303} - \frac{1}{T_{f c}} \right] }, & \text{\footnotesize desorption flow}
				\end{cases}$  \eqref{eq:j_sorp_Ge}   } \\ \hline

			\multicolumn{3}{|>{\hsize=\dimexpr3\hsize+4\tabcolsep+2\arrayrulewidth\relax}Y|}{
				\bf{Coefficients associated to liquid water in the GDL and the CL}} \\  \hline
			
			$\begin{cases}
				\texttt{e} = 3, & \text{ if $\varepsilon \in \left[ 0.1,0.4 \right] $ } \\
				\texttt{e} \in \left[ 4,5 \right], & \text{ if $\varepsilon \in \left[ 0.6,0.8 \right] $ }
			\end{cases}$
			\eqref{eq:capillary_exponent}
			&
			\multicolumn{2}{>{\hsize=\dimexpr2\hsize+2\tabcolsep+\arrayrulewidth\relax}Y|}{ 
				$K_{0}(\varepsilon) = \frac{\varepsilon}{8 \ln \left( \varepsilon \right)^{2}} \frac{\left[ \varepsilon - \varepsilon_{p} \right]^{\alpha + 2} r_{f}^{2}} {\left[ 1 - \varepsilon_{p} \right]^{\alpha} \left[ \left[ \alpha + 1 \right]  \varepsilon - \varepsilon_{p} \right]^{2}} e^{ \beta_{1} \varepsilon_{c}}$ \eqref{eq:intrinsic_permeability_TSB}} \\   & & \\
			
			\multicolumn{3}{|>{\hsize=\dimexpr3\hsize+4\tabcolsep+2\arrayrulewidth\relax}Y|}{ 
				$\sigma(T) = 235.8 \times 10^{-3} \left[ \frac{647.15 - T_{fc}}{647.15} \right]^{1.256} \left[ 1 - 0.625 \frac{647.15 - T_{fc}}{647.15} \right]$
				\eqref{eq:surface_tension}} \\   \hline

			\multicolumn{3}{|>{\hsize=\dimexpr3\hsize+4\tabcolsep+2\arrayrulewidth\relax}Y|}{
				\bf{Coefficients associated to vapour in the GDL and the CL}} \\  \hline
			
			$h_{v} = S_{h} \frac{D_{v}}{H_{gc}}
			\eqref{eq:Sherwood_number_def} $
			& \multicolumn{2}{>{\hsize=\dimexpr2\hsize+2\tabcolsep+\arrayrulewidth\relax}Y|}{
				$  D_{i/j}^{eff} = 
				\begin{cases}
					\varepsilon^{\tau}\left[ 1 - \texttt{s}\right] ^{\tau} D_{i/j}, &\text {in the CL} \\
					\varepsilon \left[ \frac{\varepsilon - \varepsilon_{p}} {1 - \varepsilon_{p}} \right]^{\alpha} \left[ 1 - \texttt{s} \right]^{2} e^{ \beta_{2} \varepsilon_{c}} D_{i/j}, &\text {in the GDL}
				\end{cases}	
				\eqref{eq:effective_diffusion_coefficient_TSB} $} \\   & & \\
			
			$S_{h} = 0.9247 \cdot \ln \left( \frac{W_{gc}}{H_{gc}} \right) + 2.3787$
			\eqref{eq:Sherwood_number_expression}
			& \multicolumn{2}{>{\hsize=\dimexpr2\hsize+2\tabcolsep+\arrayrulewidth\relax}Y|}{
				$\begin{cases}
					& D_{H_{2}O/H_{2}} = 1.644 \cdot 10^{-4} \left[ \frac{T_{fc}}{333} \right]^{2.334} \left[ \frac{101325}{P} \right] \\
					& D_{H_{2}O/O_{2}} = 3.242 \cdot 10^{-5} \left[ \frac{T_{fc}}{333} \right]^{2.334} \left[ \frac{101325}{P} \right]
				\end{cases}
				\eqref{eq:binary_diffusion_coef_Ohayre} $} \\	\hline

			\multicolumn{3}{|>{\hsize=\dimexpr3\hsize+4\tabcolsep+2\arrayrulewidth\relax}Y|}{
				\bf{Coefficients associated to $H_{2}$ and $O_{2}$ in the CL}} \\  \hline
			
			\multicolumn{3}{|>{\hsize=\dimexpr3\hsize+4\tabcolsep+2\arrayrulewidth\relax}Y|}{
				$k_{H_{2}} = 
				\begin{cases}
					\left[ 0.29 + 2.2 f_{v}\left( \lambda \right) \right] 10^{-14} \exp \left( \frac{E_{act,H_{2},v}}{R} \left[ \frac{1}{T_{ref}} - \frac{1}{T_{fc}} \right]\right) & if \lambda < \lambda_{l,eq}  \\
					1.8 \cdot 10^{-14} \exp \left( \frac{E_{act,H_{2},l}}{R} \left[ \frac{1}{T_{ref}} - \frac{1}{T_{fc}} \right]\right) & if \lambda = \lambda_{l,eq} \\
				\end{cases}$
				\eqref{eq:permeability_coefficients_crossover_H2}}  \\ 
			
			\multicolumn{3}{|>{\hsize=\dimexpr3\hsize+4\tabcolsep+2\arrayrulewidth\relax}Y|}{
				$k_{O_{2}} = 
				\begin{cases}
					\left[ 0.11 + 1.9 f_{v}\left( \lambda \right) \right] 10^{-14} \exp \left( \frac{E_{act,O_{2},v}}{R} \left[ \frac{1}{T_{ref}} - \frac{1}{T_{fc}} \right]\right) & if \lambda < \lambda_{l,eq}  \\
					1.2 \cdot 10^{-14} \exp \left( \frac{E_{act,O_{2},l}}{R} \left[ \frac{1}{T_{ref}} - \frac{1}{T_{fc}} \right]\right) & if \lambda = \lambda_{l,eq}
				\end{cases}$
				\eqref{eq:permeability_coefficients_crossover_O2}}  \\ \hline

		\end{tabularx}
		\caption{Synthesis of the spotlight flow coefficients}
		\label{table:synthesis_flow_coef}
	\end{table}
\end{landscape}

\begin{landscape}
	\begin{table}[H]
		\centering
		\linespread{1.2}
		\begin{tabularx}{\linewidth}{|Y|YYY|} \hline
			
			\multicolumn{4}{|>{\hsize=\dimexpr4\hsize+8\tabcolsep+3\arrayrulewidth\relax}Y|}{
				\bf{Spotlighted voltage polarization expressions}} \\ \hline \hline
			
			\textbf{The apparent voltage} &
			\multicolumn{3}{>{\hsize=\dimexpr3\hsize+4\tabcolsep+2\arrayrulewidth\relax}Y|}{
				$U_{cell} = U_{eq} - \eta_{c} - i_{fc}  \left[ R_{p} + R_{e} \right] $
				\eqref{eq:apparent_voltage}} \\ \hline
			
			& & & \\
			
			\textbf{The equilibrium potential} &
			\multicolumn{3}{>{\hsize=\dimexpr3\hsize+4\tabcolsep+2\arrayrulewidth\relax}Y|}{
				\multirow{2}{*}{
					$U_{eq} = E^{0} - 8.5 \cdot 10^{-4} \left[ T_{fc} - 298.15 \right] + \frac{R T_{fc}}{2 F} \left[ \ln \left( \frac{R T_{fc} C_{H_{2},\text{acl}}}{P_{\text{ref}}} \right) + \frac{1}{2} \ln \left( \frac{R T_{fc} C_{O_{2},\text{ccl}}}{P_{\text{ref}}} \right)\right]$ \eqref{eq:equilibrium_voltage}}} \\ & & & \\ \hline
			
			\multirow{3.5}{*}{
				\bf{The overpotential}}
			& \multicolumn{2}{>{\hsize=\dimexpr2\hsize+2\tabcolsep+\arrayrulewidth\relax}Y}{
				$\eta_{c} = \frac{R T_{fc}}{\alpha_{c}F}  ln \left( \frac{i_{fc} + i_{n}}{i_{0,c}^{ref} \left[ \frac{ C_{O_{2},ccl}}{C_{O_{2}}^{ref}} \right]^{\kappa_{c}}} \right) $ \eqref{eq:overpotential_fair_Tafel}}  
			& $i_{n} = i_{co,H_{2}} + i_{co,O_{2}} + i_{sc}$
			\eqref{eq:internal_current_density} \\ 
			
			& \multicolumn{2}{>{\hsize=\dimexpr2\hsize+2\tabcolsep+\arrayrulewidth\relax}Y}{
				$\begin{cases}
					i_{sc} =  \frac{U_{cell}}{r_{sc}}\\
					r_{sc} = 1.79 \cdot 10^{-2} \left[ \frac{P_{agc}}{101325} \right]^{-9.63} \left[ \frac{P_{cgc}}{101325} \right]^{0.38} 
				\end{cases}$
				\eqref{eq:i_sc}}  
			& $\begin{cases}
				i_{co,H_{2}} = 2 F k_{H_{2}} \nabla P_{H_{2}} \\
				i_{co,O_{2}} = 4 F k_{O_{2}} \nabla P_{O_{2}}
			\end{cases}$
			\eqref{eq:internal_crossover_current_density} \\ \hline
			
			\multirow{3}{*}{
				\textbf{The proton resistance}} &
			\multicolumn{3}{>{\hsize=\dimexpr3\hsize+4\tabcolsep+2\arrayrulewidth\relax}Y|}{
				$\sigma_{m} = 
				\begin{cases}
					\left[ 0.5139 \lambda - 0.326 \right] \exp \left( 1268 \left[ \frac{1}{303.15} - \frac{1}{T_{fc}} \right]\right), & \text{for $\lambda \geq 1$} \\
					
					0.1879 \exp \left( 1268 \left[ \frac{1}{303.15} - \frac{1}{T_{fc}} \right]\right), & \text{for $\lambda < 1$} \\
				\end{cases}$
				\eqref{eq:proton_conductivity_Springer}} \\ & & & \\
			
			&
			\multicolumn{3}{>{\hsize=\dimexpr3\hsize+4\tabcolsep+2\arrayrulewidth\relax}Y|}{
				$R_{p} = \int_{mem} \frac{\partial x}{\sigma_{m}} + \frac{1}{3} \int_{ccl} \frac{\partial x}{\frac{\varepsilon_{mc}}{\tau} \sigma_{m}}$ \eqref{eq:proton_conductivity_resistance}} \\ \hline
			
		\end{tabularx}
		\caption{Synthesis of the spotlighted voltage polarization expressions}
		\label{table:synthesis_spotlighted_voltage_expressions}
	\end{table}
\end{landscape}

\section{ORCID}

Raphaël Gass \url{https://orcid.org/0000-0002-9586-0884}

Zhongliang Li \url{https://orcid.org/0000-0001-7021-2103}

Rachid Outbib \url{https://orcid.org/0000-0002-9157-5269}

Daniel Hissel \url{https://orcid.org/0000-0003-0657-3320}

Samir Jemei \url{https://orcid.org/0000-0003-0195-903X}

\printnomenclature

\newpage
\appendix
\section{Other useful equations}

In this section, equations that establish connections between specific fundamental physical quantities and temperature are provided.

\subsection{Vapour saturated pressure: $P_{sat}$}

The vapour saturated pressure is expressed as \eqref{eq:vapour_saturated_pressure}. This correlation demonstrates acceptable agreement with the experimental data across the temperature range of -50 to 100 °C \cite{jiaoWaterTransportPolymer2011,yangMatchingWaterTemperature2011,wangModelingEffectsCapillary2008,fanCharacteristicsPEMFCOperating2017,hinatsuWaterUptakePerfluorosulfonic1994,wangInvestigationDryIonomer2020}.

\begin{figure*}[htb]
	\begin{equation}
		P_{sat}\left(T_{fc}\right) = 101325 \cdot 10^{- 2.1794 + 0.02953 \left[ T_{fc} - 273.15 	\right] - 9.1837 \cdot 10^{-5} \left[ T_{fc} -273.15 \right]^{2} + 1.4454 \cdot 10^{-7} \left[ T_{fc} - 273.15 \right]^{3}}
		\label{eq:vapour_saturated_pressure}
	\end{equation}
\end{figure*}

\subsection{Liquid water density: $\rho_{H_{2}O}$}

Liquid water density expression is expressed as \eqref{eq:liquid_water_density} \cite{kellDensityThermalExpansivity1975}. At 70°C, this expression yields $\rho_{H_{2}O} = 977.77$ $kg.m^{-3}$. 

\begin{figure*}[htb]
	\begin{equation}
		\begin{aligned}
			\rho_{H_{2}O} & = \frac{999.83952 + 16.945176 \left[ T_{fc} - 273.15 \right] - 	7.9870401 \cdot 10^{-3} \left[ T_{fc} - 273.15 \right]^{2} - 46.170461 \cdot 10^{-6} \left[ T_{fc} - 273.15 \right]^{3}}{1 + 16.879850 \cdot 10^{-3} \left[ T_{fc} - 273.15 \right]} \\
			& + \frac{105.56302 \cdot 10^{-9} \left[ T_{fc} - 273.15 \right]^{4} - 	280.54253 \cdot 10^{-12} \left[ T_{fc} - 273.15 \right]^{5}}{1 + 16.879850 \cdot 10^{-3} \left[ T_{fc} - 273.15 \right]}
		\end{aligned}
		\label{eq:liquid_water_density}
	\end{equation}
\end{figure*}

\subsection{Liquid water dynamic viscosity: $\mu_{l}$}

Liquid water dynamic viscosity is expressed as \eqref{eq:liquid_water_dynamic_viscosity} \cite{fanCharacteristicsPEMFCOperating2017}. 

\begin{equation}
	\mu_{l} = 2.414 \cdot 10^{-5 + \frac{247.8}{T_{fc} - 140.0}}
	\label{eq:liquid_water_dynamic_viscosity}
\end{equation}

The following table \ref{table:liquid_water_dynamic_viscosity} compares equation \eqref{eq:liquid_water_dynamic_viscosity} with data from alternative sources. Equation \eqref{eq:liquid_water_dynamic_viscosity} is evaluated there at 70°C.

\begin{table}[htb]
	\centering
	\begin{tabularx}{\linewidth}{|X|*{4}{X|}} \hline
		
		& Fan \cite{fanCharacteristicsPEMFCOperating2017}
		& Hu \cite{huAnalyticalCalculationEvaluation2016}
		& Yang \cite{yangMatchingWaterTemperature2011}
		& Bao \cite{baoTwodimensionalModelingPolymer2015} \\ \hline
		
		$\mu_{l}$ ($10^{-4}$ Pa.s)
		& {\normalsize $4.01$} 
		& {\normalsize $3.56$}   
		& {\normalsize $3.517$}
		& {\normalsize $3.508$} \\ \hline
		
	\end{tabularx}
	\caption{Comparison between the values given by the mentioned expression for the liquid water dynamic viscosity and values found in other works}
	\label{table:liquid_water_dynamic_viscosity}
\end{table}

\subsection{Liquid water kinematic viscosity: $\nu_{l}$}

Liquid water kinematic viscosity is expressed as \eqref{eq:liquid_water_kinematic_viscosity}. At 70°C, this expression yields $\nu_{l} = 4.10 \cdot 10^{-7} m^{2}.s^{-1}$, which is a close to $\nu_{l} = 3.7 \cdot 10^{-7} m^{2}.s^{-1}$ obtained from \cite{huAnalyticalCalculationEvaluation2016}.

\begin{equation}
	\nu_{l} \overset{\vartriangle}{=} \frac{\mu_{l}}{\rho_{H_{2}O}}
	\label{eq:liquid_water_kinematic_viscosity}
\end{equation}

\section{Synthesis of the constant values founded in the literature}

The objective of this appendix is to furnish a large set of constants utilized by previous researchers. These constants are presented in Tables \ref{table:constant_values_1} and \ref{table:constant_values_2}.

\begin{table*}[!htbp]
	\centering
	\begin{minipage}{\linewidth}
		\begin{tabularx}{\linewidth}{|c|*{13}{X|}} \hline
			
			References         
			& \cite{xuReduceddimensionDynamicModel2021}                        
			& \cite{wangInvestigationDryIonomer2020}  
			& \cite{fanCharacteristicsPEMFCOperating2017} 
			& \cite{huAnalyticalCalculationEvaluation2016} 
			& \cite{baoTwodimensionalModelingPolymer2015} 
			& \cite{jiaoWaterTransportPolymer2011} 
			& \cite{yangMatchingWaterTemperature2011} 
			& \cite{namMicroporousLayerWater2009} 
			& \cite{wangModelingEffectsCapillary2008}
			& \cite{yeThreeDimensionalSimulationLiquid2007} 
			& \cite{mengTwophaseNonisothermalMixeddomain2007} 
			& \cite{pasaogullariTwoPhaseModelingFlooding2005} 
			& \cite{kulikovskyQuasi3DModelingWater2003} \\
			
			Year               
			& 2021                                
			& 2020         
			& 2017         
			& 2016        
			& 2015         
			& 2011        
			& 2011         
			& 2009         
			& 2008         
			& 2007        
			& 2007         
			& 2005         
			& 2003         \\ \hline
			
			\multicolumn{14}{|c|}{Operating inputs} \\ \hline
			
			$T_{fc}$ $(K)$            
			& {\scriptsize $343$}                                 
			& {\scriptsize $353$}          
			& {\scriptsize $353$}         
			&             
			& {\scriptsize $353$}          
			&             
			& {\scriptsize $343$}          
			& {\scriptsize $343$}          
			&              
			&             
			&              
			&              
			& {\scriptsize $353$}         \\ \hline
			
			$P_{in}$ $(Pa)$           
			& {\tiny $\left[ 1.3-1.5 \right] \cdot 10^{5}$} 
			&              
			& {\scriptsize $101325$}      
			&             
			& {\scriptsize $202650$}      
			&             
			&              
			& {\scriptsize $101325$}       
			&              
			&             
			& {\scriptsize $202650$}       
			&              
			& {\scriptsize $303975$}       \\ \hline
			
			$S_{a}$                 
			& {\scriptsize $1.4$}                                 
			& {\scriptsize $2.0$}          
			& {\scriptsize $2.0$}        
			&          
			&           
			&          
			&           
			&           
			&              
			& {\scriptsize $6$ \footnote{at 1 $A.cm^{-2}$ \label{note:A.cm}}}          
			&              
			&              
			& {\scriptsize $1.5$}          \\ \hline
			
			$S_{c}$                 
			& {\scriptsize $1.8$}                                 
			& {\scriptsize $3.0$}          
			& {\scriptsize $1.5$}          
			&             
			&              
			&             
			&              
			&              
			& {\scriptsize $3$}           
			& {\scriptsize $3$ \textsuperscript{\ref{note:A.cm}}}          
			&              
			&              
			& {\scriptsize $1.5$}          \\ \hline
			
			$\Phi_{a,in}$              
			&       
			&      
			&      
			&      
			&      
			&    
			&      
			&              
			& {\scriptsize $1$}            
			& {\scriptsize $1$}           
			&             
			&              
			&              \\ \hline
			$\Phi_{c,in}$              
			&                                     
			& {\scriptsize $0.6$}       
			&     
			&      
			&        
			&       
			&        
			&              
			& {\scriptsize $1$}            
			& {\scriptsize $1$}          
			&              
			&              
			&              \\ \hline
			
			\multicolumn{14}{|c|}{Physical constants} \\ \hline
			
			$F$   $(C.mol^{-1})$     
			& \multicolumn{13}{c|}{ {\scriptsize $96485$}}  \\ \hline
			
			$R$   $(J.mol^{-1}.K^{-1})$  
			& \multicolumn{13}{c|}{ {\scriptsize $8.314$}} \\ \hline
			
			$M_{H2O}$ $(kg.mol^{-1})$    
			& \multicolumn{13}{c|}{ {\scriptsize $0.018$}} \\ \hline
			
			$\gamma_{O_{2},in}$ $(C.mol^{-1})$ 
			&\multicolumn{13}{c|}{ {\scriptsize $0.2095$}}  \\ \hline
			
			$K_{e}^{0}$                
			&   
			&              
			&              
			&             
			& {\scriptsize $6.2$}       
			&   
			&     
			&        
			&       
			&     
			&  
			&  
			&              \\ \hline
			
			$\Delta H^{0}$ $(J.mol^{-1})$    
			&       
			&              
			&              
			&             
			& {\scriptsize $5.23\cdot10^{4}$}     
			&    
			&      
			&     
			&       
			&      
			&    
			&   
			&              \\ \hline
			
			$\mu_{cg}$  $(Pa.s)$       
			&   
			&              
			&              
			& {\scriptsize $1.881\cdot10^{-5}$}   
			&              
			& {\scriptsize $2.075\cdot10^{-5}$}  
			&              
			&       
			&              
			&             
			&              
			& {\scriptsize $1.881\cdot10^{-5}$}    
			&             	\\ \hline	
			
		\end{tabularx}
	\end{minipage}
	\caption{Comparison of constant values from different sources (1/2)}
	\label{table:constant_values_1}
\end{table*}

\begin{table*}[!htbp]
	\centering
	\begin{minipage}{\linewidth}
		\begin{tabularx}{\linewidth}{|c|*{18}{X|}} \hline	
			References          
			& \cite{xuReduceddimensionDynamicModel2021}
			& \cite{wangInvestigationDryIonomer2020}  
			& \cite{dicksFuelCellSystems2018}
			& \cite{fanCharacteristicsPEMFCOperating2017}
			& \cite{huAnalyticalCalculationEvaluation2016}
			& \cite{baoTwodimensionalModelingPolymer2015}
			& \cite{xingNumericalAnalysisOptimum2015} 
			& \cite{jiaoWaterTransportPolymer2011}                 
			& \cite{yangMatchingWaterTemperature2011} 
			& \cite{namMicroporousLayerWater2009} 
			& \cite{wangModelingEffectsCapillary2008}
			& \cite{yeThreeDimensionalSimulationLiquid2007}          
			& \cite{mengTwophaseNonisothermalMixeddomain2007}           
			& \cite{pasaogullariTwoPhaseModelingFlooding2005}
			& \cite{bultelInvestigationMassTransport2005}
			& \cite{makhariaMeasurementCatalystLayer2005} 
			& \cite{kulikovskyQuasi3DModelingWater2003}
			& \cite{motupallyDiffusionWaterNafion2000} \\ 
			
			Year                
			& {\scriptsize 2021}         
			& {\scriptsize 2020}          
			& {\scriptsize 2018}         
			& {\scriptsize 2017}          
			& {\scriptsize 2016}         
			& {\scriptsize 2015}          
			& {\scriptsize 2015}          
			& {\scriptsize 2011}                           
			& {\scriptsize 2011}          
			& {\scriptsize 2009}          
			& {\scriptsize 2008}          
			& {\scriptsize 2007}                    
			& {\scriptsize 2007}                      
			& {\scriptsize 2005}          
			& {\scriptsize 2005}          
			& {\scriptsize 2005}          
			& {\scriptsize 2003}          
			& {\scriptsize 2000}          \\ \hline
			
			\multicolumn{19}{|c|}{Fuel cell physical parameters} \\ \hline
			
			$L_{gc}$ $(m)$             
			& {\scriptsize $12$}          
			& {\scriptsize $0.1$}          
			&             
			& {\scriptsize $0.1$}          
			& {\tiny $1.298$}       
			& {\tiny $0.9282$}       
			&              
			&                               
			& {\scriptsize $0.2$}          
			&              
			&              
			& {\scriptsize $0.2$}                    
			&                          
			&              
			&              
			&              
			&              
			& {\scriptsize $1.36$}         \\ \hline
			
			$H_{gc}$ $(m)$             
			& {\scriptsize $5 \cdot 10^{-4}$}       
			& {\scriptsize $10^{-3}$}       
			&             
			& {\scriptsize $10^{-3}$}       
			& {\scriptsize $10^{-3}$}        
			& {\scriptsize $10^{-3}$}         
			& {\scriptsize $5 \cdot 10^{-4}$}        
			&                               
			& {\scriptsize $5 \cdot 10^{-4}$}       
			& {\scriptsize $10^{-3}$}          
			&              
			& {\scriptsize $5 \cdot 10^{-4}$}                   
			& {\scriptsize $10^{-3}$}                    
			& {\scriptsize $10^{-3}$}         
			&              
			&              
			& {\scriptsize $2 \cdot 10^{-3}$}       
			& {\scriptsize $7.6 \cdot 10^{-4}$}      \\ \hline
			
			$W_{gc}$ $(m)$             
			& {\scriptsize $8 \cdot 10^{-4}$}        
			& {\scriptsize $10^{-3}$}         
			&             
			& {\scriptsize $8 \cdot 10^{-4}$}        
			&             
			&              
			& {\scriptsize $7.5 \cdot 10^{-4}$}      
			&                               
			& {\scriptsize $10^{-3}$}        
			&              
			&              
			& {\scriptsize $10^{-3}$}                    
			&                          
			&              
			&              
			&              
			& {\scriptsize $10^{-3}$}          
			& {\scriptsize $1.59 \cdot 10^{-3}$}     \\ \hline
			
			$H_{gdl}$ $(m)$            
			& {\scriptsize $2.3 \cdot 10^{-4}$}      
			& {\scriptsize $3 \cdot 10^{-4}$}         
			&            
			& {\scriptsize $4.2 \cdot 10^{-4}$}       
			& {\scriptsize $2.1 \cdot 10^{-4}$}      
			& {\scriptsize $3 \cdot 10^{-4}$}        
			& {\scriptsize $3.8 \cdot 10^{-4}$}       
			& {\scriptsize $2 \cdot 10^{-4}$}                          
			& {\scriptsize $2 \cdot 10^{-4}$}         
			& {\scriptsize $2.5 \cdot 10^{-4}$}       
			& {\scriptsize $2.5 \cdot 10^{-4}$}      
			& {\scriptsize $1.8 \cdot 10^{-4}$}               
			& {\scriptsize $3 \cdot 10^{-4}$}                     
			& {\scriptsize $3 \cdot 10^{-5}$}        
			&              
			&             
			&              
			&              \\ \hline
			
			$\varepsilon_{gdl}$              
			&             
			& {\scriptsize $0.7$}           
			&             
			& {\scriptsize $0.6$}           
			& {\scriptsize $0.6$}          
			& {\scriptsize $0.4$}          
			& {\scriptsize $0.4$}           
			&                               
			& {\scriptsize $0.7$}           
			&             
			& {\scriptsize $0.5$}           
			& {\scriptsize $0.7$}                     
			& {\scriptsize $0.6$}                       
			& {\scriptsize $0.5$}          
			&              
			&              
			&              
			&              \\ \hline
			
			$H_{cl}$ $(m)$             
			& {\scriptsize $10^{-5}$}         
			& {\scriptsize $10^{-5}$}         
			&             
			& {\scriptsize $10^{-5}$}         
			& {\scriptsize $10^{-5}$}        
			&              
			&              
			& {\scriptsize $10^{-5}$}                          
			& {\scriptsize $10^{-5}$}          
			& {\scriptsize $10^{-5}$}          
			& {\scriptsize $1.6 \cdot 10^{-5}$}       
			& {\scriptsize $1.5 \cdot 10^{-5}$}                
			& {\scriptsize $10^{-5}$}                    
			& {\scriptsize $10^{-5}$}         
			&             
			&              
			& {\scriptsize $5 \cdot 10^{-5}$}        
			&              \\ \hline
			
			$\varepsilon_{cl}$               
			& {\scriptsize $0.2$}          
			& {\scriptsize $0.3$}           
			&            
			& {\scriptsize $0.3$}           
			& {\scriptsize $0.6$}          
			&              
			&              
			&                               
			& {\scriptsize $0.2$}           
			& {\tiny $0.2-0.3$}       
			& {\scriptsize $0.12$}          
			& {\scriptsize $0.2$}                     
			& {\scriptsize $0.12$}                      
			&              
			&              
			&              
			&              
			&              \\ \hline
			
			$\varepsilon_{mc}$                 
			& {\scriptsize $0.2$}          
			& {\scriptsize $0.25$\footnote{optimal value according to \cite{wangInvestigationDryIonomer2020}}}         
			&             
			& {\tiny $0.22/0.27$}     
			&             
			&              
			&              
			&                              
			&              
			&              
			& {\tiny $0.393$}         
			&                        
			& {\scriptsize $0.4$}                       
			& {\scriptsize $0.2$}          
			&              
			& {\scriptsize $0.15$}          
			&              
			&              \\ \hline
			
			$s_{lim}$                
			& {\tiny $0.2735b$\footnote{value obtained with experimental fits from \cite{xuReduceddimensionDynamicModel2021}}}      
			&              
			&             
			&              
			&             
			&              
			&              
			&                              
			&              
			&              
			&              
			&                        
			&                          
			&              
			&              
			&              
			&              
			&              \\ \hline
			
			$H_{mem}$ $(m)$            
			& {\scriptsize $2.5 \cdot 10^{-5}$}      
			& {\scriptsize $2.5 \cdot 10^{-5}$}    
			&             
			& {\scriptsize $5 \cdot 10^{-5}$}      
			& {\scriptsize $2.5 \cdot 10^{-5}$}     
			& {\scriptsize $5 \cdot 10^{-5}$}        
			&              
			& {\scriptsize $5 \cdot 10^{-5}$}                        
			& {\scriptsize $2.5 \cdot 10^{-5}$}      
			&              
			& {\scriptsize $5 \cdot 10^{-5}$}       
			& {\scriptsize $5 \cdot 10^{-5}$}                  
			& {\scriptsize $2.5 \cdot 10^{-5}$}                 
			& {\scriptsize $5 \cdot 10^{-5}$}        
			&              
			&              
			& {\scriptsize $2 \cdot 10^{-4}$}        
			& {\scriptsize $1.5 \cdot 10^{-4}$}     \\ \hline
			
			$\rho_{mem}$ $(kg.m^{-3})$     
			&             
			& {\scriptsize $1980$}          
			&             
			& {\scriptsize $1980$}          
			& {\scriptsize $1980$}         
			& {\scriptsize $2000$}          
			& {\scriptsize $2000$}          
			& {\scriptsize $1980$}                           
			& {\scriptsize $1980$}         
			&              
			&              
			&                        
			& {\scriptsize $1980$}                      
			& {\scriptsize $1980$}          
			&              
			& {\scriptsize $2000$}          
			&              
			& {\scriptsize $2000$}          \\ \hline
			
			$M_{eq}$ $(kg.mol^{-1})$   
			&             
			& {\scriptsize $1.1$}          
			&             
			& {\scriptsize $1.1$}           
			& {\scriptsize $1.1$}          
			& {\scriptsize $1.1$}           
			& {\scriptsize $1.1$}           
			& {\scriptsize $1.1$}                           
			& {\scriptsize $1.1$}           
			&              
			&              
			&                        
			& {\scriptsize $1.1$}                       
			& {\scriptsize $1.1$}           
			&              
			&              
			&              
			& {\scriptsize $1.1$}           \\ \hline
			
			$A_{act}$ $(m^{2})$         
			& {\scriptsize $2.91 \cdot 10^{-2}$}    
			&              
			&             
			&             
			&             
			&             
			&              
			&                               
			&              
			&              
			&              
			&                        
			&                          
			&              
			&             
			&              
			&              
			& {\scriptsize $5 \cdot 10^{-3}$}      \\ \hline
			
			\multicolumn{19}{|c|}{Constants based on the interaction between water and the structure}   \\ \hline
			
			$\gamma_{v}$ $(s^{-1})$            
			& {\scriptsize $1.3$}          
			&              
			&             
			& {\scriptsize $1.3$}           
			&             
			&              
			&              
			& {\scriptsize $1.3$}                            
			&              
			&             
			&              
			&                        
			&                          
			&              
			&             
			&              
			&              
			&              \\ \hline
			
			$\gamma_{cond}$ $(s^{-1})$       
			&             
			& {\scriptsize $5 \cdot 10^{3}$}       
			&             
			& {\scriptsize $5 \cdot 10^{3}$}      
			& {\scriptsize $10^{4}$}       
			&              
			& {\scriptsize $10^{2}$}         
			& {\tiny $\left[ 1,10^{4} \right]$}  
			& {\scriptsize $10^{2}$}          
			&              
			& {\scriptsize $10^{2}$}          
			& {\scriptsize $1.0$}                
			& {\scriptsize $5 \cdot 10^{3}$\footnote{optimal value according to \cite{mengTwophaseNonisothermalMixeddomain2007} \label{note:optimal_Meng}}}  
			&              
			&              
			&              
			&              
			&              \\ \hline
			
			$\gamma_{evap}$ $(Pa^{-1}.s^{-1})$  
			&             
			& {\scriptsize $10^{-4}$}          
			&             
			& {\scriptsize $10^{-4}$}           
			&             
			&              
			& {\scriptsize $10^{-3}$}           
			&                              
			& {\scriptsize $10^{-3}$}        
			&              
			& {\scriptsize $10^{-3}$}        
			& {\scriptsize $5 \cdot 10^{-5}$}                
			& {\scriptsize $10^{-4}$\textsuperscript{\ref{note:optimal_Meng}}}   
			&              
			&              
			&              
			&              
			&              \\ \hline
			
			$\theta^{cl}_{c}$ $(°)$          
			& {\scriptsize $120$}          
			& {\scriptsize $95$}            
			&             
			& {\scriptsize $95$}            
			& {\scriptsize $110$}          
			&              
			& {\scriptsize $120$}           
			&                               
			& {\scriptsize $95$}            
			&              
			&              
			&                        
			& {\scriptsize $95$}                        
			& {\scriptsize $110$}           
			&              
			&              
			&              
			&              \\ \hline
			
			$\theta^{gdl}_{c}$ $(°)$        
			& {\scriptsize $120$}          
			& {\scriptsize $110$}          
			&             
			& {\scriptsize $120$}           
			& {\scriptsize $110$}          
			&              
			& {\scriptsize $120$}           
			&                               
			& {\scriptsize $110$}           
			&              
			&              
			&                        
			& {\scriptsize $110$}                       
			& {\scriptsize $110$}           
			&              
			&              
			&              
			&              \\ \hline
			
			\multicolumn{19}{|c|}{Referenced values} \\ \hline
			
			$i_{n}$ $(A.m^{-2})$        
			&             
			&              
			& {\scriptsize $20$}         
			&              
			&             
			&         
			&              
			&                               
			&              
			&         
			&          
			&                       
			&                          
			&              
			&           
			&              
			&             
			&              \\ \hline
			
			$i_{0,c}$ $(A.m^{-2})$        
			&             
			&              
			& {\scriptsize $0.67$}         
			&              
			&             
			& {\scriptsize $150$\footnote{at 353.15 K \label{note:353.15K}}}         
			&              
			&                               
			&              
			& {\scriptsize $0.1$}           
			& {\scriptsize $0.01$}          
			&                       
			&                          
			&              
			& {\scriptsize $0.42$}          
			&              
			&             
			&              \\ \hline
			
			$i_{0,c}$   $(A.m^{-3})$      
			&             
			& {\scriptsize $120$\textsuperscript{\ref{note:353.15K}}}           
			&             
			& {\scriptsize $120$\textsuperscript{\ref{note:353.15K}}}           
			&             
			&              
			&              
			&                               
			&              
			&              
			&              
			& {\scriptsize $10^{4}$\footnote{at 343 K \label{note:343K}}}                   
			&              
			& {\scriptsize $120$}           
			&              
			&              
			&    
			&          \\ \hline
			
			$i_{0,a}$   $(A.m^{-3})$      
			&             
			&              
			&             
			& {\scriptsize $10^{8}$\textsuperscript{\ref{note:353.15K}}}     
			&             
			&              
			&              
			&                               
			&              
			&              
			&              
			&                        
			& {\scriptsize $10^{9}$}                   
			&              
			& {\scriptsize $10^{8}$}         
			&              
			&              
			&              \\ \hline
			
			$\alpha_{c}$                  
			&             
			& {\scriptsize $0.5$}           
			& {\scriptsize $0.5$}          
			& {\scriptsize $0.5$}           
			&             
			& {\scriptsize $0.18$}          
			&              
			&                               
			&              
			& {\scriptsize $1$}             
			& {\scriptsize $1$}             
			& {\scriptsize $1$}                       
			& {\scriptsize $1$}                         
			& {\scriptsize $1$}            
			&             
			&              
			& {\scriptsize $1$}             
			&              \\ \hline
			
			$E_{act}$ $(J.mol^{-1})$
			&             
			& {\tiny $6.568 \cdot 10^{4}$}          
			&           
			& {\tiny $6.568 \cdot 10^{4}$}           
			&             
			& {\tiny $7.32 \cdot 10^{4}$}     
			&              
			&                               
			&              
			&          
			&              
			&                      
			&                         
			&        
			&             
			&              
			&          
			&              \\ \hline
			
			$C^{ref}_{O_{2}}$  $(mol.m^{-3})$
			&             
			& {\scriptsize $3.39$}          
			&             
			& {\scriptsize $3.39$}         
			&             
			& {\tiny $40.89$}         
			& {\scriptsize $3.39$}          
			&                               
			&              
			& {\tiny $40.89$}         
			& {\scriptsize $5.55$}          
			& {\scriptsize $5.24$\textsuperscript{\ref{note:343K}}}                  
			& {\scriptsize $40$}                        
			&              
			&              
			&              
			&              
			&              \\ \hline
			
			$C^{ref}_{H_{2}}$   $(mol.m^{-3})$
			&             
			& {\scriptsize $56.4$}          
			&             
			& {\scriptsize $56.4$}          
			&             
			& {\tiny $40.89$}         
			& {\scriptsize $56.4$}          
			&                              
			&              
			&              
			&              
			&                        
			& {\scriptsize $40$}                        
			&              
			&              
			&              
			&              
			&              \\ \hline
			
			$P_{ref}$   $(Pa)$         
			&             
			&              
			&             
			& {\scriptsize $10^{5}$}           
			&             
			& {\scriptsize $10^{5}$}         
			&              
			& {\scriptsize $10^{5}$}                          
			&              
			&              
			&              
			&                        
			&                          
			&              
			&             
			&              
			&              
			&              \\ \hline
			
			\multicolumn{19}{|c|}{Mathematical factors}  \\ \hline
			
			$K_{shape}$              
			& {\scriptsize $5$}            
			&              
			&             
			&             
			&             
			& {\scriptsize $2$}             
			&              
			&                               
			&              
			&              
			&              
			&                        
			&                          
			&              
			&              
			&              
			&              
			&             	 \\ \hline
			
		\end{tabularx}
	\end{minipage}
	\caption{Comparison of constant values from different sources (2/2)}
	\label{table:constant_values_2}
\end{table*}

\section{Synthesis of the hypothesis made in this work}

The purpose of the appendix is to succinctly summarize and categorize all hypotheses formulated in this study.

\subsection{Globally}
\begin{itemize}[itemsep = -2pt]
	\item The stack described in these equations is composed of 1 cell. 
	\item The stack temperature is considered constant and uniform (the cooling system is not represented).
	\item All the gas species behave ideally \cite{xuReduceddimensionDynamicModel2021}.
	\item The effect of gravity is ignored.
	\item The cell is operated with pure hydrogen, thus no contamination effects are considered.
	\item Nitrogen is supposed to be homogenous in all the cathode and the CGC.
\end{itemize}

\subsection{In the membrane}
\begin{itemize}[itemsep = -2pt]
	\item The experimental equations were generally measured on \Nafion-117 membrane \cite{springerPolymerElectrolyteFuel1991,hinatsuWaterUptakePerfluorosulfonic1994,baoTwodimensionalModelingPolymer2015}.
	\item Certain experiments were conducted at a fixed temperature of 30°C or 80°C. It is assumed that these data can be used at any working PEMFC temperature \cite{springerPolymerElectrolyteFuel1991,hinatsuWaterUptakePerfluorosulfonic1994,baoTwodimensionalModelingPolymer2015}.
	\item The thickness of the membrane at different water contents is assumed to be unchanged. The membrane expansion is ignored \cite{geAbsorptionDesorptionTransport2005}. 
	\item Water generated at the triple points is produced in dissolved form in the membrane \cite{jiaoWaterTransportPolymer2011}.
	\item Water that crosses the membrane to the CL is in vapour form \cite{geAbsorptionDesorptionTransport2005}.
	\item $N_{2}$ crossover is neglected. Please refer to \cite{baikCharacterizationNitrogenGas2011} for more information.
	
\end{itemize}

\subsection{In the CLs}
\begin{itemize}[itemsep = -2pt]
	\item The gas flow in the CL exhibits laminar characteristics.
	\item The electrolyte in the CL is assumed to have the same tortuosity characteristics as the CL carbon structure.
	\item The CLs are modelled as an agglomerate of packed spherical particles.
	\item The redox reactions of oxygen and hydrogen are considered to be infinitely fast.
\end{itemize}

\subsection{In the GDLs}
\begin{itemize}[itemsep = -2pt]
	\item The GDLs are modelled as a fibrous porous media composed of randomly oriented cylindrical fibers.
	\item To characterize water transport in GDLs, the Leverett function is employed. This function is derived from experimental data obtained from structures distinct from those found in PEMFCs. Nevertheless, it continues to be extensively utilized \cite{jiaoWaterTransportPolymer2011}.
	\item The gas flow in the GDL exhibits laminar characteristics.
	\item The deformation of the porous medium is considered negligible, and the water flow is sufficiently slow to result in a small Reynolds number under stationary conditions \cite{whitakerMethodVolumeAveraging1999}.
	\item Gas motions transport liquid water, which generate a convective flow denoted as $J_{l,conv}$. Nevertheless, it is neglected compared to the capillary flow $J_{l,cap}$.
\end{itemize}

\subsection{In the GCs}
\begin{itemize}[itemsep = -2pt]
	\item The gas flow in the channel is predominantly convective.
	\item Liquid water is considered nonexistent in the GC, and a Dirichlet boundary condition is imposed at the GDL/GC interface, setting the liquid water saturation variable \texttt{s} to zero.
	\item All gases have the same velocity in the gas mixture.
	\item Water phase change is ignored in the GC.
	\item The 'dividing line', or boundary between convective-dominated flow inside the core of the GC and diffusive-dominated flow inside the core of the electrodes, is assumed to occur at the interface between the GC and the GDL.
	\item Without better knowledge, it is considered that both concentrations at the two side of the GDL/GC interface are instantaneously equal. This is available for all gases : $C_{gc}^{\text{inter}} = C_{gdl}^{\text{inter}}$.
\end{itemize}

\subsection{For the voltage}
\begin{itemize}[itemsep = -2pt]
	\item It is assumed that the stack can follow the imposed current density.
	\item Anode overpotential is neglected.
	\item Anode potential is set to zero.
	\item Among the four elementary steps of the oxidation reduction reaction on the $Pt_{(111)}$ surface, OH formation reaction is the rate-limiting step \cite{baoTwodimensionalModelingPolymer2015}.
\end{itemize}

\section{Demonstrations}
\label{sec:demonstrations}

\subsection{Additional information concerning the capillary flow $J_{l,cap}$ and the convective flow $J_{l,conv}$}
\label{subsec:J_cap}

To enhance the understanding of $J_{l,cap}$ and $J_{l,conv}$, supplementary information is provided herein. 
First, an adaptation of Darcy's law incorporating the variables of this study is expressed as \ref{eq:Darcy_law} \cite{neumanTheoreticalDerivationDarcy1977}.

\begin{equation}
	\bm{J_{l}} = - \dfrac{K_{l}}{\nu_{l}} \bm{\nabla P_{l}}
	\label{eq:Darcy_law}
\end{equation}	
where $\bm{J_{l}}$ is the liquid water flow, $K_{l}$ $(m^{2})$ is the liquid phase permeability, and $P_{l}$ $(Pa)$ is the liquid-phase pressure.

Then, the standard approach for calculating $P_{l}$ involves the utilization of capillary pressure $P_{c}$ as defined in \eqref{eq:P_c_def}. This is because $P_{c}$ is exclusively influenced by pore geometry, fluid characteristics, and phase saturation. Consequently, it is a measurable quantity.

\begin{equation}
	P_{c} \overset{\vartriangle}{=} P_{g} - P_{l}
	\label{eq:P_c_def}
\end{equation}	
where $P_{g}$ $(Pa)$ is the gas-phase pressure. For information, within a liquid, the intermolecular cohesive forces (e.g., hydrogen bonding for water) compensate each other. Each molecule generates interaction forces in all directions in a isotropic manner with neighbouring molecules, the resultant of these forces is therefore zero. However, at the surface, this is not the case (interactions with gas molecules are negligible) and the resultant of the forces for the molecules at the surface is directed towards the interior of the liquid. Therefore, there is an additional force, which counterbalances the pressure of the liquid at its surface; this is the capillary pressure Pc. In fact, wherever we are in the liquid, the pressure is globally the same (if we set aside gravity) and is mainly owing to the concentration of the species and their temperature. However, the surface molecules are slowed down, which reduces the pressure at the surface. As this surface pressure is at equilibrium equal to the gas pressure, it follows that liquid water is at a higher pressure, which is logical as it is a much more condensed phase.

Subsequently, by reapplying Darcy's law to establish the relationship between the gas phase pressure $P_{g}$ and its velocity $u_{g}$, \eqref{eq:J_l} is derived. Two distinct flows are identified: the capillary flow, as discussed in \ref{subsec:liquid_water_capillary_flow}, and the convective flow, as discussed in \ref{subsec:liquid_water_convective_flow}.

\begin{equation}
	\bm{J_{l}} = \dfrac{K_{l}}{\nu_{l}} \bm{\nabla P_{c}} + \dfrac{\mu_{g}}{\nu_{l}} \dfrac{K_{l}}{K_{g}} \bm{u_{g}} = \bm{J_{l,cap}} + \bm{J_{l,conv}}
	\label{eq:J_l}
\end{equation}	
where $\mu_{g}$ $(Pa.s)$ is the gas mixture dynamic viscosity, $K_{l} = K_{0} s^{e}$ $(m^{2})$ is the liquid water phase permeability, $K_{g} = K_{0} \left( 1 - s \right)^{e}$ $(m^{2})$ is the gas mixture phase permeability, and $\bm{u_{g}}$ $(m.s^{-1})$ is the gas mixture velocity.

Capillary flow requires further refinement to become practical. Consequently, \eqref{eq:J_l_cap} is extracted from \eqref{eq:J_l}.

\begin{equation}
	\bm{J_{l,cap}} = \dfrac{K_{l}}{\nu_{l}} \bm{\nabla P_{c}}
	\label{eq:J_l_cap}
\end{equation}	

The subsequent step involves emphasizing $\texttt{s}$, representing the liquid water saturation. To achieve this, the gradient \bm{$\nabla$} is transitioned from $P_{c}$ to $\texttt{s}$, reinterpreting \eqref{eq:J_l_cap} as a Fick-like equation in \eqref{eq:j_cap_detail}, where $D_{cap}$ $(kg.m^{-1}.s^{-1})$ represents the capillary diffusion coefficient.

\begin{equation}
	\linespread{1.2}
	\begin{cases}
		\bm{J_{l,cap}} = - D_{cap} \bm{\nabla \texttt{s}} \\
		D_{cap} = - \frac{K_{l}}{\nu_{l}} \frac{\partial P_{c}}{\partial \texttt{s}}  
	\end{cases}
	\label{eq:j_cap_detail}
\end{equation}
\nomenclature[A,1]{$D_{c}$}{capillary diffusion coefficient $(kg.m^{-1}.s^{-1})$}

Next, the permeability of the liquid phase can be determined through \eqref{eq:K_l}, while the capillary pressure is correlated with the properties of porous materials as indicated in \eqref{eq:P_c} \cite{jiaoWaterTransportPolymer2011,xingNumericalAnalysisOptimum2015,wangModelingEffectsCapillary2008,yeThreeDimensionalSimulationLiquid2007,pasaogullariTwoPhaseModelingFlooding2005}.

\begin{equation}
	K_{l} = K_{0} \texttt{s}^{\texttt{e}}
	\label{eq:K_l}
\end{equation}

\begin{equation}
	P_{c} = - \sigma \left| \cos \left( \theta_{c} \right)\right| \sqrt{\frac{\varepsilon}{K_{0}}} J(\texttt{s})
	\label{eq:P_c}
\end{equation}	
\nomenclature[A,1]{$J(s)$}{Leverett function}
where $K_{0}$ $(m^{2})$ is the intrinsic permeability and $J(s)$ is the Leverett function.

For information, $k_{rl}$ can also be sourced from existing literature. It represents the relative permeability of the liquid phase and is solely a function of phase saturation, as expressed in \eqref{eq:k_rl}.

\begin{equation}
	k_{rl} = \frac{K_{l}}{K_{0}} = \texttt{s}^{\texttt{e}}
	\label{eq:k_rl}
\end{equation}	

Then, the Leverett function J, depicted in \ref{eq:Leverett_function}, relies on experimental data obtained from homogeneous soil or a sand bend with uniform wettability, which differs from the structures of GDL and CL in PEMFC. Additional experimental measurements have been undertaken in an effort to evaluate the actual conditions in PEMFC. Nonetheless, the obtained results exhibit discrepancies. As a consequence, the aforementioned equation continues to be extensively employed in PEMFC studies \cite{jiaoWaterTransportPolymer2011}.

\begin{equation}
	J \left( \texttt{s} \right) = 1.417\texttt{s} - 2.12\texttt{s}^{2} + 1.263\texttt{s}^{3}
	\label{eq:Leverett_function}
\end{equation}

Finally, given all of these considerations, it is possible to derive the mainly used expression of $\bm{J_{l,cap}}$ as shown in \ref{eq:j_cap}.

\subsection{Additional information concerning the convective-diffusive flow at the GDL/GC interface $J_{v,codi}$}
\label{subsec:additional_information_concerning_Jv_codi}

The expression of $J_{v,codi}$ in \eqref{eq:vapour_convective-diffusive_flow} needs further explanations. This flow is primary based on the diffusive theory, which rules that a diffusive flow is proportional to the gradient of its characteristic variable, which is here the vapour concentration, as shown in \eqref{eq:diffusive_flow}. 

\begin{equation}
	J_{v,dif} = D_{v} \bf{\nabla C}
	\label{eq:diffusive_flow}
\end{equation}

However, this theory is applicable only in case of a very thin volume at the GDL/GC interface at the GC side, where diffusion is the dominant flow. The thickness of this thin volume is written as $\varepsilon_{gc}$. Elsewhere in the GC, convection is dominant and leads, for simple modelling, to an homogeneous value of the concentration in the $x$ direction (see figure \ref{fig:matter_transport_in_PEMFC}). This homogeneity is only valid along the thickness. Thus, in the GC outside the mentioned thin volume, $C_{v,gc}$ is not function of $x$ anymore. Considering that $\varepsilon_{gc}$ is very small, the diffusive flow can then be rewritten as in \eqref{eq:diffusive_flow_epsilon}.

\begin{equation}
	J_{v,dif} = \pm D_{v} \frac{C_{v,gc} - C_{v,gc}^{\text{inter}}}{\varepsilon_{gc}} \bf{\imath}
	\label{eq:diffusive_flow_epsilon}
\end{equation}

$\varepsilon_{gc}$ is a variable influenced by both the GC geometry and the characteristics of the flows. The challenge in measuring $\varepsilon_{gc}$ is circumvented by introducing a dimensionless number, the Sherwood number $S_{h}$, which is defined as follows:

\begin{equation}
	S_{h} = \frac{H_{gc}}{\varepsilon_{gc}}
	\label{eq:Sh}
\end{equation}
with $H_{gc}$ the characteristic thickness of the GC. Then, equation \eqref{eq:diffusive_flow_epsilon} becomes \eqref{eq:diffusive_flow_Sh}.

\begin{equation}
	J_{v,dif} = \pm S_{h} \frac{D_{v}}{H_{gc}} \left[C_{v,gc} - C_{v,gc}^{\text{inter}} \right] \bf{\imath}
	\label{eq:diffusive_flow_Sh}
\end{equation}

As $C_{v,gc}$ is unaffected by the $x$ direction due to convection, determining its value becomes straightforward. 
Finally, these coefficients are encompassed within $h_{v}$, as elaborated in \ref{subsec:water_effective_convective-diffusive_mass_transfer_coefficient}, resulting in \eqref{eq:vapour_convective-conductive_flow_final}.

\subsection{Simplified flows at the inlet of the AGC}
\label{subsec:simplified_flows_inlet_outlet_AGC}

The consumed molar rate of hydrogen is given by equation \eqref{eq:dot{n}_{H_{2},cons}}. It is important to extract the active area from the fuel flow, considering that MEA and GC have different flow areas.

\begin{equation}
	\dot{n}_{H_{2},\text{cons}} = A_{\text{act}} J_{H_{2},c} = A_{\text{act}} \frac{i_{fc}}{2 F}
	\label{eq:dot{n}_{H_{2},cons}}
\end{equation}
where $\dot{n}$ ($mol.s^{-1}$) is the temporal derivative of the number of moles $n$.
\nomenclature[B,2]{$\dot{n}$}{temporal derivative of $n$ ($mol.s^{-1}$)}

In the simplified model, the inlet flow of hydrogen at the anode is selected to be a certain amount of time $\dot{n}_{H_{2},\text{cons}}$. This coefficient is the anode stoichiometric ratio of hydrogen : $S_{a}$.

\begin{equation}
	\dot{n}_{H_{2},\text{in}} = A_{\text{act}} \frac{S_{a} i_{fc}}{2 F}
\end{equation}

Using the ideal gas law and the definition of the relative humidity, a link between $n_{H_{2}}$ and $n_{H_{2}O}$ is obtained as \eqref{eq:frac{n_{H_{2}O}}{n_{H_{2}}}}

\begin{equation}
	\frac{n_{H_{2}O}}{n_{H_{2}}} = \frac{P_{H_{2}O}}{P_{H_{2}}} = \frac{\Phi_{a} P_{sat}}{P - \Phi_{a} P_{sat}}
	\label{eq:frac{n_{H_{2}O}}{n_{H_{2}}}}
\end{equation}

Thus, assuming the humidity of the incoming gases is automatically adjusted to the desired humidity $\Phi_{des}$ and approximating the pressure of the incoming gases as the pressure of the gases in the GC inlet $P_{agc,in}$, while neglecting pressure losses, equation \eqref{eq:dot{n}_{H_{2}O,in}} is derived.

\begin{equation}
	\dot{n}_{H_{2}O,in} = \frac{\Phi_{a,des} P_{sat}}{P_{agc,in} - \Phi_{a,des} P_{sat}} A_{act} \frac{S_{a} \left[i_{fc} + i_{n}\right]}{2 F} 
	\label{eq:dot{n}_{H_{2}O,in}}
\end{equation}

Finally, the simplified flow of water at the inlet of the AGC is given in \eqref{eq:J_{v,in}^{agc}}. 

\begin{equation}
	\begin{split}
		J_{v,in}^{agc} & = \frac{\dot{n}_{H_{2}0,in}}{H_{gc} W_{gc}} \\
		& = \frac{\Phi_{a,des} P_{sat}}{P_{agc,in} - \Phi_{a,des} P_{sat}} \frac{A_{act}}{H_{gc} W_{gc}} \frac{S_{a} \left[i_{fc} + i_{n}\right]}{2 F} 
	\end{split} 
	\label{eq:J_{v,in}^{agc}}
\end{equation}

\subsection{Simplified flows at the inlet of the CGC}
\label{subsec:simplified_flows_inlet_outlet_CGC}

The consumed molar rate of oxygen is given by equation \eqref{eq:dot{n}_{O_{2},cons}}.

\begin{equation}
	\dot{n}_{O_{2},\text{cons}} = \frac{i_{fc}}{4 F} A_{\text{act}}
	\label{eq:dot{n}_{O_{2},cons}}
\end{equation}

In this model, the cathode inlet oxygen flow is chosen to be proportional to $\dot{n}{0{2},\text{cons}}$. The proportionality coefficient is denoted as the cathode stoichiometric ratio of oxygen, represented by $S_{c}$. Thus, equation \eqref{eq:dot{n}_{O_{2},in}} is derived.

\begin{equation}
	\dot{n}_{O_{2},\text{in}} = \frac{S_{c} i_{fc}}{4 F} A_{\text{act}}
	\label{eq:dot{n}_{O_{2},in}}
\end{equation}

Using the ideal gas law and the definition of the relative humidity, a link between the dry air $n_{a}$, composed of $y_{O_{2},in} = 20.95 \% $ of $O_{2}$ and $79.05 \%$ of $N_{2}$, and $n_{H_{2}O}$ is obtained as \eqref{eq:frac{n_{H_{2}O}}{n_{a}}}.

\begin{equation}
	\frac{n_{H_{2}O}}{n_{a}} = \frac{P_{H_{2}O}}{P_{a}} = \frac{\Phi_{c} P_{sat}}{P - \Phi_{c} P_{sat}}
	\label{eq:frac{n_{H_{2}O}}{n_{a}}}
\end{equation}

Moreover, by definition of the molar fraction of oxygen in dry air, is given in \eqref{eq:y_{O_{2}}}.

\begin{equation}
	y_{O_{2}} = \frac{n_{O_{2}}}{n_{a}}
	\label{eq:y_{O_{2}}}
\end{equation}

Thus, assuming the humidity of the incoming gases is automatically adjusted to the desired humidity $\Phi_{des}$ and approximating the pressure of the incoming gases as the pressure of the gases in the GC inlet $P_{cgc,in}$, while neglecting pressure losses, equation \eqref{eq:dot{n}_{H_{2}O,in}_c} is derived.

\begin{equation}
	\dot{n}_{H_{2}O,in} = \frac{\Phi_{c,des} P_{sat}}{P_{cgc,in} - \Phi_{c,des} P_{sat}} \frac{1}{y_{O_{2},ext}} A_{act} \frac{S_{c} \left[i_{fc} + i_{n}\right]}{4 F}
	\label{eq:dot{n}_{H_{2}O,in}_c}
\end{equation}

Finally, the simplified flow of water at the inlet of the CGC is given in \eqref{eq:J_{v,in}^{cgc}}.

\begin{equation}
	\begin{split}
		J_{v,in}^{cgc} & = \frac{\dot{n}_{H_{2}0,in}}{H_{gc} W_{gc}} \\
		& = \frac{\Phi_{c,des} P_{sat}}{P_{cgc,in} - \Phi_{c,des} P_{sat}} \frac{1}{y_{O_{2},ext}} \frac{A_{act}}{H_{gc} W_{gc}} \frac{S_{c} \left[i_{fc} + i_{n}\right]}{4 F}
	\end{split}
	\label{eq:J_{v,in}^{cgc}}
\end{equation}


\bibliographystyle{elsarticle-num}
\bibliography{PEMFC_review_material_transport}

\begin{thebibliography}{10}
\expandafter\ifx\csname url\endcsname\relax
  \def\url#1{\texttt{#1}}\fi
\expandafter\ifx\csname urlprefix\endcsname\relax\def\urlprefix{URL }\fi
\expandafter\ifx\csname href\endcsname\relax
  \def\href#1#2{#2} \def\path#1{#1}\fi

\bibitem{FutureHydrogenSeizing2019}
The {{Future}} of {{Hydrogen}}. {{Seizing Today}}'s {{Opportunities}}, Tech. rep., IEA (Jun. 2019).

\bibitem{wangReviewPolymerElectrolyte2011}
Y.~Wang, K.~S. Chen, J.~Mishler, S.~C. Cho, X.~C. Adroher, A {{Review}} of {{Polymer Electrolyte Membrane Fuel Cells}}: {{Technology}}, {{Applications}}, and {{Needs}} on {{Fundamental Research}}, Applied Energy 88~(4) (2011) 981--1007.
\newblock \href {https://doi.org/10.1016/j.apenergy.2010.09.030} {\path{doi:10.1016/j.apenergy.2010.09.030}}.

\bibitem{ParisAgreementUnited}
The {{Paris Agreement}}, Tech. rep., United Nations (Dec. 2015).

\bibitem{StrategieNationalePour2020}
Strat{\'e}gie {{Nationale Pour Le D{\'e}veloppement}} de l'hydrog{\`e}ne {{D{\'e}carbon{\'e} En France}}, Dossier de {{Presse}}, Gouvernement fran{\c c}ais (Sep. 2020).

\bibitem{dicksFuelCellSystems2018}
A.~Dicks, D.~A.~J. Rand, Fuel {{Cell Systems Explained}}, third edition Edition, Wiley, Hoboken, NJ, 2018.

\bibitem{jiaoDesigningNextGeneration2021}
K.~Jiao, J.~Xuan, Q.~Du, Z.~Bao, B.~Xie, B.~Wang, Y.~Zhao, L.~Fan, H.~Wang, Z.~Hou, S.~Huo, N.~P. Brandon, Y.~Yin, M.~D. Guiver, Designing the next {{Generation}} of {{Proton-Exchange Membrane Fuel Cells}}, Nature 595~(7867) (2021) 361--369.
\newblock \href {https://doi.org/10.1038/s41586-021-03482-7} {\path{doi:10.1038/s41586-021-03482-7}}.

\bibitem{jiaoWaterTransportPolymer2011}
K.~Jiao, X.~Li, Water {{Transport}} in {{Polymer Electrolyte Membrane Fuel Cells}}, Progress in Energy and Combustion Science 37~(3) (2011) 221--291.
\newblock \href {https://doi.org/10.1016/j.pecs.2010.06.002} {\path{doi:10.1016/j.pecs.2010.06.002}}.

\bibitem{ohayreFuelCellFundamentals2016}
R.~P. O'Hayre, S.-W. Cha, W.~G. Colella, F.~B. Prinz, Fuel {{Cell Fundamentals}}, third edition Edition, Wiley, Hoboken, New Jersey, 2016.

\bibitem{pukrushpanControlOrientedModelingAnalysis2004}
J.~T. Pukrushpan, H.~Peng, A.~G. Stefanopoulou, Control-{{Oriented Modeling}} and {{Analysis}} for {{Automotive Fuel Cell Systems}}, Journal of Dynamic Systems, Measurement, and Control 126~(1) (2004) 14--25.
\newblock \href {https://doi.org/10.1115/1.1648308} {\path{doi:10.1115/1.1648308}}.

\bibitem{weberTransportPolymerElectrolyteMembranes2004}
A.~Z. Weber, J.~Newman, Transport in {{Polymer-Electrolyte Membranes}}, Journal of The Electrochemical Society 151~(2) (2004) A311.
\newblock \href {https://doi.org/10.1149/1.1639157} {\path{doi:10.1149/1.1639157}}.

\bibitem{futterPhysicalModelingPolymerelectrolyte2018}
G.~A. Futter, P.~Gazdzicki, K.~A. Friedrich, A.~Latz, T.~Jahnke, Physical {{Modeling}} of {{Polymer-Electrolyte Membrane Fuel Cells}}: {{Understanding Water Management}} and {{Impedance Spectra}}, Journal of Power Sources 391 (2018) 148--161.
\newblock \href {https://doi.org/10.1016/j.jpowsour.2018.04.070} {\path{doi:10.1016/j.jpowsour.2018.04.070}}.

\bibitem{newmanElectrochemicalSystems3rd2004}
J.~Newman, K.~E. {Thomas-Alyea}, Electrochemical {{Systems}}, 3rd {{Edition}}, John Wiley \& Sons, 2004.

\bibitem{springerPolymerElectrolyteFuel1991}
T.~E. Springer, T.~A. Zawodzinski, S.~Gottesfeld, Polymer {{Electrolyte Fuel Cell Model}}, Journal of The Electrochemical Society 138~(8) (1991) 2334--2342.
\newblock \href {https://doi.org/10.1149/1.2085971} {\path{doi:10.1149/1.2085971}}.

\bibitem{kulikovskyQuasi3DModelingWater2003}
A.~A. Kulikovsky, Quasi-{{3D Modeling}} of {{Water Transport}} in {{Polymer Electrolyte Fuel Cells}}, Journal of The Electrochemical Society 150~(11) (2003) A1432.
\newblock \href {https://doi.org/10.1149/1.1611489} {\path{doi:10.1149/1.1611489}}.

\bibitem{xuReduceddimensionDynamicModel2021}
L.~Xu, Z.~Hu, C.~Fang, L.~Xu, J.~Li, M.~Ouyang, A {{Reduced-dimension Dynamic Model}} of a {{Proton-exchange Membrane Fuel Cell}}, International Journal of Energy Research 45~(12) (2021) 18002--18017.
\newblock \href {https://doi.org/10.1002/er.6945} {\path{doi:10.1002/er.6945}}.

\bibitem{huAnalyticalCalculationEvaluation2016}
J.~Hu, J.~Li, L.~Xu, F.~Huang, M.~Ouyang, Analytical {{Calculation}} and {{Evaluation}} of {{Water Transport}} through a {{Proton Exchange Membrane Fuel Cell Based}} on a {{One-Dimensional Model}}, Energy 111 (2016) 869--883.
\newblock \href {https://doi.org/10.1016/j.energy.2016.06.020} {\path{doi:10.1016/j.energy.2016.06.020}}.

\bibitem{xingNumericalAnalysisOptimum2015}
L.~Xing, P.~K. Das, X.~Song, M.~Mamlouk, K.~Scott, Numerical {{Analysis}} of the {{Optimum Membrane}}/{{Ionomer Water Content}} of {{PEMFCs}}: {{The Interaction}} of {{Nafion}}{\textregistered} {{Ionomer Content}} and {{Cathode Relative Humidity}}, Applied Energy 138 (2015) 242--257.
\newblock \href {https://doi.org/10.1016/j.apenergy.2014.10.011} {\path{doi:10.1016/j.apenergy.2014.10.011}}.

\bibitem{yangMatchingWaterTemperature2011}
X.-G. Yang, Q.~Ye, P.~Cheng, Matching of {{Water}} and {{Temperature Fields}} in {{Proton Exchange Membrane Fuel Cells}} with {{Non-Uniform Distributions}}, International Journal of Hydrogen Energy 36~(19) (2011) 12524--12537.
\newblock \href {https://doi.org/10.1016/j.ijhydene.2011.07.014} {\path{doi:10.1016/j.ijhydene.2011.07.014}}.

\bibitem{karimiRecentApproachesImprove2019}
M.~B. Karimi, F.~Mohammadi, K.~Hooshyari, Recent {{Approaches}} to {{Improve Nafion Performance}} for {{Fuel Cell Applications}}: {{A Review}}, International Journal of Hydrogen Energy 44~(54) (2019) 28919--28938.
\newblock \href {https://doi.org/10.1016/j.ijhydene.2019.09.096} {\path{doi:10.1016/j.ijhydene.2019.09.096}}.

\bibitem{dickinsonModellingProtonConductiveMembrane2020}
E.~J.~F. Dickinson, G.~Smith, Modelling the {{Proton-Conductive Membrane}} in {{Practical Polymer Electrolyte Membrane Fuel Cell}} ({{PEMFC}}) {{Simulation}}: {{A Review}}, Membranes 10~(11) (2020) 310.
\newblock \href {https://doi.org/10.3390/membranes10110310} {\path{doi:10.3390/membranes10110310}}.

\bibitem{zawodzinskiDeterminationWaterDiffusion1991}
T.~A. Zawodzinski, M.~Neeman, L.~O. Sillerud, S.~Gottesfeld, Determination of {{Water Diffusion Coefficients}} in {{Perfluorosulfonate Ionomeric Membranes}}, The Journal of Physical Chemistry 95~(15) (1991) 6040--6044.
\newblock \href {https://doi.org/10.1021/j100168a060} {\path{doi:10.1021/j100168a060}}.

\bibitem{motupallyDiffusionWaterNafion2000}
S.~Motupally, A.~J. Becker, J.~W. Weidner, Diffusion of {{Water}} in {{Nafion}} 115 {{Membranes}}, Journal of The Electrochemical Society 147~(9) (2000) 3171.
\newblock \href {https://doi.org/10.1149/1.1393879} {\path{doi:10.1149/1.1393879}}.

\bibitem{wangModelingEffectsCapillary2008}
X.~Wang, T.~V. Nguyen, Modeling the {{Effects}} of {{Capillary Property}} of {{Porous Media}} on the {{Performance}} of the {{Cathode}} of a {{PEMFC}}, Journal of The Electrochemical Society 155~(11) (2008) B1085.
\newblock \href {https://doi.org/10.1149/1.2965512} {\path{doi:10.1149/1.2965512}}.

\bibitem{fanCharacteristicsPEMFCOperating2017}
L.~Fan, G.~Zhang, K.~Jiao, Characteristics of {{PEMFC Operating}} at {{High Current Density}} with {{Low External Humidification}}, Energy Conversion and Management 150 (2017) 763--774.
\newblock \href {https://doi.org/10.1016/j.enconman.2017.08.034} {\path{doi:10.1016/j.enconman.2017.08.034}}.

\bibitem{vanbusselDynamicModelSolid1998}
H.~P. L.~H. {van Bussel}, F.~G.~H. Koene, R.~K. A.~M. Mallant, Dynamic {{Model}} of {{Solid Polymer Fuel Cell Water Management}}, Journal of Power Sources 71~(1) (1998) 218--222.
\newblock \href {https://doi.org/10.1016/S0378-7753(97)02744-4} {\path{doi:10.1016/S0378-7753(97)02744-4}}.

\bibitem{pasaogullariTwoPhaseModelingFlooding2005}
U.~Pasaogullari, C.-Y. Wang, Two-{{Phase Modeling}} and {{Flooding Prediction}} of {{Polymer Electrolyte Fuel Cells}}, Journal of The Electrochemical Society 152~(2) (2005) A380.
\newblock \href {https://doi.org/10.1149/1.1850339} {\path{doi:10.1149/1.1850339}}.

\bibitem{hinatsuWaterUptakePerfluorosulfonic1994}
J.~T. Hinatsu, M.~Mizuhata, H.~Takenaka, Water {{Uptake}} of {{Perfluorosulfonic Acid Membranes}} from {{Liquid Water}} and {{Water Vapor}}, Journal of The Electrochemical Society 141~(6) (1994) 1493--1498.
\newblock \href {https://doi.org/10.1149/1.2054951} {\path{doi:10.1149/1.2054951}}.

\bibitem{geAbsorptionDesorptionTransport2005}
S.~Ge, X.~Li, B.~Yi, I.-M. Hsing, Absorption, {{Desorption}}, and {{Transport}} of {{Water}} in {{Polymer Electrolyte Membranes}} for {{Fuel Cells}}, Journal of The Electrochemical Society 152~(6) (2005) A1149.
\newblock \href {https://doi.org/10.1149/1.1899263} {\path{doi:10.1149/1.1899263}}.

\bibitem{zawodzinskijrCharacterizationPolymerElectrolytes1993}
T.~Zawodzinskijr, T.~Springer, F.~Uribe, S.~Gottesfeld, Characterization of {{Polymer Electrolytes}} for {{Fuel Cell Applications}}, Solid State Ionics 60~(1-3) (1993) 199--211.
\newblock \href {https://doi.org/10.1016/0167-2738(93)90295-E} {\path{doi:10.1016/0167-2738(93)90295-E}}.

\bibitem{yeThreeDimensionalSimulationLiquid2007}
Q.~Ye, T.~V. Nguyen, Three-{{Dimensional Simulation}} of {{Liquid Water Distribution}} in a {{PEMFC}} with {{Experimentally Measured Capillary Functions}}, Journal of The Electrochemical Society 154~(12) (2007) B1242.
\newblock \href {https://doi.org/10.1149/1.2783775} {\path{doi:10.1149/1.2783775}}.

\bibitem{baoTwodimensionalModelingPolymer2015}
C.~Bao, W.~G. Bessler, Two-{{Dimensional Modeling}} of a {{Polymer Electrolyte Membrane Fuel Cell}} with {{Long Flow Channel}}. {{Part I}}. {{Model Development}}, Journal of Power Sources 275 (2015) 922--934.
\newblock \href {https://doi.org/10.1016/j.jpowsour.2014.11.058} {\path{doi:10.1016/j.jpowsour.2014.11.058}}.

\bibitem{wuModelingPEMFCTransients2010}
H.~Wu, P.~Berg, X.~Li, Modeling of {{PEMFC Transients}} with {{Finite-Rate Phase-Transfer Processes}}, Journal of The Electrochemical Society 157~(1) (2010) B1.
\newblock \href {https://doi.org/10.1149/1.3248005} {\path{doi:10.1149/1.3248005}}.

\bibitem{mengTwophaseNonisothermalMixeddomain2007}
H.~Meng, A {{Two-Phase Non-Isothermal Mixed-Domain PEM Fuel Cell Model}} and {{Its Application}} to {{Two-Dimensional Simulations}}, Journal of Power Sources 168~(1) (2007) 218--228.
\newblock \href {https://doi.org/10.1016/j.jpowsour.2007.03.012} {\path{doi:10.1016/j.jpowsour.2007.03.012}}.

\bibitem{wuMathematicalModelingTransient2009}
H.~Wu, Mathematical {{Modeling}} of {{Transient Transport Phenomena}} in {{PEM Fuel Cells}}, Ph.D. thesis, University of Waterloo, Waterloo, Ontario, Canada (2009).

\bibitem{namNumericalAnalysisGas2010}
J.~Nam, P.~Chippar, W.~Kim, H.~Ju, Numerical {{Analysis}} of {{Gas Crossover Effects}} in {{Polymer Electrolyte Fuel Cells}} ({{PEFCs}}), Applied Energy 87~(12) (2010) 3699--3709.
\newblock \href {https://doi.org/10.1016/j.apenergy.2010.05.023} {\path{doi:10.1016/j.apenergy.2010.05.023}}.

\bibitem{whitakerMethodVolumeAveraging1999}
S.~Whitaker, The {{Method}} of {{Volume Averaging}}, Vol.~13 of Theory and {{Applications}} of {{Transport}} in {{Porous Media}}, Springer Netherlands, Dordrecht, 1999.
\newblock \href {https://doi.org/10.1007/978-94-017-3389-2} {\path{doi:10.1007/978-94-017-3389-2}}.

\bibitem{kimModelingTwophaseFlow2017}
J.~Kim, G.~Luo, C.-Y. Wang, Modeling {{Two-Phase Flow}} in {{Three-Dimensional Complex Flow-Fields}} of {{Proton Exchange Membrane Fuel Cells}}, Journal of Power Sources 365 (2017) 419--429.
\newblock \href {https://doi.org/10.1016/j.jpowsour.2017.09.003} {\path{doi:10.1016/j.jpowsour.2017.09.003}}.

\bibitem{tomadakisViscousPermeabilityRandom2005}
M.~M. Tomadakis, T.~J. Robertson, Viscous {{Permeability}} of {{Random Fiber Structures}}: {{Comparison}} of {{Electrical}} and {{Diffusional Estimates}} with {{Experimental}} and {{Analytical Results}}, Journal of Composite Materials 39~(2) (2005) 163--188.
\newblock \href {https://doi.org/10.1177/0021998305046438} {\path{doi:10.1177/0021998305046438}}.

\bibitem{fishmanHeterogeneousThroughPlaneDistributions2011}
Z.~Fishman, A.~Bazylak, Heterogeneous {{Through-Plane Distributions}} of {{Tortuosity}}, {{Effective Diffusivity}}, and {{Permeability}} for {{PEMFC GDLs}}, Journal of The Electrochemical Society 158~(2) (2011) B247.
\newblock \href {https://doi.org/10.1149/1.3524284} {\path{doi:10.1149/1.3524284}}.

\bibitem{baoTransportPropertiesGas2021}
Z.~Bao, Y.~Li, X.~Zhou, F.~Gao, Q.~Du, K.~Jiao, Transport {{Properties}} of {{Gas Diffusion Layer}} of {{Proton Exchange Membrane Fuel Cells}}: {{Effects}} of {{Compression}}, International Journal of Heat and Mass Transfer 178 (2021) 121608.
\newblock \href {https://doi.org/10.1016/j.ijheatmasstransfer.2021.121608} {\path{doi:10.1016/j.ijheatmasstransfer.2021.121608}}.

\bibitem{zamelEffectiveTransportProperties2013}
N.~Zamel, X.~Li, Effective {{Transport Properties}} for {{Polymer Electrolyte Membrane Fuel Cells}} -- {{With}} a {{Focus}} on the {{Gas Diffusion Layer}}, Progress in Energy and Combustion Science 39~(1) (2013) 111--146.
\newblock \href {https://doi.org/10.1016/j.pecs.2012.07.002} {\path{doi:10.1016/j.pecs.2012.07.002}}.

\bibitem{yimInfluenceStackClamping2010}
S.-D. Yim, B.-J. Kim, Y.-J. Sohn, Y.-G. Yoon, G.-G. Park, W.-Y. Lee, C.-S. Kim, Y.~C. Kim, The {{Influence}} of {{Stack Clamping Pressure}} on the {{Performance}} of {{PEM Fuel Cell Stack}}, Current Applied Physics 10~(2) (2010) S59--S61.
\newblock \href {https://doi.org/10.1016/j.cap.2009.11.042} {\path{doi:10.1016/j.cap.2009.11.042}}.

\bibitem{wangInvestigationDryIonomer2020}
Y.~Wang, T.~Liu, H.~Sun, W.~He, Y.~Fan, S.~Wang, Investigation of {{Dry Ionomer Volume Fraction}} in {{Cathode Catalyst Layer}} under {{Different Relative Humilities}} and {{Nonuniform Ionomer-Gradient Distributions}} for {{PEM Fuel Cells}}, Electrochimica Acta 353 (2020) 136491.
\newblock \href {https://doi.org/10.1016/j.electacta.2020.136491} {\path{doi:10.1016/j.electacta.2020.136491}}.

\bibitem{vargaftikInternationalTablesSurface1983}
N.~B. Vargaftik, B.~N. Volkov, L.~D. Voljak, International {{Tables}} of the {{Surface Tension}} of {{Water}}, Journal of Physical and Chemical Reference Data 12~(3) (1983) 817--820.
\newblock \href {https://doi.org/10.1063/1.555688} {\path{doi:10.1063/1.555688}}.

\bibitem{pasaogullariTwophaseTransportRole2004}
U.~Pasaogullari, C.-Y. Wang, Two-phase transport and the role of micro-porous layer in polymer electrolyte fuel cells, Electrochimica Acta 49~(25) (2004) 4359--4369.
\newblock \href {https://doi.org/10.1016/j.electacta.2004.04.027} {\path{doi:10.1016/j.electacta.2004.04.027}}.

\bibitem{wangMultiphaseMixtureModel}
C.~Y. Wang, P.~Cheng, A {{Multiphase Mixture Model}} for {{Multiphase}}, {{Multicomponent Transport}} in {{Capillary Porous Media I}}. {{Model Development}}, International Journal of Heat and Mass Transfer 39 (1996) 3607.
\newblock \href {https://doi.org/10.1016/0017-9310(96)00036-1} {\path{doi:10.1016/0017-9310(96)00036-1}}.

\bibitem{kotakaImpactInterfacialWater2014}
T.~Kotaka, Y.~Tabuchi, U.~Pasaogullari, C.-Y. Wang, Impact of {{Interfacial Water Transport}} in {{PEMFCs}} on {{Cell Performance}}, Electrochimica Acta 146 (2014) 618--629.
\newblock \href {https://doi.org/10.1016/j.electacta.2014.08.148} {\path{doi:10.1016/j.electacta.2014.08.148}}.

\bibitem{haoModelingExperimentalValidation2015}
L.~Hao, K.~Moriyama, W.~Gu, C.-Y. Wang, Modeling and {{Experimental Validation}} of {{Pt Loading}} and {{Electrode Composition Effects}} in {{PEM Fuel Cells}}, Journal of The Electrochemical Society 162~(8) (2015) F854--F867.
\newblock \href {https://doi.org/10.1149/2.0221508jes} {\path{doi:10.1149/2.0221508jes}}.

\bibitem{wangModelingTwophaseFlow2008}
Y.~Wang, S.~Basu, C.-Y. Wang, Modeling {{Two-Phase Flow}} in {{PEM Fuel Cell Channels}}, Journal of Power Sources 179~(2) (2008) 603--617.
\newblock \href {https://doi.org/10.1016/j.jpowsour.2008.01.047} {\path{doi:10.1016/j.jpowsour.2008.01.047}}.

\bibitem{jiangNumericalModelingLiquid2014}
F.~Jiang, C.-Y. Wang, Numerical {{Modeling}} of {{Liquid Water Motion}} in a {{Polymer Electrolyte Fuel Cell}}, International Journal of Hydrogen Energy 39~(2) (2014) 942--950.
\newblock \href {https://doi.org/10.1016/j.ijhydene.2013.10.113} {\path{doi:10.1016/j.ijhydene.2013.10.113}}.

\bibitem{liPorousMediaModeling2015}
Y.~Li, S.-C. Yao, Porous {{Media Modeling}} of {{Microchannel Cooled Electronic Chips}} with {{Nonuniform Heating}}, Journal of Thermophysics and Heat Transfer 29~(4) (2015) 695--704.
\newblock \href {https://doi.org/10.2514/1.T4509} {\path{doi:10.2514/1.T4509}}.

\bibitem{jieliuPorousMediaModeling2014}
J.~Jie~Liu, H.~Zhang, S.~C. Yao, Y.~Li, Porous {{Media Modeling}} of {{Two-Phase Microchannel Cooling}} of {{Electronic Chips With Nonuniform Power Distribution}}, Journal of Electronic Packaging 136~(2) (2014) 021008.
\newblock \href {https://doi.org/10.1115/1.4027420} {\path{doi:10.1115/1.4027420}}.

\bibitem{imkePorousMediaSimplified2004}
U.~Imke, Porous {{Media Simplified Simulation}} of {{Single-}} and {{Two-Phase Flow Heat Transfer}} in {{Micro-Channel Heat Exchangers}}, Chemical Engineering Journal 101~(1-3) (2004) 295--302.
\newblock \href {https://doi.org/10.1016/j.cej.2003.10.012} {\path{doi:10.1016/j.cej.2003.10.012}}.

\bibitem{wangPorousMediumModel1994}
C.-Y. Wang, M.~Groll, S.~R{\"o}sler, C.-J. Tu, Porous {{Medium Model}} for {{Two-Phase Flow}} in {{Mini Channels}} with {{Applications}} to {{Micro Heat Pipes}}, Heat Recovery Systems and CHP 14~(4) (1994) 377--389.
\newblock \href {https://doi.org/10.1016/0890-4332(94)90041-8} {\path{doi:10.1016/0890-4332(94)90041-8}}.

\bibitem{sugumarThermalAnalysisInclined2006}
D.~Sugumar, K.-K. Tio, Thermal {{Analysis}} of {{Inclined Micro Heat Pipes}}, Journal of Heat Transfer 128~(2) (2006) 198--202.
\newblock \href {https://doi.org/10.1115/1.2137763} {\path{doi:10.1115/1.2137763}}.

\bibitem{tioThermalAnalysisMicro2000}
K.-K. Tio, C.~Y. Liu, K.~C. Toh, Thermal {{Analysis}} of {{Micro Heat Pipes Using}} a {{Porous-Medium Model}}, Heat and Mass Transfer 36~(1) (2000) 21--28.
\newblock \href {https://doi.org/10.1007/s002310050359} {\path{doi:10.1007/s002310050359}}.

\bibitem{basuTwophaseFlowMaldistribution2009}
S.~Basu, J.~Li, C.-Y. Wang, Two-{{Phase Flow}} and {{Maldistribution}} in {{Gas Channels}} of a {{Polymer Electrolyte Fuel Cell}}, Journal of Power Sources 187~(2) (2009) 431--443.
\newblock \href {https://doi.org/10.1016/j.jpowsour.2008.11.039} {\path{doi:10.1016/j.jpowsour.2008.11.039}}.

\bibitem{wangModelingPolymerElectrolyte2005}
Y.~Wang, C.-Y. Wang, Modeling {{Polymer Electrolyte Fuel Cells}} with {{Large Density}} and {{Velocity Changes}}, Journal of The Electrochemical Society 152~(2) (2005) A445.
\newblock \href {https://doi.org/10.1149/1.1851059} {\path{doi:10.1149/1.1851059}}.

\bibitem{namMicroporousLayerWater2009}
J.~H. Nam, K.-J. Lee, G.-S. Hwang, C.-J. Kim, M.~Kaviany, Microporous {{Layer}} for {{Water Morphology Control}} in {{PEMFC}}, International Journal of Heat and Mass Transfer 52~(11-12) (2009) 2779--2791.
\newblock \href {https://doi.org/10.1016/j.ijheatmasstransfer.2009.01.002} {\path{doi:10.1016/j.ijheatmasstransfer.2009.01.002}}.

\bibitem{xieValidationMethodologyPEM2022}
B.~Xie, M.~Ni, G.~Zhang, X.~Sheng, H.~Tang, Y.~Xu, G.~Zhai, K.~Jiao, Validation {{Methodology}} for {{PEM Fuel Cell Three-Dimensional Simulation}}, International Journal of Heat and Mass Transfer 189 (2022) 122705.
\newblock \href {https://doi.org/10.1016/j.ijheatmasstransfer.2022.122705} {\path{doi:10.1016/j.ijheatmasstransfer.2022.122705}}.

\bibitem{zamelCorrelationEffectiveGas2009}
N.~Zamel, X.~Li, J.~Shen, Correlation for the {{Effective Gas Diffusion Coefficient}} in {{Carbon Paper Diffusion Media}}, Energy \& Fuels 23~(12) (2009) 6070--6078.
\newblock \href {https://doi.org/10.1021/ef900653x} {\path{doi:10.1021/ef900653x}}.

\bibitem{bultelInvestigationMassTransport2005}
Y.~Bultel, K.~Wiezell, F.~Jaouen, P.~Ozil, G.~Lindbergh, Investigation of {{Mass Transport}} in {{Gas Diffusion Layer}} at the {{Air Cathode}} of a {{PEMFC}}, Electrochimica Acta 51~(3) (2005) 474--488.
\newblock \href {https://doi.org/10.1016/j.electacta.2005.05.007} {\path{doi:10.1016/j.electacta.2005.05.007}}.

\bibitem{taineTransfertsThermiquesIntroduction2014}
J.~Taine, F.~Enguehard, E.~Iacona, {Transferts thermiques - Introduction aux transferts d'{\'e}nergie}, Dunod, 2014.

\bibitem{xuRobustControlInternal2017}
L.~Xu, J.~Hu, S.~Cheng, C.~Fang, J.~Li, M.~Ouyang, W.~Lehnert, Robust {{Control}} of {{Internal States}} in a {{Polymer Electrolyte Membrane Fuel Cell Air-Feed System}} by {{Considering Actuator Properties}}, International Journal of Hydrogen Energy 42~(18) (2017) 13171--13191.
\newblock \href {https://doi.org/10.1016/j.ijhydene.2017.03.191} {\path{doi:10.1016/j.ijhydene.2017.03.191}}.

\bibitem{shaoComparisonSelfHumidificationEffect2020}
Y.~Shao, L.~Xu, X.~Zhao, J.~Li, Z.~Hu, C.~Fang, J.~Hu, D.~Guo, M.~Ouyang, Comparison of {{Self-Humidification Effect}} on {{Polymer Electrolyte Membrane Fuel Cell}} with {{Anodic}} and {{Cathodic Exhaust Gas Recirculation}}, International Journal of Hydrogen Energy 45~(4) (2020) 3108--3122.
\newblock \href {https://doi.org/10.1016/j.ijhydene.2019.11.150} {\path{doi:10.1016/j.ijhydene.2019.11.150}}.

\bibitem{giner-sanzHydrogenCrossoverInternal2014}
J.~{Giner-Sanz}, E.~Ortega, V.~{P{\'e}rez-Herranz}, Hydrogen {{Crossover}} and {{Internal Short-Circuit Currents Experimental Characterization}} and {{Modelling}} in a {{Proton Exchange Membrane Fuel Cell}}, International Journal of Hydrogen Energy 39~(25) (2014) 13206--13216.
\newblock \href {https://doi.org/10.1016/j.ijhydene.2014.06.157} {\path{doi:10.1016/j.ijhydene.2014.06.157}}.

\bibitem{trucNumericalExperimentalInvestigation2018}
N.~T. Truc, S.~Ito, K.~Fushinobu, Numerical and {{Experimental Investigation}} on the {{Reactant Gas Crossover}} in a {{PEM Fuel Cell}}, International Journal of Heat and Mass Transfer 127 (2018) 447--456.
\newblock \href {https://doi.org/10.1016/j.ijheatmasstransfer.2018.07.092} {\path{doi:10.1016/j.ijheatmasstransfer.2018.07.092}}.

\bibitem{ahluwaliaBuildupNitrogenDirect2007}
R.~Ahluwalia, X.~Wang, Buildup of {{Nitrogen}} in {{Direct Hydrogen Polymer-Electrolyte Fuel Cell Stacks}}, Journal of Power Sources 171~(1) (2007) 63--71.
\newblock \href {https://doi.org/10.1016/j.jpowsour.2007.01.032} {\path{doi:10.1016/j.jpowsour.2007.01.032}}.

\bibitem{kochaCharacterizationGasCrossover2006}
S.~S. Kocha, J.~Deliang~Yang, J.~S. Yi, Characterization of {{Gas Crossover}} and {{Its Implications}} in {{PEM Fuel Cells}}, AIChE Journal 52~(5) (2006) 1916--1925.
\newblock \href {https://doi.org/10.1002/aic.10780} {\path{doi:10.1002/aic.10780}}.

\bibitem{baikCharacterizationNitrogenGas2011}
K.~D. Baik, M.~S. Kim, Characterization of {{Nitrogen Gas Crossover}} through the {{Membrane}} in {{Proton-Exchange Membrane Fuel Cells}}, International Journal of Hydrogen Energy 36~(1) (2011) 732--739.
\newblock \href {https://doi.org/10.1016/j.ijhydene.2010.09.046} {\path{doi:10.1016/j.ijhydene.2010.09.046}}.

\bibitem{yangEffectsOperatingConditions2019}
Z.~Yang, Q.~Du, Z.~Jia, C.~Yang, K.~Jiao, Effects of {{Operating Conditions}} on {{Water}} and {{Heat Management}} by a {{Transient Multi-Dimensional PEMFC System Model}}, Energy 183 (2019) 462--476.
\newblock \href {https://doi.org/10.1016/j.energy.2019.06.148} {\path{doi:10.1016/j.energy.2019.06.148}}.

\bibitem{zihrulVoltageCyclingInduced2016}
P.~Zihrul, I.~Hartung, S.~Kirsch, G.~Huebner, F.~Hasch{\'e}, H.~A. Gasteiger, Voltage {{Cycling Induced Losses}} in {{Electrochemically Active Surface Area}} and in {{H}} {\textsubscript{2}} /{{Air-Performance}} of {{PEM Fuel Cells}}, Journal of The Electrochemical Society 163~(6) (2016) F492--F498.
\newblock \href {https://doi.org/10.1149/2.0561606jes} {\path{doi:10.1149/2.0561606jes}}.

\bibitem{bardElectrochemicalMethodsFundamentals2001}
A.~J. Bard, L.~R. Faulkner, Electrochemical Methods: Fundamentals and Applications, 2nd Edition, Wiley, New York, 2001.

\bibitem{dickinsonButlerVolmerEquationPolymer2019}
E.~J.~F. Dickinson, G.~Hinds, The {{Butler-Volmer Equation}} for {{Polymer Electrolyte Membrane Fuel Cell}} ({{PEMFC}}) {{Electrode Kinetics}}: {{A Critical Discussion}}, Journal of The Electrochemical Society 166~(4) (2019) F221--F231.
\newblock \href {https://doi.org/10.1149/2.0361904jes} {\path{doi:10.1149/2.0361904jes}}.

\bibitem{kimDegradationModelingOperational2014}
J.~Kim, M.~Kim, T.~Kang, Y.-J. Sohn, T.~Song, K.~H. Choi, Degradation modeling and operational optimization for improving the lifetime of high-temperature {{PEM}} (proton exchange membrane) fuel cells, Energy 66 (2014) 41--49.
\newblock \href {https://doi.org/10.1016/j.energy.2013.08.053} {\path{doi:10.1016/j.energy.2013.08.053}}.

\bibitem{wangQuasi2DTransientModel2018}
B.~Wang, K.~Wu, Z.~Yang, K.~Jiao, A {{Quasi-2D Transient Model}} of {{Proton Exchange Membrane Fuel Cell}} with {{Anode Recirculation}}, Energy Conversion and Management 171 (2018) 1463--1475.
\newblock \href {https://doi.org/10.1016/j.enconman.2018.06.091} {\path{doi:10.1016/j.enconman.2018.06.091}}.

\bibitem{khajeh-hosseini-dalasmParametricStudyCathode2010}
N.~{Khajeh-Hosseini-Dalasm}, M.~J. Kermani, D.~G. Moghaddam, J.~M. Stockie, A parametric study of cathode catalyst layer structural parameters on the performance of a {{PEM}} fuel cell, international journal of hydrogen energy (2010).

\bibitem{hamnettComponentsElectrochemicalCell2010}
A.~Hamnett, The Components of an Electrochemical Cell, 1st Edition, Wiley, 2010.
\newblock \href {https://doi.org/10.1002/9780470974001.f101001} {\path{doi:10.1002/9780470974001.f101001}}.

\bibitem{zhaoReviewPhysicsbasedDatadriven2021}
J.~Zhao, X.~Li, C.~Shum, J.~McPhee, A {{Review}} of physics-based and data-driven models for real-time control of polymer electrolyte membrane fuel cells, Energy and AI 6 (2021) 100114.
\newblock \href {https://doi.org/10.1016/j.egyai.2021.100114} {\path{doi:10.1016/j.egyai.2021.100114}}.

\bibitem{darlingKineticModelPlatinum2003}
R.~M. Darling, J.~P. Meyers, Kinetic {{Model}} of {{Platinum Dissolution}} in {{PEMFCs}}, Journal of The Electrochemical Society 150~(11) (2003) A1523.
\newblock \href {https://doi.org/10.1149/1.1613669} {\path{doi:10.1149/1.1613669}}.

\bibitem{rinaldoCatalystDegradationNanoparticle2013}
S.~G. Rinaldo, W.~Lee, J.~Stumper, M.~Eikerling, Catalyst {{Degradation}}: {{Nanoparticle Population Dynamics}} and {{Kinetic Processes}}, ECS Transactions 50~(2) (2013) 1505--1513.
\newblock \href {https://doi.org/10.1149/05002.1505ecst} {\path{doi:10.1149/05002.1505ecst}}.

\bibitem{baroodyPredictingPlatinumDissolution2021}
H.~A. Baroody, E.~Kjeang, Predicting {{Platinum Dissolution}} and {{Performance Degradation}} under {{Drive Cycle Operation}} of {{Polymer Electrolyte Fuel Cells}}, Journal of The Electrochemical Society 168~(4) (2021) 044524.
\newblock \href {https://doi.org/10.1149/1945-7111/abf5aa} {\path{doi:10.1149/1945-7111/abf5aa}}.

\bibitem{santarelliParametersEstimationPEM2006}
M.~Santarelli, M.~Torchio, P.~Cochis, Parameters {{Estimation}} of a {{PEM Fuel Cell Polarization Curve}} and {{Analysis}} of {{Their Behavior}} with {{Temperature}}, Journal of Power Sources 159~(2) (2006) 824--835.
\newblock \href {https://doi.org/10.1016/j.jpowsour.2005.11.099} {\path{doi:10.1016/j.jpowsour.2005.11.099}}.

\bibitem{williamsAnalysisPolarizationCurves}
M.~V. Williams, H.~R. Kunz, J.~M. Fenton, Analysis of {{Polarization Curves}} to {{Evaluate Polarization Sources}} in {{Hydrogen}}/{{Air PEM Fuel Cells}}, Journal of The Electrochemical Society 152~(3) (2005) A635.
\newblock \href {https://doi.org/10.1149/1.1860034} {\path{doi:10.1149/1.1860034}}.

\bibitem{ramousseModellingHeatMass2005}
J.~Ramousse, J.~Deseure, O.~Lottin, S.~Didierjean, D.~Maillet, Modelling of {{Heat}}, {{Mass}} and {{Charge Transfer}} in a {{PEMFC Single Cell}}, Journal of Power Sources 145~(2) (2005) 416--427.
\newblock \href {https://doi.org/10.1016/j.jpowsour.2005.01.067} {\path{doi:10.1016/j.jpowsour.2005.01.067}}.

\bibitem{neyerlinCathodeCatalystUtilization2007}
K.~C. Neyerlin, W.~Gu, J.~Jorne, A.~Clark, H.~A. Gasteiger, Cathode {{Catalyst Utilization}} for the {{ORR}} in a {{PEMFC}}, Journal of The Electrochemical Society 154~(2) (2007) B279.
\newblock \href {https://doi.org/10.1149/1.2400626} {\path{doi:10.1149/1.2400626}}.

\bibitem{makhariaMeasurementCatalystLayer2005}
R.~Makharia, M.~F. Mathias, D.~R. Baker, Measurement of {{Catalyst Layer Electrolyte Resistance}} in {{PEFCs Using Electrochemical Impedance Spectroscopy}}, Journal of The Electrochemical Society 152~(5) (2005) A970.
\newblock \href {https://doi.org/10.1149/1.1888367} {\path{doi:10.1149/1.1888367}}.

\bibitem{meyerSituOperandoCharacterization2019}
Q.~Meyer, Y.~Zeng, C.~Zhao, In {{Situ}} and {{Operando Characterization}} of {{Proton Exchange Membrane Fuel Cells}}, Advanced Materials (2019).

\bibitem{weberCriticalReviewModeling2014}
A.~Z. Weber, R.~L. Borup, R.~M. Darling, P.~K. Das, T.~J. Dursch, W.~Gu, D.~Harvey, A.~Kusoglu, S.~Litster, M.~M. Mench, R.~Mukundan, J.~P. Owejan, J.~G. Pharoah, M.~Secanell, I.~V. Zenyuk, A {{Critical Review}} of {{Modeling Transport Phenomena}} in {{Polymer-Electrolyte Fuel Cells}}, Journal of The Electrochemical Society 161~(12) (2014) F1254.
\newblock \href {https://doi.org/10.1149/2.0751412jes} {\path{doi:10.1149/2.0751412jes}}.

\bibitem{zhangMultiphaseModelsWater2018}
G.~Zhang, K.~Jiao, Multi-phase models for water and thermal management of proton exchange membrane fuel cell: {{A}} review, Journal of Power Sources 391 (2018) 120--133.
\newblock \href {https://doi.org/10.1016/j.jpowsour.2018.04.071} {\path{doi:10.1016/j.jpowsour.2018.04.071}}.

\bibitem{karniadakisPhysicsinformedMachineLearning2021}
G.~E. Karniadakis, I.~G. Kevrekidis, L.~Lu, P.~Perdikaris, S.~Wang, L.~Yang, Physics-informed machine learning, Nature Reviews Physics 3~(6) (2021) 422--440.
\newblock \href {https://doi.org/10.1038/s42254-021-00314-5} {\path{doi:10.1038/s42254-021-00314-5}}.

\bibitem{shuklaPhysicsinformedNeuralNetwork2020}
K.~Shukla, P.~C. Di~Leoni, J.~Blackshire, D.~Sparkman, G.~E. Karniadakis, Physics-{{Informed Neural Network}} for {{Ultrasound Nondestructive Quantification}} of {{Surface Breaking Cracks}}, Journal of Nondestructive Evaluation 39~(3) (2020) 61.
\newblock \href {https://doi.org/10.1007/s10921-020-00705-1} {\path{doi:10.1007/s10921-020-00705-1}}.

\bibitem{chenPhysicsinformedMachineLearning2021}
W.~Chen, Q.~Wang, J.~S. Hesthaven, C.~Zhang, Physics-{{Informed Machine Learning}} for {{Reduced-Order Modeling}} of {{Nonlinear Problems}}, Journal of Computational Physics 446 (2021) 110666.
\newblock \href {https://doi.org/10.1016/j.jcp.2021.110666} {\path{doi:10.1016/j.jcp.2021.110666}}.

\bibitem{dingApplicationMachineLearning2022}
R.~Ding, S.~Zhang, Y.~Chen, Z.~Rui, K.~Hua, Y.~Wu, X.~Li, X.~Duan, X.~Wang, J.~Li, J.~Liu, Application of {{Machine Learning}} in {{Optimizing Proton Exchange Membrane Fuel Cells}}: {{A Review}}, Energy and AI 9 (2022) 100170.
\newblock \href {https://doi.org/10.1016/j.egyai.2022.100170} {\path{doi:10.1016/j.egyai.2022.100170}}.

\bibitem{mayurLifetimePredictionPolymer2018}
M.~Mayur, M.~Gerard, P.~Schott, W.~G. Bessler, Lifetime {{Prediction}} of a {{Polymer Electrolyte Membrane Fuel Cell}} under {{Automotive Load Cycling Using}} a {{Physically-Based Catalyst Degradation Model}}, Energies 11~(8) (2018) 2054.
\newblock \href {https://doi.org/10.3390/en11082054} {\path{doi:10.3390/en11082054}}.

\bibitem{jahnkePerformanceDegradationProton2016}
T.~Jahnke, Performance and {{Degradation}} of {{Proton Exchange Membrane Fuel Cells}}: {{State}} of the {{Art}} in {{Modeling}} from {{Atomistic}} to {{System Scale}}, Journal of Power Sources (2016).

\bibitem{FuelCellsHydrogen2018}
Fuel {{Cells}} and {{Hydrogen}} 2 {{Joint Undertaking}} ({{FCH}} 2 {{JU}}). {{Addendum}} to the {{Multi}} - {{Annual Work Plan}} 2014 - 2020 ({{FCH}} 2 {{JU}}, 2018), Tech. rep., European Union (2018).

\bibitem{kellDensityThermalExpansivity1975}
G.~S. Kell, Density, {{Thermal Expansivity}}, and {{Compressibility}} of {{Liquid Water}} from 0.{{Deg}}. to 150.{{Deg}}.. {{Correlations}} and {{Tables}} for {{Atmospheric Pressure}} and {{Saturation Reviewed}} and {{Expressed}} on 1968 {{Temperature Scale}}, Journal of Chemical \& Engineering Data 20~(1) (1975) 97--105.
\newblock \href {https://doi.org/10.1021/je60064a005} {\path{doi:10.1021/je60064a005}}.

\bibitem{neumanTheoreticalDerivationDarcy1977}
S.~P. Neuman, Theoretical {{Derivation}} of {{Darcy}}'s {{Law}}, Acta Mechanica 25~(3-4) (1977) 153--170.
\newblock \href {https://doi.org/10.1007/BF01376989} {\path{doi:10.1007/BF01376989}}.

\end{thebibliography}
\biboptions{sort&compress}

\end{document}